\documentclass[useAMS,usenatbib]{mn2e}

\usepackage{amsmath}
\usepackage{amsfonts}
\usepackage{graphics}
\usepackage{graphicx}
\usepackage{subfigure}
\usepackage{relsize}
\usepackage{xcolor}
\usepackage{framed}

\newcommand{\etal}{et al.}
\newcommand{\be}{\begin{equation}}
\newcommand{\ee}{\end{equation}}
\newcommand{\ba}{\begin{eqnarray}}
\newcommand{\ea}{\end{eqnarray}}
\newcommand{\eg}{e.g.}
\newcommand{\ie}{i.e.~}

\renewcommand{\vec}[1]{\boldsymbol{#1}}
\renewcommand{\d}{\textrm{d}}
\newcommand{\grad}{\vec{\nabla}}

\newcommand{\gradq}{\grad_{\!\!q}}
\newcommand{\D}{D_{\!\mathsmaller{\mathsmaller{+} } }}

\title[The Zeldovich approximation: \\ \ key to understanding Cosmic Web complexity]{The Zeldovich approximation: \\ key to understanding 
Cosmic Web complexity}
\author[Johan Hidding \etal]
{Johan Hidding$^*$, Sergei F. Shandarin$^\dagger$ and Rien van de Weygaert$^*$\\
$^*$Kapteyn Astronomical Institute, University of Groningen, Groningen, The Netherlands\\
$^\dagger$Department of Physics and Astronomy, University 
of Kansas, KS 66045, U.S.A.\\
emails: johannes.hidding@gmail.com\,;  sergei@ku.edu\,; weygaert@astro.rug.nl}

\begin{document}

\date{Accepted 2013 November 4.  Received 2013 November 2; in original form 2013 September 23}
\pagerange{\pageref{firstpage}--\pageref{lastpage}} \pubyear{2013}

\maketitle

\label{firstpage}

\begin{abstract}
  We describe how the dynamics of cosmic structure formation defines the
intricate geometric structure of the spine of the cosmic web. The Zel'dovich
approximation is used to model the backbone of the cosmic web in terms of its
singularity structure. The description by Arnold et al. (1982) in terms of
catastrophe theory forms the basis of our analysis.\\

  This two-dimensional analysis involves a profound assessment of the
Lagrangian and Eulerian projections of the gravitationally evolving
four-dimensional phase-space manifold. It involves the identification of the
complete family of singularity classes, and the corresponding caustics that we
see emerging as structure in Eulerian space evolves. In particular, as it is
instrumental in outlining the spatial network of the cosmic web, we investigate
the nature of spatial connections between these singularities.\\

  The major finding of our study is that all singularities are located on a set
of lines in Lagrangian space. All dynamical processes related to the caustics
are concentrated near these lines. We demonstrate and discuss extensively how
all 2D singularities are to be found on these lines. When mapping this spatial
pattern of lines to Eulerian space, we find a growing connectedness between
initially disjoint lines, resulting in a percolating network. In other words,
the lines form the blueprint for the global geometric evolution of the cosmic
web.

\end{abstract}


\section{Introduction}
Over the last three decades, the study of the large-scale structure has thrived 
in both observations and theory. This was the consequence of the new insights obtained 
via large new galaxy redshift surveys, the tremendous increase of computing power, 
the related emergence and advances of computer simulations, and the development 
of new analysis methods and sophisticated theoretical models of gravitationally 
driven structure formation. However, while an overall theoretical framework for 
the emergence and evolution of the large scale structure of the universe has been 
gradually taking shape, there is not yet a comprehensive theory for the formation 
and evolution of the intricate and complex structure of the cosmic web. 

\subsection{Historical overview}

\begin{figure*}
	\centering
	\centerline{\includegraphics[width=\textwidth]{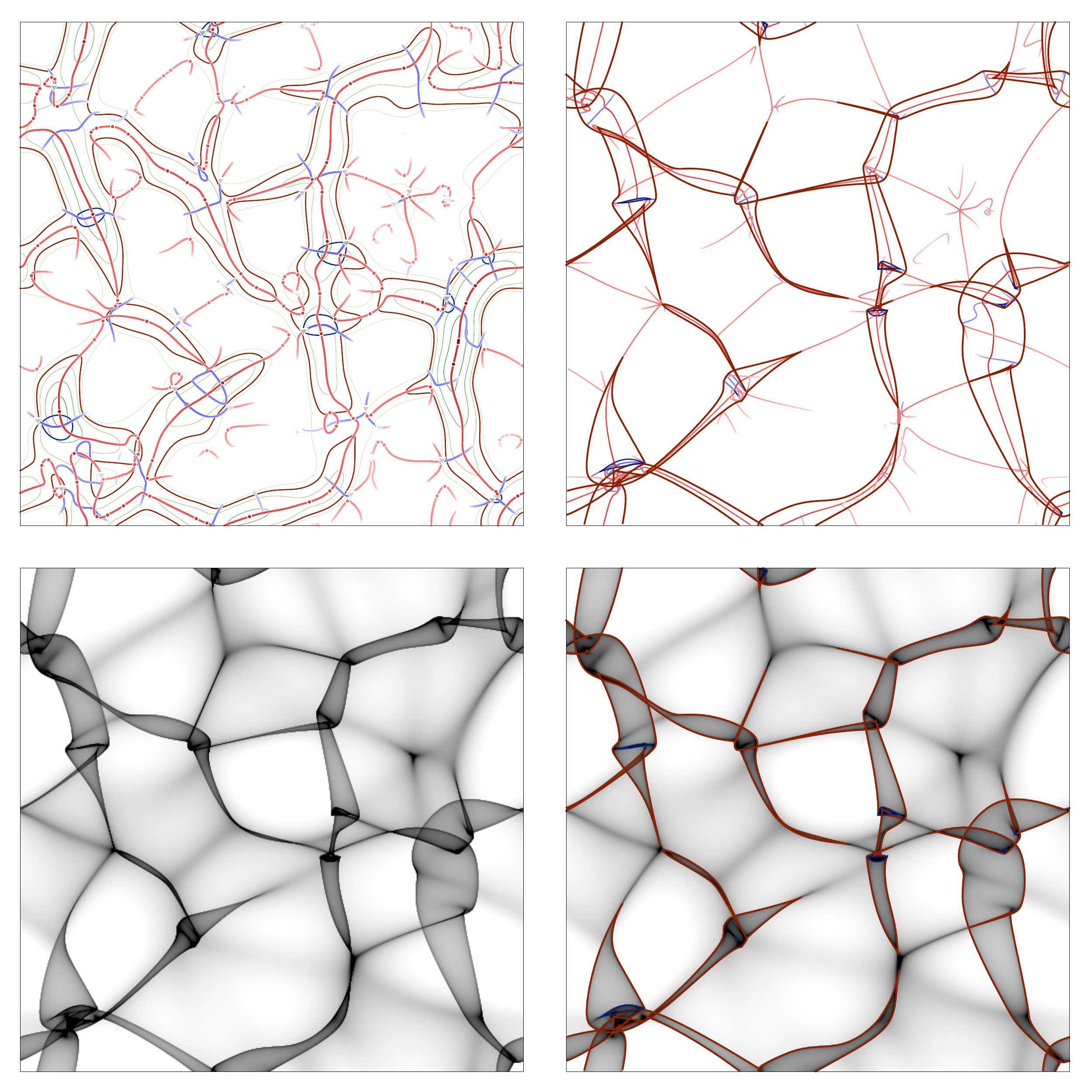}}
	\caption{Zeldovich Approximation: Lagrangian \& Eulerian Singularity
	Structure.  First panel: the Lagrangian skeleton that we find using the
	method presented in this paper.  In dark-red is the contour of the
	major eigenvalue corresponding to the time at which we plotted the
	Eulerian images that occupy the other three panels. Top-right panel:
	structural and singularity information from the first panel mapped to
	Eulerian space, as proscribed by the ZA.  Bottom lefthand panel:
	density distribution in Eulerian space. Bottom righthand panel: density
field in Eulerian space, with the singularity outline superimposed (red
edges).}
	\label{fig:overview}
\end{figure*}

Currently, there are a few approximate analytical models describing a range of 
features and characteristics of the cosmic web. It is in particular the theoretical 
framework of \cite{bkp-96}, responsible for coining the name "cosmic web", which has 
succeeded in combining the aspects of the anisotropic nature of gravitational collapse 
with the hierarchical nature of the process \citep[for a review, see][]{wb-08}. It incorporates 
the statistics of peaks in primordial Gaussian fields \cite{bbks-86} with a sophisticated nonlocal 
description of the hierarchical evolution of peaks by means of the peak-patch 
formalism \citep{bm-96a,bm-96b,bm-96c}. The latter models the nonlinear anisotropic 
collapse by means of the homogeneous ellipsoid model \citep{lin1965,icke1972,icke1973,whitesilk1979}, 
linking its linear stage to the Zel'dovich approximation (hereafter the ZA).

The former is supposed to work on small scales where density perturbations have
reached the nonlinear stage of structure formation, while on large scales the density
perturbations remain linear or quasilinear. 
The boundary separating the two approaches  is defined by a heuristic condition requiring that
the  multistream flows in the ZA involve only a small fraction of mass and 
thus can be neglected.

Zeldovich himself (1970) \citep[see also][]{sz-89} suggested to use the linear 
Lagrangian model until the multistream flows are well -- but not  overly --
developed  in the velocity field. The studies of \citet{cms-93} and \citet{mps-94} 
showed that the ZA can be safely used on all scales down to the scale of nonlinearity, 
$R_{nl}$, defined by the condition  $\sigma_{\delta}(R_{nl}) \approx 1$. 
By this stage most of pancakes have emerged, and still remain fairly thin. Up to 
this stage, the Zeldovich approximation describes conceptually correct, be it crudely, the 
emergence of multistream flow regions, caustics and the structural skeleton 
of the Cosmic Web \citep{drs-73,sz-89,buchert89,b-92,cms-93,mps-94,msw-94,cs-95,mbw-95,
ymm-98,ymgm-06}.

An additional prominent application of the Zeldovich approximation is that of its use as 
a tool for generating initial conditions in the simulations of cosmological 
structures \citep{drs-73,ds-73,d-etal-80,ks-83}. Starting with these 
studies shortly after its inception, the Zeldovich approximation has become 
the major generator of the initial conditions in practically every cosmological 
N-body simulation since 1983.  
It also has been used in many theoretical studies of the
non-perturbative nonlinear stage.  However, its complete description and
explanation of what it exactly predicts has not been provided yet.  The first
step  in this direction  was made twelve years after Zeldovich had published
his paper. \citet{asz-82} described the topology of all generic singularities
formed in the two-dimensional case and \citet{a-82,a-83} found the  normal forms of
all generic singularities in three dimensions.  The two-dimensional study was
focused mostly on the local properties of archetypical singularities while the
global picture was described rather schematically.  The authors obviously did not
realize  the computational complexity of the problem when promised to do a
similar analysis in 3D in the following paper. However, impressive progress in
computational geometry and  computational graphics observed in the last couple
of decades allows to hope for a significant step forward in understanding
the geometry and topology of the structure in the ZA. 

\subsection{the Cosmic Skeleton}
One of the major new results of this study is the demonstration that a particular set of lines
and points in Lagrangian space form a progenitor of the skeleton of the structure
at the non-linear stage\footnote{The skeleton becomes a set of surfaces, edges and 
vertices in 3D.}. The first attempt to reduce the structure to a geometrical skeleton 
was done in the adhesion approximation \citep{gss-89,kps-90,gss-12} (also see \cite{hidding2013}). 
In the adhesion description, the skeleton was constructed from the linear gravitational potential 
by a geometrical procedure based on the approximate gravitational dynamics of the ZA. Later, 
\citet{ncd-06} and \citet{s-etal-08}, as well as \cite{aragon2010}, also discussed the skeleton or 
spine of the large-scale structure, be it in a rather different context. They constructed a skeleton 
from the known nonlinear density field on the basis of Morse-Smale theory. According to this 
description, the skeleton is traced by the separatrices that connect singularities in the 
density field.  While the former derives the skeleton from the linear state by using gravitational 
dynamics as approximated by the ZA, the latter is purely a topological construction derived from 
the given density field. 

Following a philosophy similar to that of our study, \cite{pogosyan2009} followed up on the work by 
\citet{s-etal-08} with a profound analysis of the critical lines in a density field. These 
were identified with those of the skeleton of the cosmic web. Our study sheds a different light on the 
skeleton of the cosmic web, by connecting it to the dynamical development of the 
density field. It acccomplishes this by focussing on the linear deformation field in the 
primordial density field. Our study demonstrates that the spatial structure of the cosmic web, 
as it emerges out of the primordial universe, should be identified with critical lines in 
Lagrangian space. These lines mark the regions near which all decisive dynamical processes  
take place.

\subsection{This paper}
The principal incentive for our study is the identification of the spine of the 
dynamically evolving cosmic web in the primordial density field. We argue that this 
is entirely determined by the set of singularity lines, along which we may trace 
the caustics that form in the matter distribution as its phase-space structure 
attains an increasingly complex multistream character. 

It is hardly feasible, or practical, to analyze the full nonlinear gravitationally 
evolved mass distribution such as modelled by cosmological N-body simulations. However, 
for our understanding of the geometric structure of the cosmic web we are greatly helped 
by the observation that the fundamental singularity and caustic structure of the cosmic 
web is basically entailed in the quasi-linear description of the Zeldovich approximation. 
Although the dynamics in the ZA is substantially simpler than that in cosmological 
N-body simulations, the ZA provides all the necessary concepts and language required for 
the analysis of the realistic dynamics of cosmological N-body simulations. It involves 
the concepts of phase space, Lagrangian submanifolds, multistream flows and caustics. 
This fact that the Zeldovich approximation combines its mathematical flexibility and 
simplicity with the full richness of concepts that we find in the more complex reality 
makes it unique among all theoretical models that seek to describe the structure and evolution 
of the cosmic web.


The main rationale behind our study is therefore that of an attempt to provide a 
comprehensive description of what the ZA actually predicts in terms of the dynamically 
evolving spine of the cosmic web. It may seem puzzling that in more than forty years
after its inception, the ZA would still require further discussions and explanations. 
Instead of presenting another discussion of the accuracy of the Zeldovich 
approximation, this paper has the intention to reveal new interesting and useful aspects 
of the ZA and demonstrate that the seemingly simplistic mathematical form of the ZA 
harbours a rich geometry and complex topology for the emerging structures it describes. 

There are a few characteristics of the ZA that makes its analysis and understanding
drastically different and more difficult than that of linearly evolving spatial random 
fields. A first important aspect is that the evaluation of the Eulerian density field in 
the ZA requires both the deformation tensor and the displacement vector fields (see 
sect.~\ref{sec:zeldovich}). Eulerian 
linear theory only requires one scalar field, the initial density field. The geometry 
and topology of vector and tensor fields are considerably more complex than that of 
scalar fields \citep{tricoche2002}. An important second aspect is that of the stochastic 
properties of the eigenvalues of the deformation tensor field. Even when its 
individual components have a Gaussian distribution, the eigenvalues have a non-Gaussian 
distribution in all senses \citep{dorosh1970}: geometrical, topological and in 
terms of their probability distribution function. An additional major complication is 
related to the generation of multistream flows. This requires a nonlocal analysis. 
We will address these issues in far greater detail than previous studies have done. 

In this study we restrict our analysis to the two-dimensional case. While it is considerably 
simpler than in 3D, the two-dimensional analysis still reveals many nontrivial features that also 
emerge in three dimensions. All basic concepts of two-dimensional theory remain the same in 3D, 
which renders the 2D analysis a useful and crucial step towards understanding the 
full three-dimensional case. The latter is the subject of an accompanying paper in 
preparation.

Aiming for the cosmological applications, we consider the evolution of a system
consisting of collisionless particles interacting only via gravity and having
an extremely low temperature. The examples of currently popular hypothetical 
dark matter particles are neutralinos and axions. For instance, if the mass of the universe 
is dominated by a weakly interacting massive particle with mass of 100 GeV, then there would 
be $\sim10^{67}$ such particles in the Milky Way alone. This implies that the evolution of the 
dark matter distribution in the universe can be accurately described by a model of a cold 
collisionless continuous medium. 

Reading this paper does not require previous knowledge of catastrophe theory or even the ZA. We 
provide a brief summary of the ZA in section 2, follow by an informal introduction to the theory of 
Lagrangian singularities in section \ref{sec:theoryofsing}. In the rest of the manuscript, we 
analyze the role of singularities in outlining the cosmic skeleton. Section~\ref{sec:singinza} 
describes the role of every generic singularity in the Zeldovich approximation. Subsequently, 
in section \ref{sec:blueprint} we explain how the cosmic web is assembled around the progenitor 
of the skeleton. Finally, in section \ref{sec:summ} we summarize and discuss the major results. 

\section{The Zel'dovich approximation}
\label{sec:zeldovich}
The Zeldovich approximation is particularly simple in comoving coordinates 
$\vec{x} = \vec{r}/a(t)$, where $a(t)$ is the scale factor and $\vec{r}$ are physical coordinates. 
It relates the initial Lagrangian coordinates $\vec{q}$ at $t \rightarrow 0$ 
and Eulerian coordinates $\vec{x}$ at time $t$ by an explicit relation
\begin{equation}
\label{eq:za}              
\vec{x}(\vec{q}, \D(t)) = \vec{q} + \D(t) \vec{s}(\vec{q})
\end{equation}
where the vector field $\vec{s}(\vec{q})$  is determined by the initial density
perturbations and $\D(t)$ is the linear density growth factor fully specified by the
cosmological parameters. 
The Zeldovich approximation assumes that the initial displacement field $\vec{s}(\vec{q})$ 
is a potential vector field, 
\begin{equation}
\vec{s}(\vec{q}) = - \gradq \Psi (\vec{q})\,,
\end{equation}
in which the scalar displacement potential $\Psi$ is directly proportional to the linearly 
extrapolated gravitational perturbation, 
\begin{equation}
\Psi({\vec{q}})=\frac{2}{3D a^2 H^2 \Omega}\,\phi_{lin}({\vec{q}})\,.
\end{equation}
The potential character of the deformation vector is a direct consequence of the potential 
character of the growing mode of the gravitational instability in an expanding universe. 
An additionally important aspect of the Zeldovich approximation, reflecting its Lagrangian 
nature, is the deformation of mass elements. The deformation is specified by the deformation 
tensor, 
\begin{equation}
d_{ij} = -\partial s_i/ \partial q_j\,.
\end{equation}

\bigskip
From equation~(\ref{eq:za}) one may easily infer an explicit expression for the density as a 
function of Lagrangian coordinates and time. If we consider the conservation of mass in 
differential form, 
\begin{equation}
\label{eq:mass_conserv}
\rho(\vec{x},t) \d\vec{x} = \bar{\rho}(t) \d\vec{q},
\end{equation}
the density evolution directly follows from 
\begin{equation}
\label{eq:J}
\rho(\vec{x},t) = \bar{\rho}(t)\, J\left[\frac{\partial \vec{x}}{\partial \vec{q}}\right]^{-1},
\end{equation}
where $J[\partial \vec{x}/\partial \vec{q}]$ is the Jacobian determinant of the map  given by 
equation \ref{eq:za}. It is convenient to write equation \ref{eq:J} in terms of the eigen values 
$\lambda_i=(\alpha,\beta)$ of the deformation tensor $d_{ij}$, 
\begin{equation}
\label{eq:den_eigen_val}   %
\rho(\vec{q},t) = {\bar{\rho}(t) \over (1-\D(t)\alpha(\vec{q}))(1-\D(t)\beta(\vec{q}))},
\end{equation}
where a commonly used ordering of the eigen values $\alpha \ge \beta$ is assumed. On the basis of 
this equation, we may immediately infer two key features of structure formation described by the 
Zeldovich approximation. The first one is that the density becomes infinite as soon as 
$D(t)\alpha =1$ or $D(t)\beta =1$. In addition, this collapse is nearly always anisotropic 
as in general the eigenvalues\footnote{In this study we will demonstrate that 
the equality of the eigenvalues $\alpha$ and $\beta$ only occurs at isolated points.}
$\alpha \ne \beta$.

It is easy to see that Eulerian linear perturbation theory (ELPT) is a limiting case of the 
ZA. Expanding the expression for the Lagrangian density (eqn.~\ref{eq:den_eigen_val}) 
in a Taylor series,
\begin{equation}
\rho(\vec{q},t) = \bar{\rho}(t)\,+\,\bar{\rho}(t)\D(t)\left(\alpha+\beta\right)\,+\,\ldots\,,
\end{equation}
clarifies the relation between the known linear expression for the density contrast, $\delta_{lin}$ , 
\begin{equation}
\label{eq:denelpt}
\delta_{\rm lin} := \frac{(\rho_{\rm lin} - \bar{\rho})}{\bar{\rho}} = \D(t)(\alpha+\beta). 
\end{equation}

Formally speaking, the ZA is valid only in the linear regime when both $|D\alpha| \ll 1$ and $|D \beta|\ll 1$. 
However, Zeldovich made the bold prediction that it might be a good qualitative and arguably even a quantitative 
approximation up to the beginning of the nonlinear stage, i.e. up to the stage at which $|D\alpha| \approx 1$. 
Numerous studies confirmed the conjecture and elaborated on the quantitative accuracy and limitations of the 
approximation \citep[][, also see references therein]{drs-73,sz-89,b-92,cms-93,mps-94,msw-94,cs-95,mbw-95,ymm-98,
ymgm-06}.

Technically speaking, the ZA is an extrapolation of Lagrangian linear perturbation theory (LLPT) beyond the range of 
its formal applicability. There are two fundamental differences of LLPT from its Eulerian counterpart ELPT. As may be 
inferred from equations~\ref{eq:den_eigen_val} and ~\ref{eq:denelpt}, the calculation of the density LLPT uses the 
full deformation tensor while ELPT relies only on its trace. The 
second difference is due to the necessity in LLPT of mapping from Lagrangian to Eulerian space in order to
evaluate the density field in Eulerian space. Even at small $\sigma_{\delta}$, where $\delta = \Delta \rho/\bar\rho$, 
the difference between LLPT and ELPT can be quite noticeable if the scale of the initial \ie linear velocity field is
considerably greater than that of density fluctuations \citep{p-80}. In this, we assume that $\sigma_{\delta}$ is 
evaluated in ELPT. The difference between LLPT and ELPT becomes considerable when both are extrapolated to 
a larger $\sigma_{\delta}$. A particularly obvious problem occurring in ELPT is the emergence of negative 
densities, i.e. $\rho <0$, in regions with a large initial density deficit. For example, if ELPT is extrapolated 
to  $\sigma_{\delta} = 0.5$, the regions with negative densities occupy approximately 2.3\% of the volume. 
This fraction increases to almost 15\% at $\sigma_{\delta} =1$. Evidently, for a physical model this is 
an unacceptable circumstance. LLPT is completely free of this problem: at all times it predicts $\rho > 0$, 
regardless of the magnitude of $\sigma_{\delta}$.

Equation~(\ref{eq:za}) encapsulates the mapping by the LLPT from Lagrangian space to Eulerian space. Its ramifications can 
be explored analytically before an overwhelming fraction of mass elements starts to experience shell crossing. The mathematical 
complications increase rapidly as the number and extent of the multistream flow regions proliferates. In particular interesting are 
those mass elements which are separated by finite distances in Lagrangian space and end up at the same places 
in Eulerian space. These caustics are key manifestations of the dynamically evolving mass distribution and mark the 
emerging cosmic web. To understand and assess this aspect of the Zeldovich approximation, we require numerical 
modeling, in particular for the cosmologically relevant situation of random initial conditions. 

\section{Theory of singularities}
\label{sec:theoryofsing}
Before we start our investigation of the spatial network defined by the 
Zeldovich approximation, we need an inventory of the individual 
singularities that may emerge as the cosmic mass distribution evolves 
into a weblike pattern. These singularities define the outline of the 
principal components of the cosmic web, while others arrange the spatial 
connections that line these components into a pervasive spatial network.

To be able to appreciate this structure, we need to build an inventory 
of the different caustic geometries that we encounter in the Zeldovich 
approximation. Having obtained insight into the character of the individual 
singularities, we will be equipped to appreciate their embedding in an 
extensive and more realistic network of structures. 

For a complete inventory of all possible singularities we need to resort 
to the mathematical branch of catastrophe theory. In this section we will 
present a systematic, yet schematic, inventory of 2D singularities. While 
involving a complex mathematical formalism, within the limited context 
of the present study we invoke images to elucidate the key aspects 
of catastrophe theory and to appreciate the basic character of each 
singularity. 

In the subsequent sections we will find how these caustic objects are the 
building stones of the evolving network of filaments and nodes. 

\subsection{Singularities and the Zeldovich approximation}
When considering the displacement of mass elements from their Lagrangian location 
$\vec{q}$ to their Eulerian position $\vec{x}$, according to the Zeldovich displacement 
equation \ref{eq:za}, we may imagine this in terms of the Zel'dovich approximation 
outlining a (hyper)-surface in a four-dimensional \emph{phase-space}. In this case, 
the four-dimensional phase-space $({\vec q},{\vec x})$ is defined by four coordinates, the 
Eulerian space $\vec{x}$ coordinates and the Lagrangian space $\vec{q}$ coordinates. 
The hypersurface ${\vec x}({\vec q})$ is called a Lagrangian submanifold. Note that this
definition of phase-space is different than the classic position-momentum phase-space.
The non-disipative nature of dark matter dynamics allows us to define a diffeomorphism 
(being a smooth and differentiable mapping that is both on-to and one-to-one) between 
these two concepts of phase-space throughout even the full non-linear dynamical evolution.
In this mapping all the properties of the projection to Eulerian space are completely
preserved.

The Lagrangian submanifold is a smooth single-valued function ${\vec x}({\vec q})$ of Lagrangian 
coordinates ${\vec q}$.  While our brains are generally unable to imagine such a four-dimensional 
manifold, for the sake of insight we may illustrate some of the basic qualities for the equivalent 
situation of one-dimensional space, i.e. for the equivalent 
two-dimensional phase-space. This you may find in appendix~A. 

When looking at the evolving structure in the Universe, we are looking at the 
projection of the full four-dimensional submanifold on to Eulerian space ${\vec x}$. The 
evolving structure is a direct reflection of the changing geometry of the Lagrangian submanifold, 
in terms the gradual formation of folds in the submanifold and the orientation of the manifold. 
The folding is a direct manifestation of the emergence of multistream regions, where the local 
number of folds is determined by the number of streams at that location. 

Locally, we may determine the resulting Eulerian density (perturbation) $\delta(\vec{x},t)$ 
by integrating over the various folds. If at location ${\vec x}$ we find an $N$ stream 
region, each corresponding to a solution ${\vec q}^{\star}$ for which 
${\vec x}={\vec x}({\vec q}^{\star},t)$, the Eulerian density is 
\begin{equation} 
\delta(\vec{x}, t) + 1 = \sum_{x = x(q^{\star}, t)} \left|\ \det \left(\frac{\partial 
\vec{x}}{\partial \vec{q}}\right)\ \right|^{-1}_{q=q^{\star}}. 
\end{equation}
where the sum is taken over all $N$ solutions. It is in the projection of the 
four-dimensional Lagrangian submanifold that we find the appearance of singularities. 
These appear if one or more solutions $\vec{q}={\vec q}^{\star}$ involve an accumulation 
of (Lagrangian) mass elements at the Eulerian location ${\vec x}$. This means that 
$\det \left(\frac{\partial \vec{x}}{\partial \vec{q}}\right)=0$, implying an infinite 
local mass density, $\delta(\vec{x}, t)=\infty$. 

The nature of the singularities that appear in the evolving cosmic mass distribution may 
vary. There is a limited number of different classes of singularities. These appear in 
different circumstances and configurations, and each play their role in outlining the 
spine of the cosmic web. 

In the following subsections we will present a systematic census of the singularities 
that may emerge in the two-dimensional reality of the Zeldovich approximation. We will 
follow a mostly visual approach. After passing the basic mathematical concepts of 
Morse theory and catastrophe theory, we illustrate the various singularities and 
caustics as projections of a higher dimensional \emph{phase-space} sheet onto Eulerian 
space. 

\subsection{Morse Theory}
Catastrophe theory \citep{thom72,zeeman77,arnold92} describes how discrete events and objects 
emerge in a smoothly behaving system. It is built on Morse theory \cite{milnor1963}. 

Morse theory describes the general case of a - smooth and differentiable - function $f(\vec{q})$,  
in a mapping from $\mathbb{R}^n$ to $\mathbb{R}$. Specific example of a Morse function are the 
heights of a mountain landscape, or the stochastic spatial random field defined by the mass 
distribution in the universe. 

A key aspect of a Morse function $f(\vec{q})$ is that at each location ${\vec q}$ we can define 
its gradient, $\nabla f$. Within this gradient field, an exceptional and special position is taken by the 
\emph{critical points}. These points are the ones where the derivative vanishes, $\nabla f=0$. Singularities 
may have one of $d+1$ different identities. In the 2D situation we recognize three classes: 
maximum, saddle point and minimum. This identity is determined by the signature of the eigenvalues of 
the second derivative $\nabla_{ij} f$ of the Morse function, following the listing in 
table~\ref{tab:signature}. 

The \emph{critical points} play an essential role in determining the topology of a landscape. 
Considering the example of a cosmological density field, we may consider the density 
superlevel sets. These consist of the regions with a density superseding the defining 
threshold value and which have the corresponding density isocontour as boundary. As we 
lower the threshold value, the superlevel sets change topology as the threshold 
isocontour touches a \emph{critical point}. 
\ \\
\begin{table}
\begin{center}
\begin{tabular}{|p{6.0cm}|r|}
\hline 
$\nabla_{ij} f$ & singularity\\
\hline
($-$ $-$)& maximum \\
($-$ $+$)& saddle\\
($+$ $+$)& minimum\\
\hline
\end{tabular}
\caption{Singularity character \& Hessian eigenvalue signature}
\label{tab:signature}
\end{center}
\end{table}

\begin{figure*}
	\centering
	\centerline{\includegraphics[width=\textwidth]{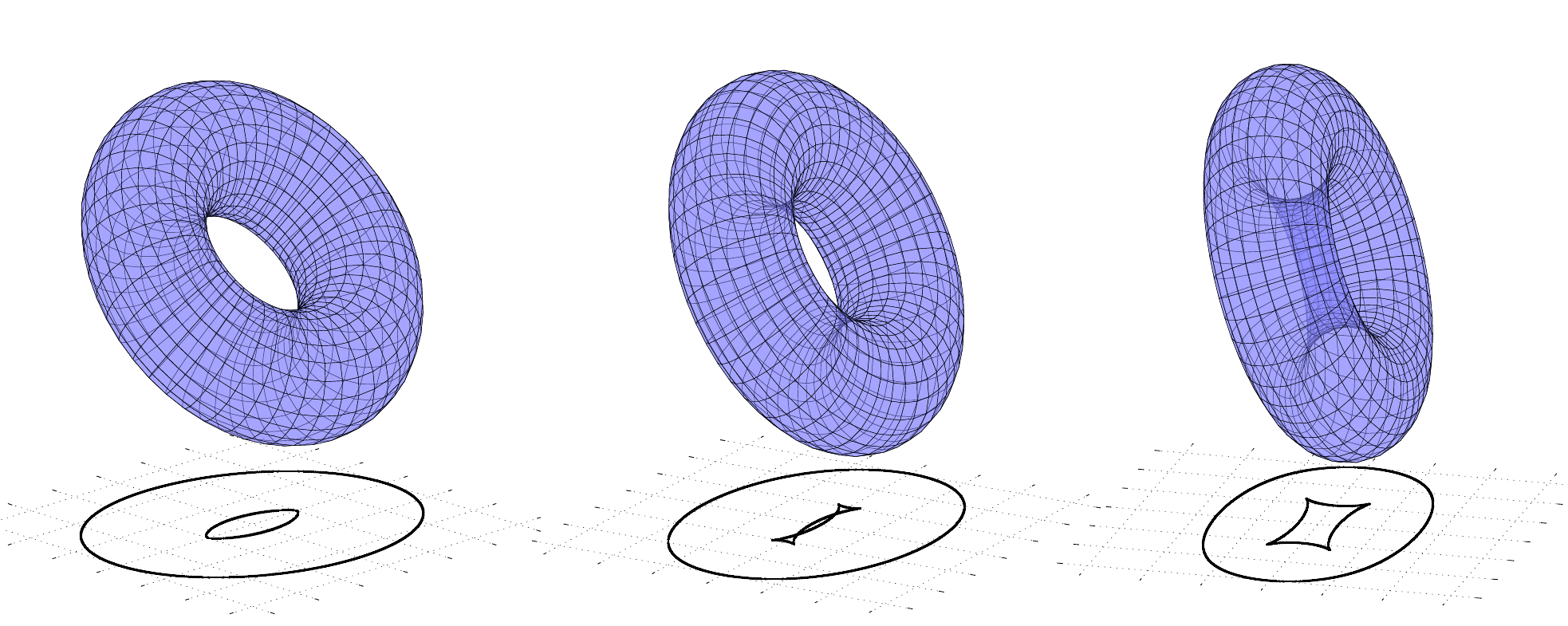}}
	\caption{Caustic formation by projection. The figure illustrates how we may 
                 observe different caustics through the projection of a higher 
                 dimensional structure. The image shows a torus at different orientations in 3-D space. 
                 As we rotate the non-singular torus from a face-on to an edge-on view, 
                 we observe the changing morphology of the projection of the torus on the 
                 2-D plane at the bottom of the figure panels change their morphology. 
                 In three snapshots, we show the metamorphosis of the resulting caustics.
                 Left: instantaneous development of a caustic with two swallow-tail points ($A_4$). 
                 Centre: splitting of central caustic into two cusps, a $A_3$ singularity. Right: as 
                 the torus rotates further, the cusp points move to the edge.}
	\label{fig:torus}
\end{figure*}

\subsection{Catastrophes}
Around its critical points, a Morse function may always be transformed into a quadratic form of local 
coordinates, following a transformation 
\begin{equation}
\vec{q}=(q_1,q_2) \rightarrow (\xi_1,\xi_2)\,, 
\end{equation}
for which
\begin{equation}
\zeta(\xi_1,\xi_2) = f(q_1,q_2) = \pm \xi_1^2 \pm \xi_2^2\,.
\end{equation}
Any point for which the function cannot be transformed into such a {\it normal form} is 
degenerate. In general, the degeneracy of these points will disappear by expanding the 
function in a Taylor series in small perturbations of its parameters. 

As the function becomes more complex and more parameters are needed 
for its description, a growing family of functions is generated. Most of these 
functions will not have degenerate points. However, at certain values of 
the parameters, degenerate points may appear. At these parameter values, the 
function will make a transition to new, very different behaviour. The 
parameter set with these values defines a bifurcation point, and these 
events are called \emph{catastrophes} \citep{thom72,zeeman77,arnold92}.  

As more parameters are added, and the functions become more complex, the family 
of different types of catastrophes grows. Theory teaches us that there are only a 
few topologically different types of singularities possible, and that these are stable 
under perturbation \citep{thom72,zeeman77,a-82,a-83}. In this paper we will 
follow the ADE classification of catastrophes introduced by Arnold. The name reflects a 
deep connection of the theory of singularities with simple Lie groups. In total, five 
point--like singularities can be identified, the cusps $A_3^+$ and $A_3^-$, the swallow 
tail $A_4$, and the umbilics $D_4^+$ and $D_4^-$. 

Within the context of the Zeldovich approximation, we are interested in the branch of 
catastrophe theory that deals with Lagrangian singularities. Given the obvervation 
that the limited family of possible singularities is stable under perturbation, 
we expect that the singularities identified in the ZA evolved mass distribution will 
also be present in the nonlinear case. Within the limited context of the present study, 
we restrict elucidation of its key aspects through visualization by means of images.\\*

\begin{table*}
\centering
\begin{tabular}{llll}
\hline
name	    & ADE & $F(\vec{q})$ normal form & time \\
\hline \hline
\vspace{1mm}
Cusp 	    & $A_3$ & $F(q_1) = q_1^4 + \lambda q_1^2$ & $t = \pm \lambda \mp \tau^2$ \\
\hspace{5mm} creation & $A_3^+$ & \multicolumn{2}{l}{birth of a pancake} \\
\hspace{5mm} annihilation & $A_3^-$ & \multicolumn{2}{l}{merger of two pancakes} \\
\hline
\vspace{1mm}
Swallowtail & $A_4$ & $F(q_1) = q_1^5 + \lambda q_1^3 + \mu q_1^2$ & $t = \pm \lambda.$ \\
            & & \multicolumn{2}{l}{branching of a pancake} \\
\hline
\vspace{1mm}
Umbilic     & $D_4$ & $F(q_1, q_2) = 3q_1^2 q_2 \pm q_2^3 + 3\lambda (q_1^2 \mp q_2^2)$ & $t = \pm \lambda + x_1 + a x_2.$ \\
\hspace{5mm} Pyramid (elliptic) & $D_4^-$ & \multicolumn{2}{l}{change in associated eigenvalue for three cusps} \\
\hspace{5mm} Purse (hyperbolic) & $D_4^+$ & \multicolumn{2}{l}{swap of associated eigenvalue between cusp and fold} \\
\hline
\end{tabular}
\caption{Possible catastrophe metamorphoses (or perestroikas) in the 2-dimensional Zeldovich approximation.
These are the normal forms as given in \citet{a-83}. The formulas give the potentials belonging to each normal
form, for a time parameter $t$. Each normal form has one or more free parameters, represented by $\tau$ and $\mu$. 
In the context of the singularities in the Zel'dovich approximation, $\tau$ and $\mu$ are identified with the Lagrangian 
coordinate $q_2$. The parameter $\lambda$ inserts the time dependence into the normal form expressions.}
\label{table:nf}
\end{table*}

\subsection{Planar torus projection: \\ \ \ \ \ \ \ \ catastrophe by example}
For the purpose of obtaining an intuitive appreciation of the situations in 
which singularities may appear, we may turn towards the illustration in 
figure~\ref{fig:torus}. The image shows a torus at different orientations in 
3-D space. As we rotate the non-singular torus from a face-on to an edge-on view, 
we observe the changing morphology of the projection of the torus - the black solid contour 
lines - on the plane at the bottom of the figure panels. 

The planar projection is shown for three instances. The projection in the lefthand 
panel contains $A_2$ and $A_4$ singularities, while the central panel also contains 
$A_3$ singularities. The simplest form of singularity is the fold singularity, 
the $A_2$ singularity in Arnold's ADE classification. We observe it when we look 
right through the hole at the center of the torus. The structure seen in the 
lefthand panel emerges when the torus is tilted upward and at one particular instant 
passes through a stage where the center caustic assumes to swallow-tail points at its 
edge. These are the $A_4$ singularities. The central panel shows how these swallow tail 
singularities have split almost instantaneously into two cusps, $A_3$ singularities. As 
we rotate his further, these move further outward (righthand panel). The accompanying 
loss of two crosspoints by the central caustic is the result of a global change in 
the projection, and cannot be predicted from local considerations.  

Although the illustration provides us with some visual intuition of the character and 
nature of singularities, we should also note that the configuration of a static and robust 
torus is not really representative for the situation we are addressing in this study. Within 
the context of the evolving cosmic mass distribution, we will study the projection of the 
Lagrangian submanifold. A direct manifestation of the dynamical evolution of the 
cosmic mass distribution is that this four-dimensional manifold itself is 
continuously changing, and it is the corresponding evolution of the manifold's 
projection that will result in the emergence of singularities. 

\subsection{Normal forms}
To formalize our understanding of the different singularities, we describe them in terms of a \emph{normal form}. 
Loosely speaking, a \emph{normal form} may be defined as the minimal or simplest polynomial that captures the topology 
of a singularity. The normal forms represent a complete description of each singularity: via a transformation 
each stable singularity can be smoothly tranformed into one of the normal forms. 

We may formalize the concept through the definition of a corresponding family of potential functions, $F_{\mu}(\vec{q})$. 
The manifold $\vec{x}(\vec{q})$, i.e. the Eulerian position, is the gradient of the potential function, 
\begin{equation}
\vec{x}\,=\,\vec{x}(\vec{q})\,=\, \gradq F\,.
\end{equation}
The potential function is parameterized by $\mu$. These are a set parameters that smoothly change as a 
function of space or time. For example, in the context of the Zeldovich approximation the displacement field, 
$\vec{x} = \vec{q} + \D\gradq \phi_{\rm lin}$, can be related to the potential function, 
\begin{equation}
F(\vec{q}) = \frac{\vec{q}^2}{2} + \D \phi_{\rm lin}(\vec{q})\,.
\end{equation}
The linear density growth factor $\D$ functions as parameter. This expression establishes 
the link between the ZA and its formulation in terms of a normal form. 

For understanding the topology and geometry of a singularity, the first derivative of the 
potential function $F_{\mu}(\vec{q})$ is not particularly interesting. Via a trivial 
translation, the first order derivative may be transformed away. It are the higher order 
derivatives of the potential function $F_{\mu}(\vec{q})$ which determine the type of a 
catastrophe. 

From Morse theory we know that the critical points of the phase-space manifold ${\vec x}({\vec q})$ 
are the locations at which the first derivative of the surface ${\vec x}(\vec{q})$, and hence the second 
derivative of $F_{\mu}(\vec{q})$, is zero. Via a Taylor expansion we may study the local potential function 
at every Lagrangian point ${\vec q}$ in the domain around these critical points. In the literature, these 
local functions are called the {\it germs} of the catastrophe geometries. 

By means of a suitable coordinate transformation, we may subsequently - without any loss of generality - 
write the local potential functions $F({\vec q})$ in one of the normal form expressions given 
by Arnold \citep{a-83} (see table~\ref{table:nf}). Following this formalism, we may identify any of the 
catastrophes found in the ZA modelled mass distribution with one of the basic catastrophe classes. 

\begin{figure*}
	\centerline{
	\subfigure[$A_3^+$, the birth of a pancake. The linear term in the
	surface counters the cubic term, generating the singularity when the
	magnitude of their derivates matches. This linear term grows with time, faster
	in the middle.]{
        \includegraphics[width=0.9\textwidth]{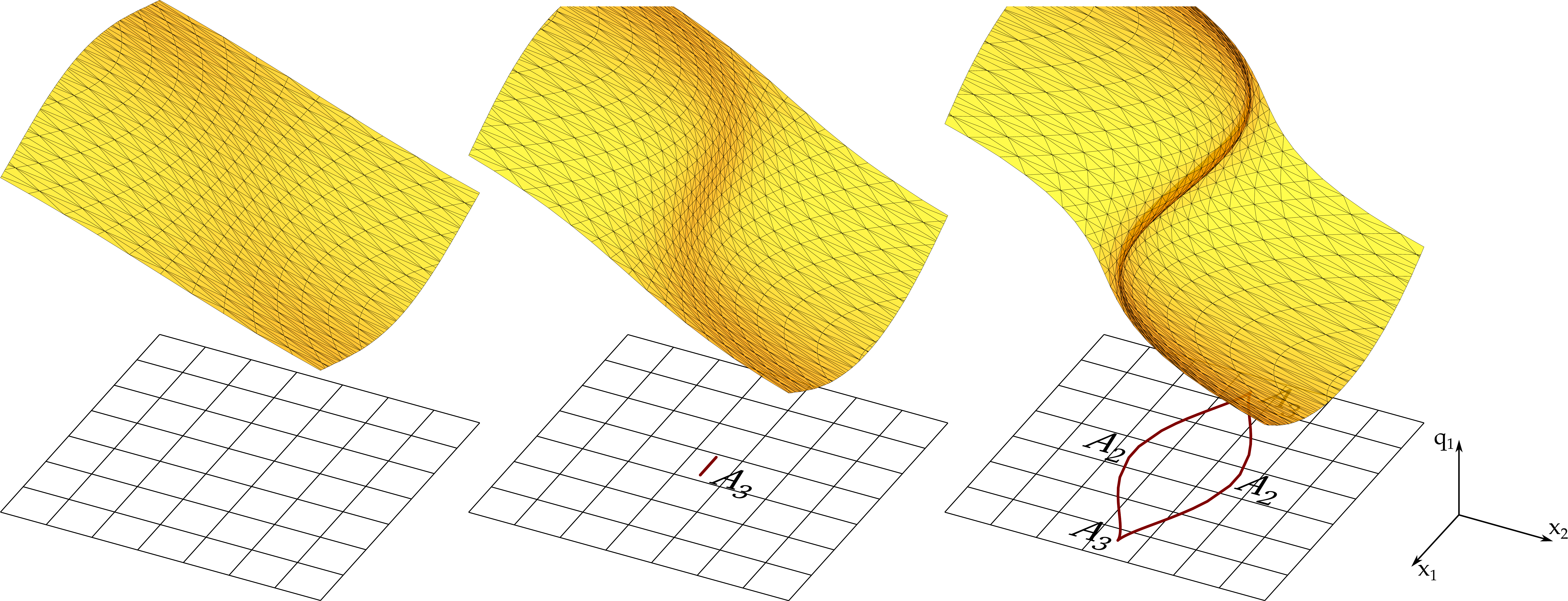}}
	\label{fig:a3p}
        }
	\centerline{
	\subfigure[$A_3^-$ singularity and the merging of two pancakes. The
	linear term in surface has a minimum in the middle. So this is where
	the surface tangent goes singular last, joining the two three-stream regions
	that are already present.]{
	\includegraphics[width=0.9\textwidth]{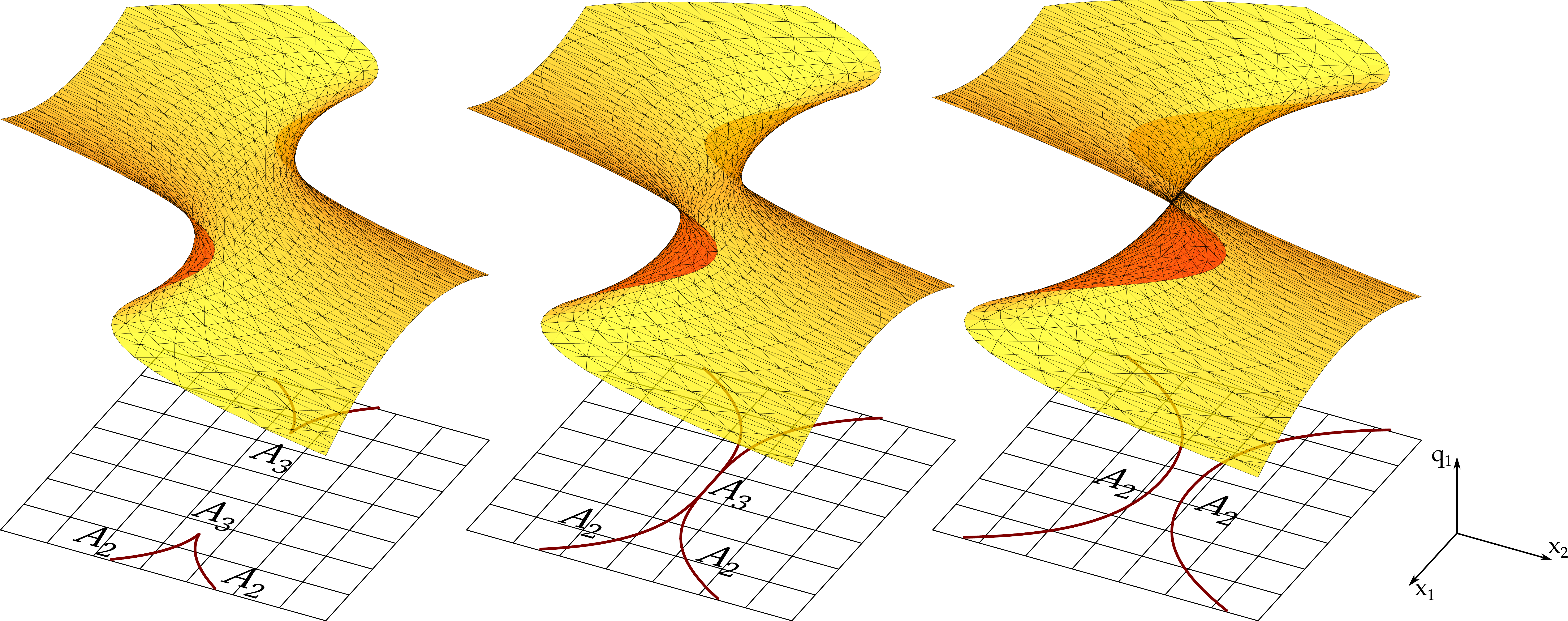}}
	\label{fig:a3m}
        }
	\centerline{
	\subfigure[The $A_4$ swallow-tail singularity and the folding of a
	pancake. In the first image we see the edge of a pancake. This pancake develops
	a singularity, that then unfolds into two cusps moving away from each other.]{
	\includegraphics[width=0.9\textwidth]{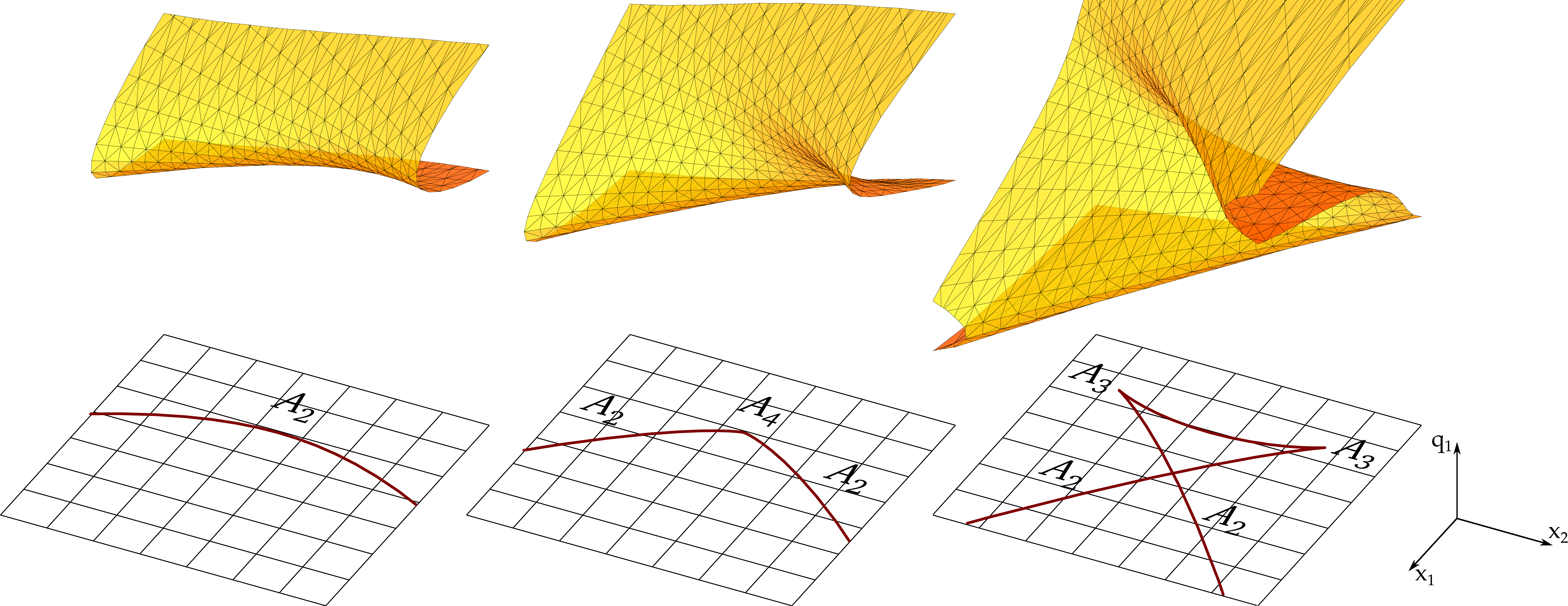}}
	\label{fig:a3}c}
	\caption{A-series of singularities. The figures shows the 3D projection of the Euler-Lagrange 
        phase-space sheet on to the configuration space $(x_1,x_2,q_1)$, with $q_1$ along the vertical direction
        and the Eulerian plane $(x_1,x_2)$ in the horizontal direction. The smooth yellow surface is 
        the 3D projection of the phase-space sheet. Each singularity is shown at three different timesteps.  
        Each image shows the phase-space sheet at one particular instant, along with the 
        caustic structure - the black solid lines - in the 2-D projection on the Eulerian $(x_1,x_2)$ plane. 
        Although the images are rather schematic, these illustrations are fully representative for the general case 
        that we encounter in the evolving cosmic mass distribution. }
	\label{fig:a3}
\end{figure*}

\begin{figure*}
	\centerline{
	\subfigure[The $D_4^+$ singularity: the hyperbolic umbilic. The
	structure of this singularity cannot be embedded in 3D without having
	self-intersecting surfaces. Note that the center image shows a symmetric
	surface, but the deprojection of it resembles an earlier (or later) stage of
	development. The projection of two folds coincides. This will always happen
	locally at a $D_4^+$ point. \citet{t-69} called this singularity a 'purse'.]{
	\includegraphics[width=0.9\textwidth]{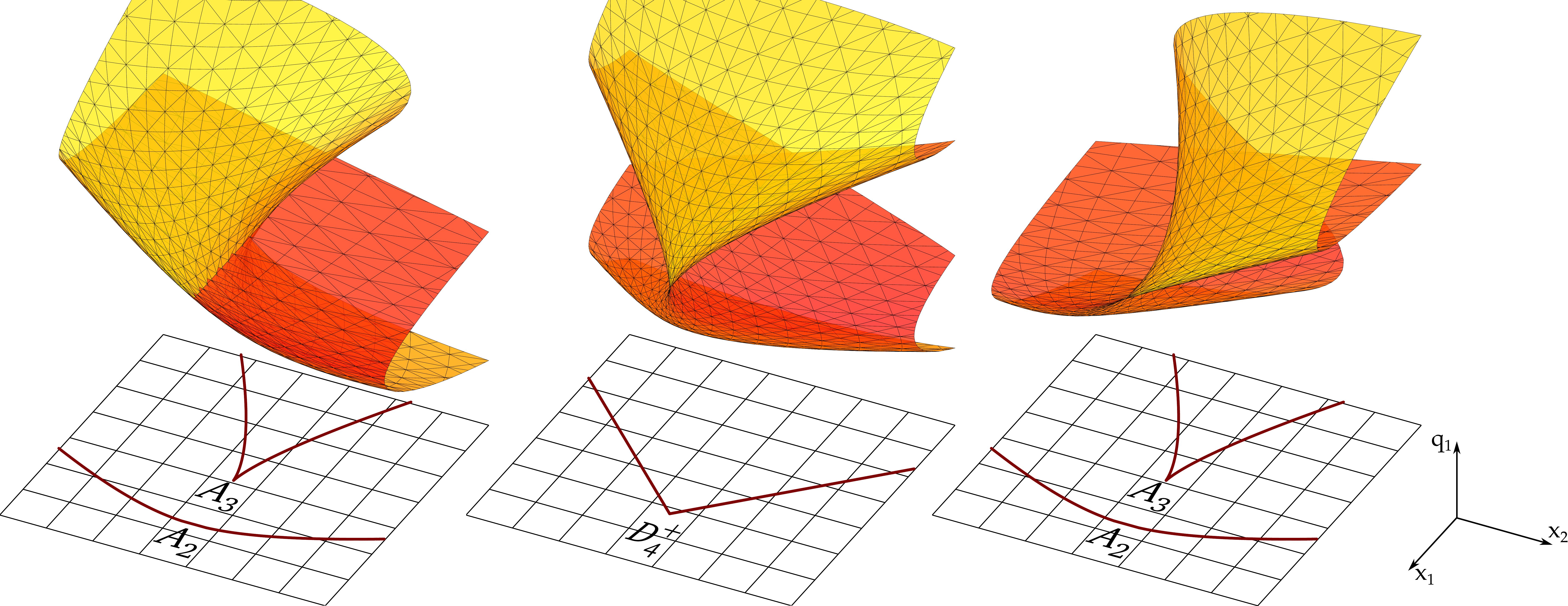}}
	}

	\centerline{
	\subfigure[The $D_4^-$ singularity: the elliptic umbilic. This singularity produces the
	steepest density profile possible in 2D. \citet{t-69} called this singularity a 'pyramid'.]{
	\includegraphics[width=0.9\textwidth]{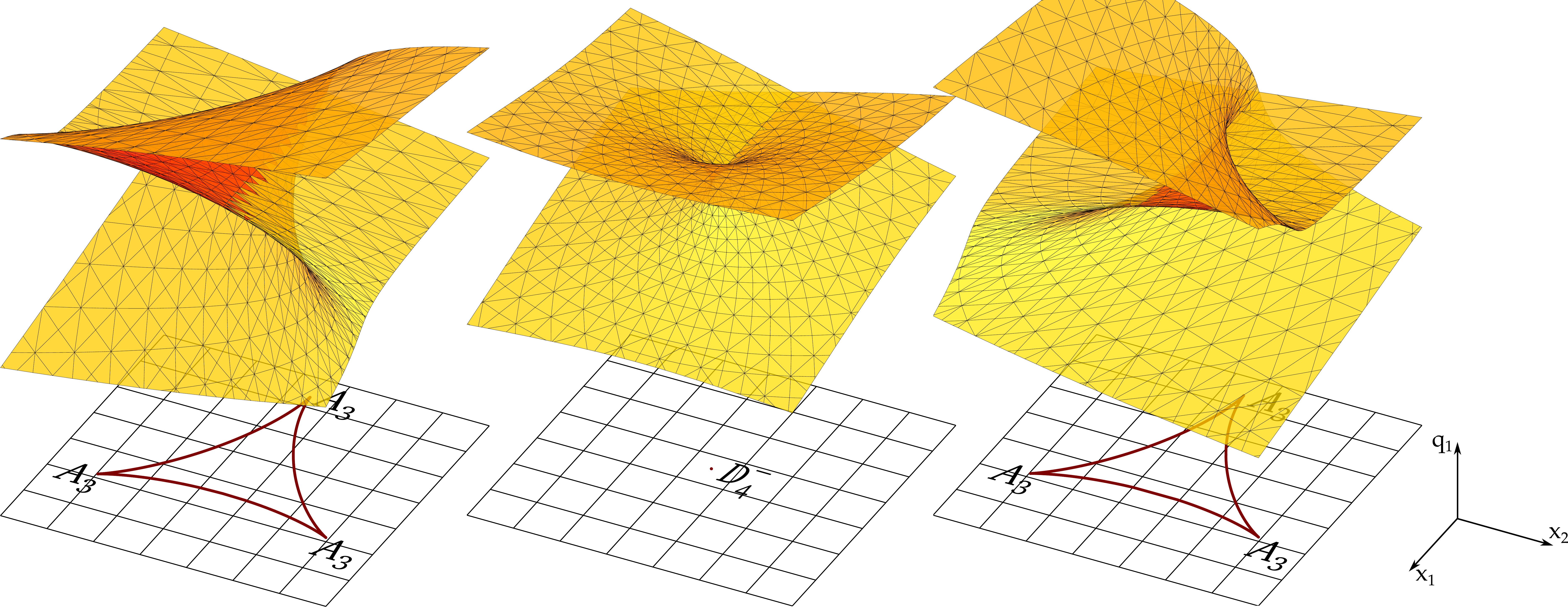}}}
        \caption{The D-series of singularities, limited to the $D_4$ singularities. The figures shows the 3D projection of the Euler-Lagrange 
        phase-space sheet on to the configuration space $(x_1,x_2,q_1)$, with $q_1$ along the vertical direction
        and the Eulerian plane $(x_1,x_2)$ in the horizontal direecction. The smooth yellow surface is 
        the 3D projection of the phase-space sheet. Each singularity is shown at three different timesteps.  
        Each image shows the phase-space sheet at one particular instant, along with the 
        caustic structure - the black solid lines - in the 2-D projection on the Eulerian $(x_1,x_2)$ plane. 
        Although the images are rather schematic, these illustrations are fully representative for the general case 
        that we encounter in the evolving cosmic mass distribution. }
	\label{fig:d4}
\end{figure*}

\subsection{Inventory of 2-D singularities}
In the following subsections, we provide an inventory of illustrations of each of the 
fundamental singularities. To this end, figure~\ref{fig:a3} until figure~\ref{fig:d4} present 
impressions of evolving three-dimensional projections of the evolving phase-space manifold 
$({\vec q},{\vec x})$ in the vicinity of the five fundamental classes of 2-D catastrophes listed 
in table~\ref{table:nf}. The table~\ref{table:nf} lists the normal forms of these  
catastrophes following the formulation given by \citet{a-83}. For a more formal description 
of the table's content, we refer to the next subsection~\ref{sec:manspace}

The images are based on the corresponding normal forms from table~\ref{table:nf}. For practical 
considerations\footnote{Taking into account, as we stated above, that the human brain is not equipped 
for processing four-dimensional images.}, we have to limit ourselves to showing a lower-dimensional projection 
of the phase-space sheet. The figures therefore depict the singularities in the form of three-dimensional projections of 
the Euler-Lagrange phase-space sheet on to the configuration space $(x_1,x_2,q_1)$, consisting of one Lagrangian 
coordinate - along the vertical direction - and two Eulerian coordinates along the horizontal plane. This 
projection of the phase-space sheet is the smooth yellow surface. 

Each singularity is shown at three different timesteps, such that a row of three images shows evolution of the 
phase-space sheet along with the emergence of the caustic structure in the 2-D projection on 
the Eulerian $(x_1,x_2)$ plane. A caustic appears where the phase-space sheet is perpendicular to the 
Eulerian plane $E$. Although the images are rather schematic, these illustrations are 
fully representative for the general case that we will encounter in the evolving cosmic mass distribution. 

\subsubsection{A-series: folds, cusps and swallow-tails}
Figure~\ref{fig:a3} shows the 3D phase-space sheet projections around the $A_3^+$ (top row), $A_3^-$ (central row) 
and $A_4$ (bottom row), each with a sequence of three timessteps. 

A \emph{fold} is the edge of a pancake. As we can see in figure \ref{fig:a3}a, the first pancakes form
at $A_3^+$ points. They are the transient state where two \emph{cusps} are created. Figure~\ref{fig:a3}b shows 
how two cusps can meet to merge into two pancakes. 

The two operations of cusp creation and annihilation will only build a network of filaments if there is a way 
to split a pancake. Figure \ref{fig:a3}c shows how this happens in the transient $A_4$ singularity.
We see how the edge of a pancake develops a `kink'\footnote{one is tempted to say `cusp', but that 
would be technically wrong}, which immediately bifurcates into two cusps showing the characteristic 
`swallow-tail' pattern (hence the name).  In practice, one of these cusps will join another pancake, thereby 
forming a tripod structure (see figure \ref{fig:evolution}). The subsequent evolution depends on the 
surrounding environment, and is explained in section \ref{sec:singinza}.

\subsubsection{D-series: purses and pyramids}
The most complicated type of singularities are the \emph{umbilics}, or $D_4$
singularities. In terms of the deformation tensor there is a $D_4$ singularity if
two eigenvalues are  equal. They are presented in figure~\ref{fig:a3}. 
The $D_4$ singularity involves two types, $D_4^+$  and $D_4^-$. They are shown in 
figures~\ref{fig:d4}a and \ref{fig:d4}b respectively. 

In the $D_4^+$ case, the cusp corresponding to the second eigenvalue meets with the fold 
corresponding to the largest eigenvalue at the middle panel for one instant. The cusp 
then becomes a fold, and the fold becomes a cusp. 

In the $D_4^-$ case a curvilinear caustic triangle corresponding to the largest eigen value 
collapses in the middle panel of figure~\ref{fig:d4}b. At the same instant, a similar triangle 
corresponding to the lowest eigenvalue emerges. This can be seen in the right panel. 

Among all singularities in 2D, both types of $D_4$ result in the highest concentration of mass. 
Note that the figures seem to show self-intersecting surfaces. This is an artefact of the 
lower-dimensional projection of the phase-space sheet. If we were to present them
in 4D, we would see this that is no longer the case. 

\subsection{Lagrangian submanifold geometry and Arnold's catastrophe formulation.}
\label{sec:manspace}
With respect to the geometry of the catastrophes, and the corresponding phase-space 
Lagrangian submanifold in four-dimensional phase-space $({\vec q},{\vec x})$, we should 
add a few explanatory remarks with respect to normal form formulation listed 
in table~\ref{table:nf}.

The geometric structure  of the catastrophes live in the Eulerian-Lagrangian phase-space $E \times L$. 
For a potential $F(\vec{q},t)$, we may define the Lagrangian submanifold as the implicit surface
\[M(t) = \left\{(\vec{x},\vec{q}) \in E \times L \Big| \vec{x} = \grad F(\vec{q},t)\right\}.\]
In the projection of $M(t)$ on to Eulerian space $E$, we find caustics $M_c \subset M$ at those places 
where the surface $M$ is perpendicular to $E$. We may parameterize the submanifold by means of the 
vanishing determinant of the Hessian of $F$, 
\[M_c(t) = \left\{(\vec{x},\vec{q}) \in M(t) \Big| \det \mathcal{H}(F) = 0 \right\}\]
At any time $t$, this gives us a set of
caustic curves. Over a span of time these curves trace a surface called the \emph{big caustic} \citep{a-83}.

To infer the time evolution from the big caustic, we need to define a second surface intersecting the
big caustic. In general this results in a set of curves. It is the parametrisation of this surface
that is given in the last column of table \ref{table:nf}. It allows us to get both cusp creation
and annihilation behaviour from a single definition of the big caustic. 

The general formulation includes a few extra parameters. To arrive at the results in figure \ref{fig:a3},  
the remaining parameter $\tau$ may be identified with $q_2$. Following the same procedure for the 
case of the swallow tail catastrophe, in addition to the substitution of the time 
parameterization, we are still left with the free parameter $\mu$. We identified $\mu$ with the 
remaining Lagrangian coordinate $q_2$ to arrive at the result in figure \ref{fig:a3}c.  

\begin{figure*}
\centering
\centerline{\includegraphics[width=\textwidth]{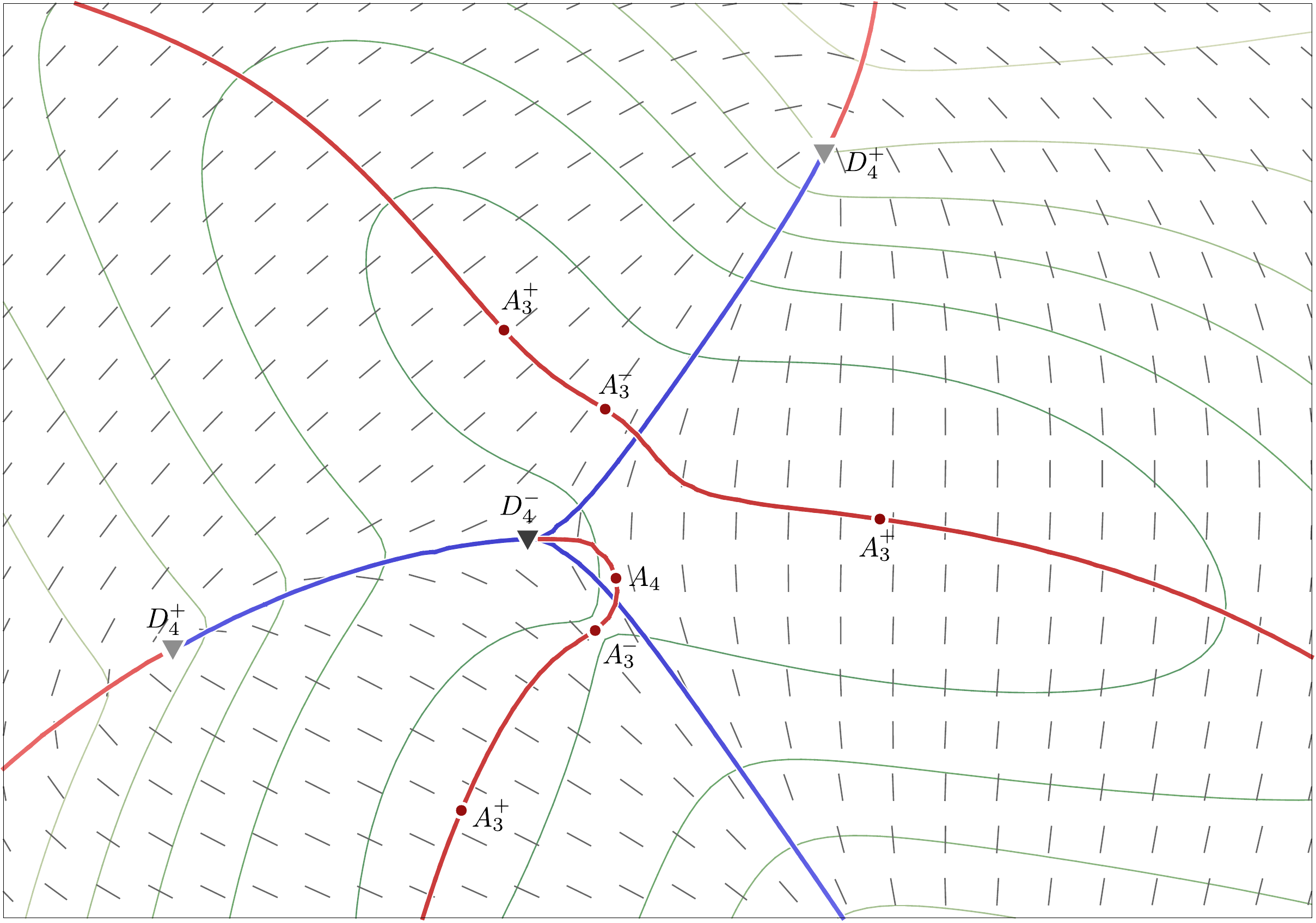}}
\caption{Deformation field: a Lagrangian visualisation. The contours show
the value of the largest eigenvalue and the segments the direction of the
corresponding eigenvector. The thick lines are the $A_3$ lines that track the
development of the caustics through time, red belonging to the first
eigenvalue, blue to the second. Note that the $A_3$ lines trace the places
where the eigenvector is tangent to the contour. The maxima and minima of the
eigenvalues also lie on such a line, denoted by $A_3^+$ and $A_3^-$
respectively.  At the $A_4$ point the $A_3$ line has a local maximum, and the
line is tangent to the contour of the eigenvalue. The $A_3$ line terminates in
a $D_4$ point, continuing as an $A_3$ line of the second eigenvalue. There are
two types of $D_4$ points, more about that is explained in section
\ref{subs:d4}. Note: there is a tiny loop of the $A_3^{\alpha}$ line (red) 
covered by $D_4^-$ symbol;
two generic structures of $D_4$ are plotted in figure \ref{fig:D4} and an additional 
explanation is given in Appendix B}
\label{fig:dfield}
\end{figure*}
\section{Zeldovich Geometry: \\ \ \ \ \ \ generic singularities \\ \ \ \ \ \ as principal elements}
\label{sec:singinza}
The normal forms discussed in the previous section provide a useful and insightful guide to the local topology 
of individual singularities. However, the geometric structure emerging from the gravitational evolution 
of the random primordial conditions is far more complex. Different types of singularities emerge in a variety 
of combinations. 

In this section we extend the image of the individual singularities to more generic circumstances. 
We will find how this involves different types of caustics, the conditions and environment in which 
they arise, and how these different caustic singularities connect into a pervading geometric network. 
The identification of the different classes of singularities and caustics within their generic 
cosmological context will form the basis for the development of a full geometric understanding of 
the way in which the Zeldovich approximation outlines the weblike spine of the Megaparsec scale 
matter distribution. 

In most examples in the following subsections, we select - by eye - the simplest possible realizations 
out of a larger field realization. The singularities in these situations are usually located in relatively 
isolated environments. Singularities of other types are far removed from these locations. This usually involves 
rather artificial situations. In more generic (random) primordial conditions the singularities typically emerge 
in groups. These groups tend to exist of various types of singularities, generally clustered and at relatively 
close distances to each other. One particular illustration of the latter is figure~\ref{fig:D4pyramid}, 
which shows an example where almost all kinds of singularities are packed in a small region. 

\begin{figure*} 
\vspace{-0.2cm}
\centering
\centerline{
	\includegraphics[width=\textwidth]{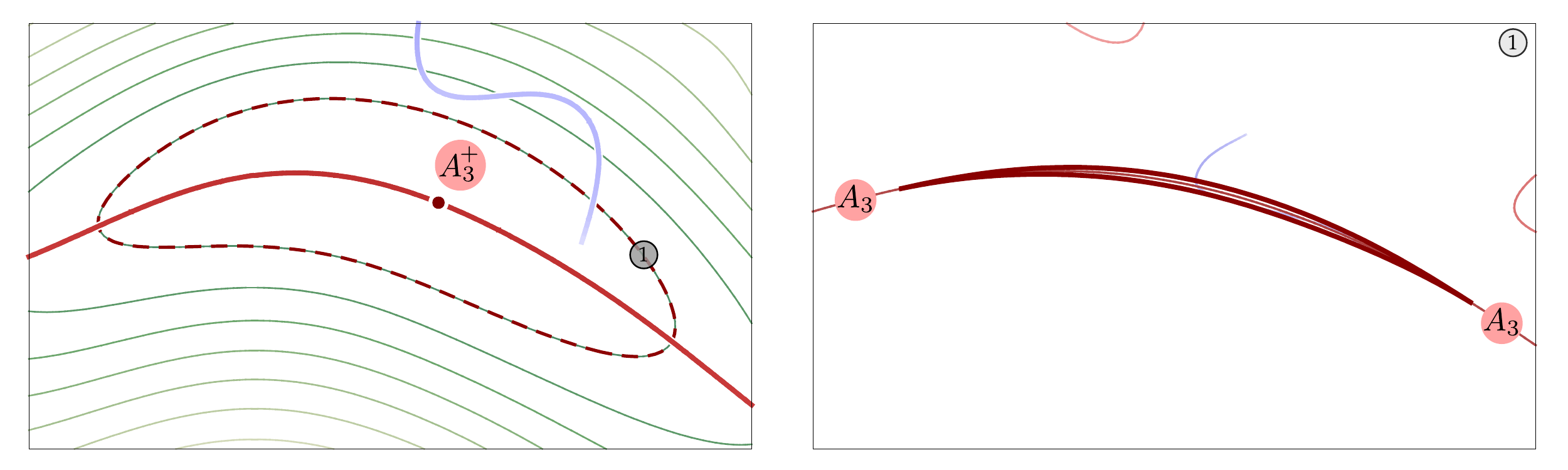}}
  	\caption{$A_3$ cusp singularities and related $A_2$ folds. Left panel: the progenitor of the 
        caustic in Lagrangian space -- red dashed line with label '1', $A_3$ -- thick red line, 
	a maximum of $\alpha$ -- a red  dot next to $A_3^+$ label.   
  Right panel: an $\alpha$-caustic in Eulerian space with two cusps marked as $A_3$.
}
\label{fig:alphaA3p}
\end{figure*} 

\subsection{Caustics and the deformation field}
The two-dimensional version of the cosmic mass distribution evolving according to the Zeldovich 
approximation is equivalent to that encountered in geometric optics \citep[see \eg~][]{zms-83}. The most 
prominent aspect is that of the anisotropy of the collapsing structures and the resulting caustic 
structure, a direct reflection of the fact that in general the deformation tensor eigenvalues $\alpha\ne\beta$. 

While it represents a complete and correct description of the patterns encountered in optics, the predicted structure is 
only an approximation of the structure forming in a gravitating medium. This was of course recognized by 
Zeldovich himself. In this respect, one might argue that the ZA becomes qualitatively incorrect when a 
significant fraction of mass experiences turns inside out more than two times in 2D, or three times in 3D.

\bigskip
\noindent To locate singularities in a generic situation, we need to study the deformation tensor 
\begin{equation}
d_{ij} := -\frac{\partial s_i}{\partial q_j}\,,
\end{equation}
of the displacement field
\begin{equation}
\vec{s}(\vec{q}, t) = \vec{x}(\vec{q}, t) - \vec{q}\,, 
\end{equation}
In the ZA, the displacement vector is factorized in a temporal and a spatial part, 
\begin{equation}
\vec{s}(\vec{q}, \D(t)) = \D \vec{s}(\vec{q})\,.
\end{equation} 
As long as the growth of structure is linear or quasi-linear, we may assume that the flow is a potential 
flow, $\vec{s}= -\gradq \phi_{\rm lin}$. In this situation, the deformation tensor will be symmetric. Our 
discussion of caustic formation is based on this situation and therefore remains valid into the nonlinear 
flows, for as long as the flow remains vorticity free. 

\medskip
The study of singularities in the displacement field is based on the characteristics of the deformation tensor 
field: critical points in the displacement field correspond to zero points in the deformation tensor $d_{ij}$. 
For obtaining insight into the nature and position of critical points, the most instructive approach is to 
focus on the study of the deformation field in terms of its eigenvalue fields $\alpha(\vec{q})$ and $\beta(\vec{q})$, 
and the corresponding eigenvector fields. These fields completely characterize the deformation field, without loss of 
generality: the nature of the critical points is dependent on the signature of the eigenvalues $\alpha$ and $\beta$. 
Important to note is that, by definition, we order the eigenvalues such that $\alpha>\beta$. 

For appreciating the Lagrangian basis of our study, figure~\ref{fig:dfield} illustrates the spatial distribution, in 
a Lagrangian volume, of the $\alpha$--component of the deformation field. For a specific realization of a random initial 
field we have depicted the $\alpha$ isocontours  as thin lines. In the same figure, we have indicated the (normalized) 
eigenvectors $\vec{n}^{\alpha}$ of the $\alpha$ eigenvalue field. They are depicted by means of line segments that are aligned along 
the direction of the eigenvectors. Also, the figure indicates the location of special lines and points - singularities - that 
will be the subject of extensive discussion in later subsections. These play a key role in defining the skeleton of the cosmic web. 

The ultimate fate of mass elements ${\vec q}$ depends on the signature of the eigenvalues $\alpha({\vec q})$ and 
$\beta({\vec q})$. As may be immediately inferred from the Zeldovich approximation (cf. eqn.~\ref{eq:den_eigen_val}, mass elements 
for which both eigenvalues $\alpha$ and $\beta$ are negative never reach a singularity. Those which have only one positive eigenvalue 
pass through a singularity once, while the ones with two positive eigenvalues do this twice. When it passes through a singularity, 
a mass element is turned inside out along the direction of the eigenvector. 

By definition, structure in the universe is the result of a collective response of groups of mass elements to the 
gravitational force field responsible for the displacement of mass. In order to understand the spatial patterns in the 
evolving mass distribution, we therefore should not only concentrate on individual singularities, but also on their  
surroundings and on the complexes in which they are embedded. In the following subsections, we will seek to identify 
the spatial setting and grouping of the different classes of singularities that were identified in 
section~\ref{sec:theoryofsing}. 

\subsection{$A_2$ lines \& pancakes}
It is not hard to see that if the evolution of the system starts from smooth initial condition, the location where collapse starts 
first are those around the maxima of $\alpha(\vec{q})$ in Lagrangian space. These are the $A_3^{+}$ points in the field, which in 
figure.~\ref{fig:dfield} have been marked accordingly. The nature of these singularities will be discussed in more detail 
in the next subsection. 

The subsequent evolution of cosmic structure consists of a gradual progression of 
Lagrangian regions experiencing collapse. At any given cosmic time $t$, these regions 
consist of the points on the isolines $\alpha(\vec{q}) = 1/D(t)$ or $\beta(\vec{q}) = 1/D(t)$. 
In Lagrangian space, they form a set of lines, $A_2^{\alpha}$ and $A_2^{\beta}$. The $A_2^{\alpha}$ 
lines are the isocontours of the eigenvalue $\alpha(\vec{q}) = 1/D(t)$ or, in other terms, its 
eigenvalue level set. 

\begin{figure*} 
\vspace{0.0cm}
\centerline{
	\includegraphics[width=1.00\textwidth]{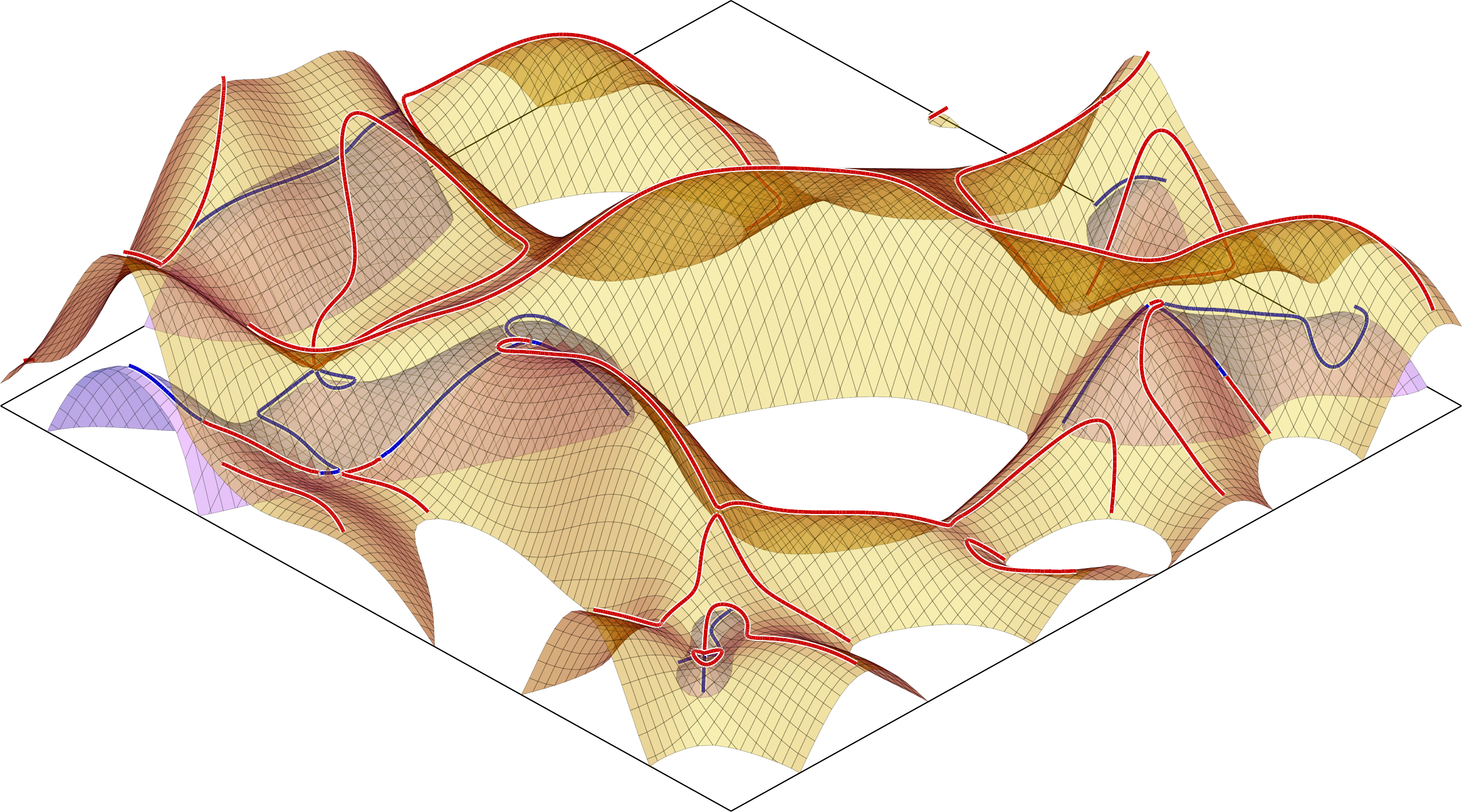}}
\vspace{0.0cm}
\caption{$A_3$ lines and the eigenvalue landscape. The landscape of the eigenvalues $\alpha$ and $\beta$ 
of the deformation tensor is shown for values above a threshold value. The brown landscape concerns the 
field of $\alpha$ values within a certain region of space. The blue landscape shows the run of the 
second eigenvalue in the same region of space, as far as the value of $\beta$ is higher than the 
threshold. The $A_3^{\alpha}$ lines are indicated as thick red lines, the $A_3^{\beta}$ lines as 
blue lines.}
\label{fig:a3landscape}
\end{figure*}
\begin{figure*} 
\vspace{0.2cm}
\centerline{
	\includegraphics[width=0.87\textwidth]{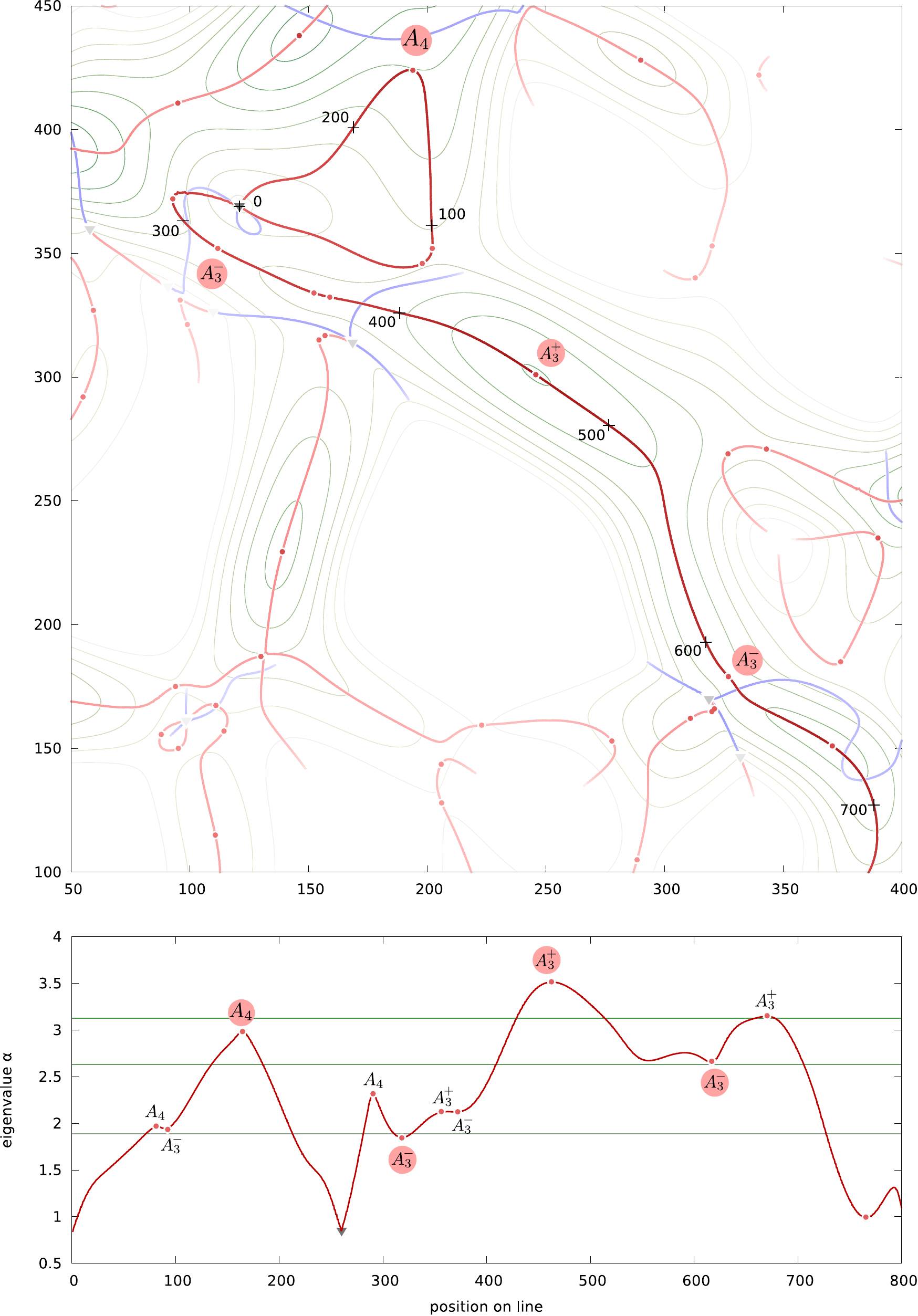}}
\caption{$A_3$ eigenvalue profile. The profile of the first eigenvalue $\alpha$ along the top A$_3$ line in 
figure~\ref{fig:alphaA3p}. This is the line which runs from the upper box edge to the righthand box edge. Along the 
profile we have marked the location and identity of the various singularity points along the  $A_3$ line.}
\label{fig:a3transv}
\end{figure*}

Points residing on an $A_2^{\alpha}$ line pass through an $A_2$ {\it fold} singularity (see fig.~\ref{fig:alphaA3p}). 
When they are mapped to Eulerian space we see them emerge as caustics. In the mapping from 
Lagrangian to Eulerian space, the lower edge of the $A_2$ line is mapped into the upper 
caustic, while the upper edge moves towards the lower caustic (fig.~\ref{fig:alphaA3p}). As the matter 
streams in and around the $A_2$ contour cross each other's path. The resulting pancake region 
demarcates a three-stream region, enclosed by the {\it fold} caustics. At any location within 
its interior we find three velocities, each corresponding with one of the streams. In such 
multistream regions, mass elements of a different origin are crossing each other. One of the  
streams in the pancake region corresponds to fluid elements that have already experienced a 
singular collapse along the direction of the eigenvector $\vec{n}^{\alpha}$. We refer to 
appendix~(\ref{appendix:1d}) for a discussion of the more more transparent equivalent one-dimensional 
situation. 

The two families of emerging caustics are called $\alpha$- and $\beta$-{\it caustics}, while the corresponding 
isocontours in Lagrangian space will be referred to as {\it progenitors of caustics}. Because nearly all 
properties of $\alpha$-- and $\beta$--caustics are equivalent, in order to prevent unnecessary repetitions 
we refer to both by means of $\lambda$, which may either mean $\alpha$ or $\beta$. 

The resulting image is one in which the $A_3^{+}$ points are the buds of the emerging cosmic structure. 
At their Eulerian locations we observe the first stages of the formation of the cosmic web. Subsequently, we 
witness a gradual progression of $\alpha$ isocontours attaining the state of $A_2^{\alpha}$ line. As this happens,  
in Eulerian space we see the expansion of the corresponding {\it fold} caustics. 

\subsection{$A_3$--lines: progenitors of cusps}
\label{sec:A3line}
A key element for our understanding of spatial pattern formation in the Zeldovich approximation is 
that of the $A_3$-lines in Lagrangian space. The $A_3$ lines in the Lagrangian deformation tensor field 
form the backbone of the emerging structure: they are the progenitors of the cosmic skeleton in 
Eulerian space. To develop one's intuition for the spatial context and significance of $A_3$ lines with 
respect to the underlying deformation field, figure~\ref{fig:a3landscape} depicts a 3D impression of the 
outline of the $A_3$ paths along the crests and slopes of the eigenvalue $\alpha$ (and $\beta$) landscape. 
The height of this landscape also depicts the time at which the local mass element collapses. At any single
instance of time the mass element where the $\lambda$-contour and $A_3$-line cross, corresponds to a point
in Eulerian space where we can find a cusp singularity. As time progresses, and the $\lambda$-contour
descends on the landscape, the cusps travel outward along the $A_3$-lines. In this manner the $A_3$-lines
trace the formation of multistream regions through time, forming the blueprint for the structure of the cosmic web.

The $A_3$ lines delineate the points where the eigenvalues $\alpha$ and $\beta$ are maximal along the 
direction of the corresponding local eigenvector. At these points the gradient of the eigenvalue 
$\lambda$ along the eigenvector direction is zero, i.e. they are the points where the eigenvector 
$\vec{n}^{\lambda}$ is orthogonal to the eigenvalue gradient vector $\nabla_q \lambda(\vec{q})$ (see fig.~\ref{fig:dfield}), 
\begin{equation}  
\vec{n}^{\lambda} \cdot \nabla_q{\lambda} = 0.
\label{eq:a3}
\end{equation}
Relating this to the eigenvalue isocontours, we observe that these 
are the points where the deformation tensor eigenvector $\vec{n}^{\lambda}$ is 
running tangential to the contour lines of the corresponding eigenvalue $\lambda$. 
In other words, the $A_3^{\lambda}$ points on a $A_2^{\lambda}$ line are those points 
where the eigenvector $\vec{n}^{\lambda}$ is tangential to the contour $A_2^{\lambda}$. 
Moreover, as the extrema of $\lambda(\vec{q})$ are defined by the condition 
$\nabla_q \lambda = 0$, they are always located on $A_3^{\lambda}$ lines. 

The physical significance of the $A_3$ ridge defined by equation~\ref{eq:a3} may be 
understood when looking at the way in which a mass element on the line develops a 
singularity. To this end, we need to take along that the deformation of 
mass elements occurs along the eigenvector direction $\vec{n}^{\lambda}$. Because of the 
eigenvalue gradient's zero value in that same direction, there is a line-up 
and accumulation of neighbouring mass elements which simultaneously pass through the 
singularity. When mapped to Eulerian space, this evokes the formation of an  
$A_3$ {\it cusp}. They are located at the tip of the evolving $A_2$ {\it fold} 
singularity, which represent the Eulerian manifestation of the corresponding $\lambda$ 
eigenvalue isocontour (see fig.~\ref{fig:alphaA3p} and fig.~\ref{fig:dfield}). 

Hence, following this observation, we see that the $A_3$ lines trace the location of the 
progenitors of $A_3$ {\it cusp} singularities. Dependent on whether it concerns the lines for the first, 
$\alpha$, eigenvalue or the second, $\beta$, eigenvalue, we distinguish between the $A_3^{\alpha}$ lines and 
the $A_3^{\beta}$ lines. In fig.~\ref{fig:dfield} we have indicated the $A_3$ lines by means of heavy solid lines, 
where we have made a distinction between the $A_3^{\alpha}$ lines (red) and the $A_3^{\beta}$ lines. Note that the 
defining equation~\ref{eq:a3} generates a set of lines $A_3^{\lambda}(\vec{q})$ that identify all points 
$A_3^{\lambda}$, regardless of the time when they become singular. 

While the solution of equation~\ref{eq:a3} defines the $A_3$ lines in Lagrangian space, for their full  
outline we also need to consider the points where they terminate. Interestingly, while the $A_3$ lines 
are continuous smooth lines marking the location of progenitors of cusps, the lines terminate 
at isolated points. 

One class of termination points relates to the trivial condition that the formal solution of 
equation~\ref{eq:a3} should not involve negative eigenvalues $\lambda$. These are excluded from 
consideration because they correspond to an expansion along the corresponding eigen direction, and will not 
be able to form a caustic. As a result, the $A_3$ lines are truncated at the level set $\lambda(\vec{q}) = 0$. 
In reality, the $\lambda=0$ level set would never be reached: it would require an infinite time 
to reach the singularity. For this reason, in all our figures we fade the $A_3$ lines as 
$\lambda$ nears zero. Note that this condition occurs more frequently for the $A_3^{\beta}$ lines: 
because $\beta<\alpha$, it has a negative value of a larger swath of Lagrangian space. A nice  
example of this can be seen in figure~\ref{fig:alphaA3p}, in which the blue fading line indicates 
the location of the $A_3^{\beta}$ line in the same area as the progenitor of an $\alpha$-caustic. 

An even more profound and fundamental termination point of $A_3$-lines occurs in more 
subtle circumstances. While rare, there are points ${\vec q}$ where both eigenvalues have the 
same value, $\alpha({\vec q})=\beta({\vec q})$. At these points the eigenvectors have no prefered direction, 
rendering equation~\ref{eq:a3} meaningless. In subsection~\ref{subs:d4}, we will elaborate on this class 
of points and demonstrate that they concern an isolated and discrete set of special points. They generate 
another class of caustics, the umbilics $D_4$ (of which there are two types). Here, it is important 
to note that they do not form crossing points of $A_3^{\alpha}$ and $A_3^{\beta}$ lines, but rather 
their connecting tips. Figure~\ref{fig:dfield} marks the location of the $D_4$ singularities by 
means of a triangle. 

Ideally, one would like to show the lines in a three-dimensional diagram, with the added dimension being time or, 
equivalently, the eigenvalue level $\alpha$. We would find that no $A_3$ line intersects another $A_3$ line: 
no Lagrangian point would ever pass at any one time through more than one singularity. Note that in 
fig.~\ref{fig:dfield} we only show the Lagrangian plane. Even while the $A_3$-lines for the major and minor
eigenvalues do seem to intersect at places, this has no physical meaning or implication. 


\begin{figure*} 
\vspace{0.2cm}
\centerline{
	\includegraphics[width=0.95\textwidth]{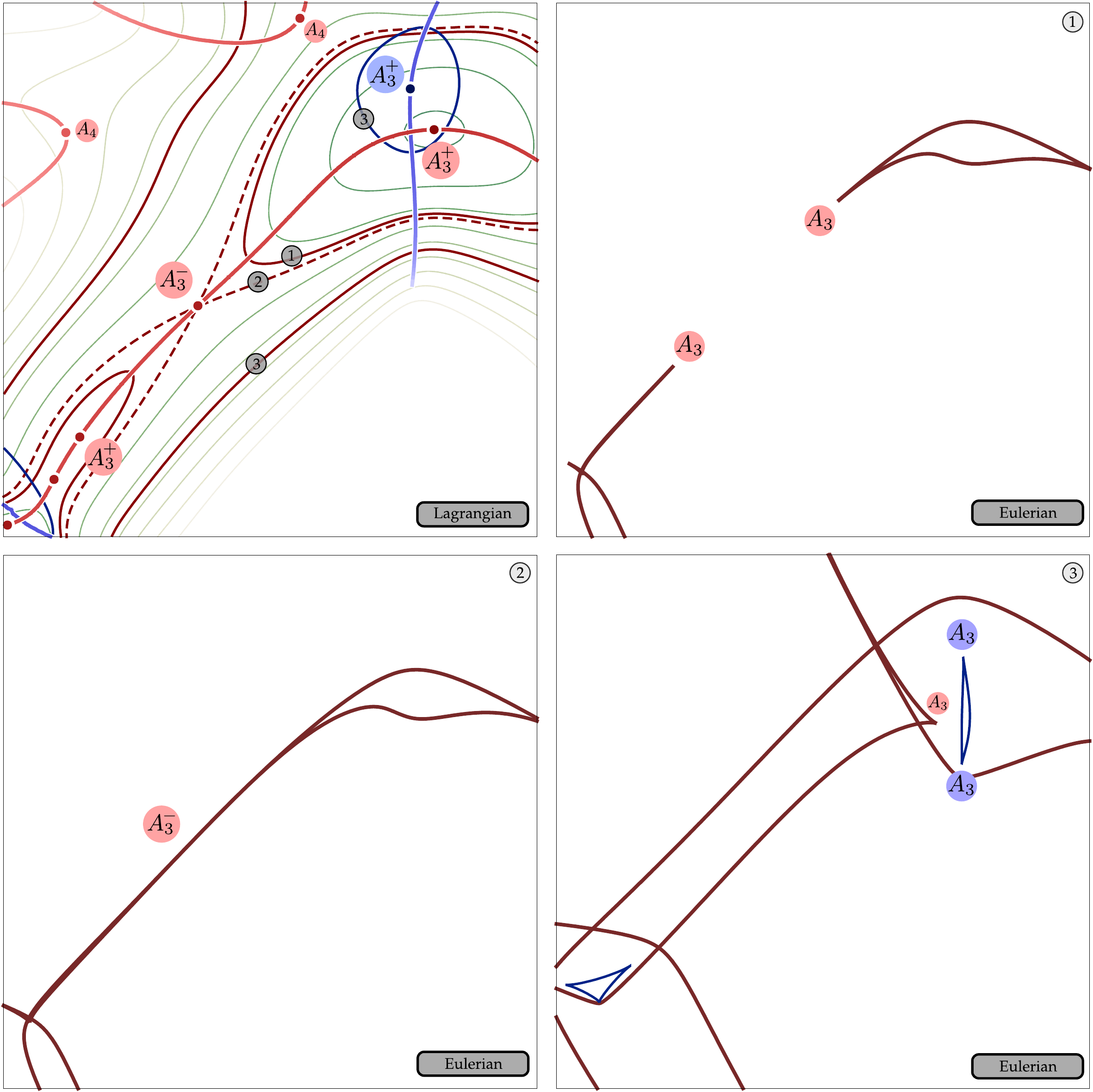}}
\caption{The merger of two pancakes. Top-left panel: Lagrangian space. Other 
panels show the evolution of caustics in Eulerian space. The progenitors of caustics
in Lagrangian space are marked in correspondence with the stages shown in Eulerian
space. Eulerian panel '2' shows exactly the instant of annihilation of two cusps
at $A_3^-$ point. Notations are similar to figure \ref{fig:alphaA3p}. }
\label{fig:alphaA3m}
\end{figure*}
\begin{figure*} 
\centering
\centerline{
  \includegraphics[width=0.95\textwidth]{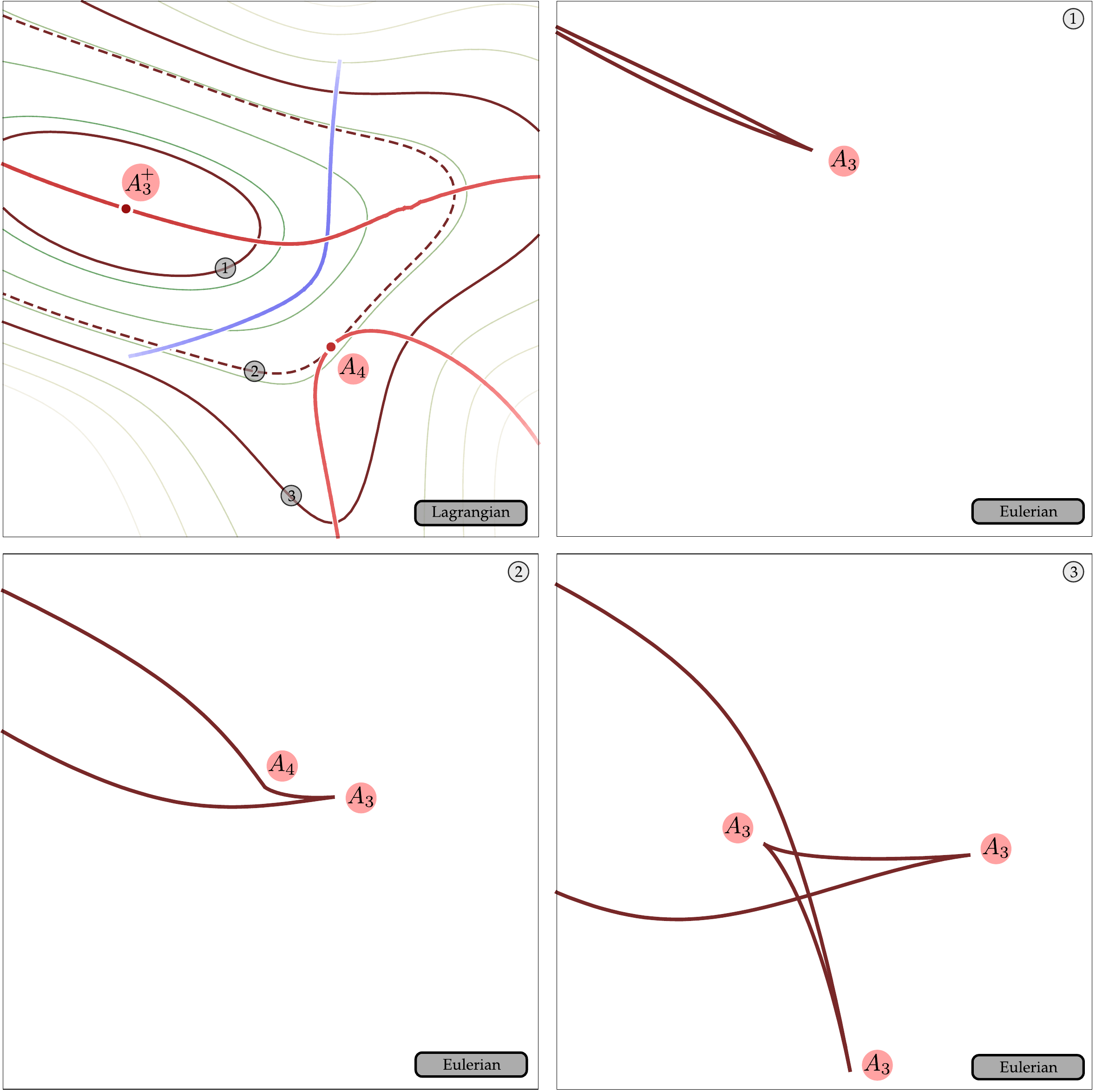}}
\caption{Formation of an $A_4$ swallow-tail singularity. Top--left panel: Lagrangian 
view of the formation of an $A_4$ singularity.
The thick contours descend down $\alpha$ as time increases. The dotted contour
shows the moment at which the `swallow tail' is formed. The three time steps
indicated by these contours are shown in Eulerian space in sub figures 1, 2 and
3. The lines in those figures show the location of the caustics, by mapping
the contours in the Lagrangian image to their Eulerian position.}
\label{fig:A4}
\end{figure*}

\subsection{$A_3$ lines: \\ \ \ \ \ \ \ \ the emergence \& merging of pancakes}   
\label{sec:A3}
In figure~\ref{fig:a3transv}, we have plotted the run of the eigenvalue $\alpha$ along the large $A_s^{\alpha}$ 
line (blue) running from top left to bottom right in figure~\ref{fig:dfield}. We can clearly distinguish the 
maxima of $\alpha$ along this ridge. These points are marked as $A_3^{+}$ and they are the Lagrangian progenitors of 
the first cusps emerging in the cosmic density field. The landscape of $\alpha$ isocontours around one of 
these $A_3^{+}$ points is illustrated in the lefthand panel of figure~\ref{fig:a3}. 

Once the cusp at the maximum $A_3^+$ has emerged, the value of $\alpha$ at which 
a point reaches singularity decreases as time proceeds, $\alpha({\vec q})=1/D(t)$. At time $t$,  
all points on the corresponding $\alpha$ isocontour become singular. These points pass through 
an $A_2$ fold singularity and in Eulerian space become visible as $\alpha$-caustics that enclose a pancake. 
Two points stand out, the ones at the intersection of the $A_3$ line and the $\alpha$ isocontour. They are 
the ones that in Eulerian space will be visible as $A_3$ cusps (fig.~\ref{fig:alphaA3p}, righthand frame). 

A particular example of a pancake forming around a maximum $A_3^{+}$ is shown in figure~\ref{fig:alphaA3p}. 
It shows a simple example of an $A_3^{+}$ point and its immediate Lagrangian surroundings (lefthand panel), 
and the corresponding $\alpha$--caustic (pancake) in Eulerian space (righthand panel): the isocontours in 
the lefthand panel are the progenitors of the caustic in the righthand panel.  The Eulerian image is one 
of an expanding and growing pancake, enclosed by its $\alpha$-caustic and with two $A_3$ cusp singularities at 
its tips. In its Lagrangian setting, we may appreciate how the corresponding singularity state - the 
progenitor of the caustic - gradually descends from the hill around $A_3^{+}$. The $A_3$ cusps are clearly special 
of the caustic progenitor: they are the points on the Lagrangian progenitor's outline that traverse along the 
$A_3$ line in an opposite direction. 

An essential process for the development of the cosmic web is that of linking different 
structural elements into a connecting network. One important representative of this 
is the linking up between two growing pancakes. With pancakes emerging around the 
maxima $A_3^{+}$ on $A_3$ lines, their subsequent merging is intimately related to 
saddle points in the $\alpha$ eigenvalue field. These saddles are the local minima $A_3^{-}$ 
along the $A_3$ lines (see fig.~\ref{fig:a3transv}). Two pancakes that grow from two maxima $A_3^{+}$ on 
the same $A_3$ line, may merge when their cusps meet and annihilate at the saddle point, $A_3^{-}$ 
(see fig.~\ref{fig:a3} for the corresponding normal form illustration). 

\begin{figure*}    
\centering
\centerline{
  \includegraphics[width=0.95\textwidth]{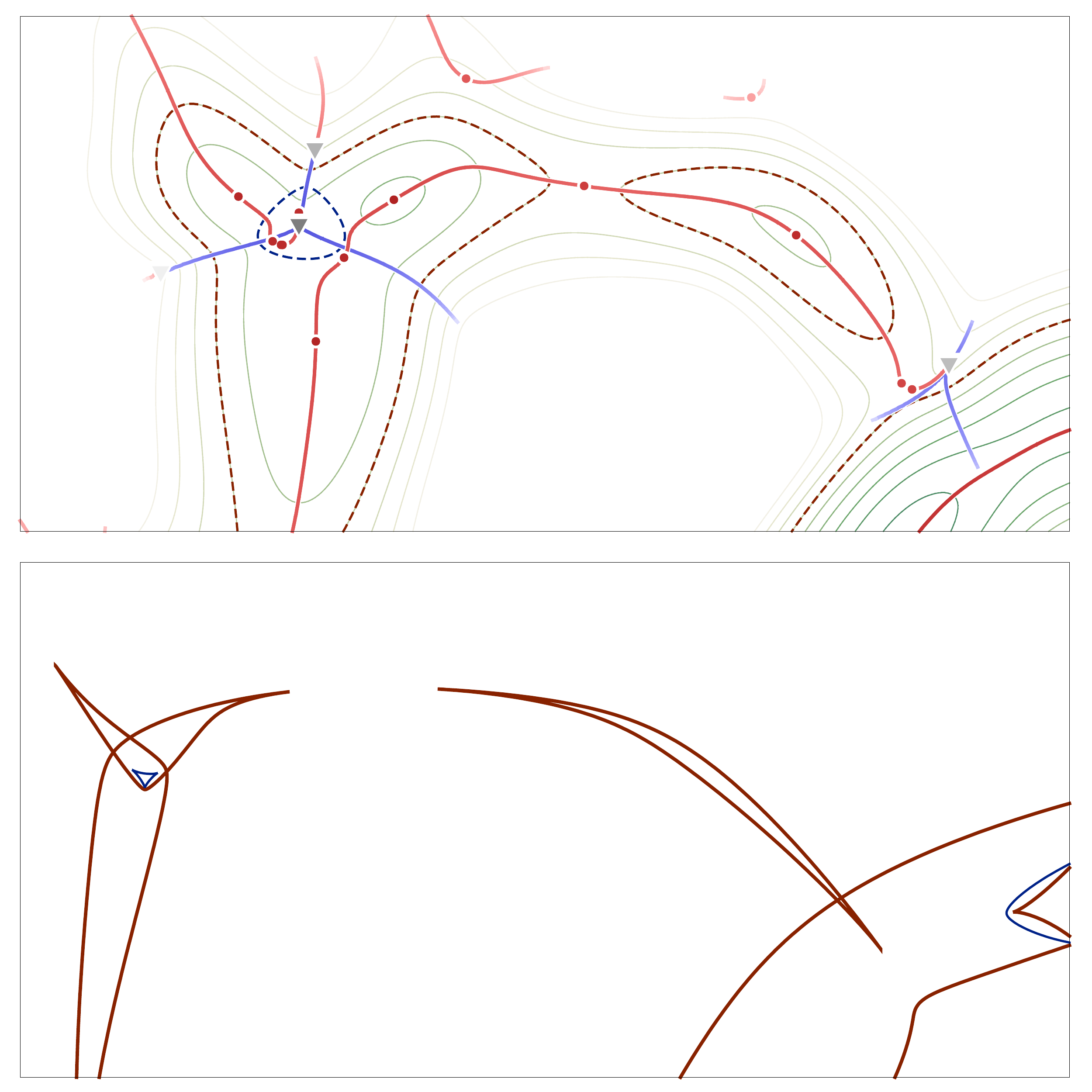}}
\caption{Connection of caustic features and network formation. Top panel: deformation field, 
singularity location and $A_3$ lines around a Lagrangian region with three disjoint 
caustic progenitors. Bottom: corresponding Eulerian region, in which the caustics have 
emerged. The caustic in the form of a simple pancake crosses the bottom caustic due to 
mapping. Further details are described in the text, sect.~\ref{sec:A4}.}
\label{fig:A4connectivity}
\end{figure*}

Figure~\ref{fig:alphaA3m} shows the merger of two pancakes. The top left panel shows the 
eigenvalue landscape in Lagrangian space. The remaining three panels illustrate the dynamics 
of the merging process, with the panel sequence marked from (1) to (3) in the right
corner of each panel. In the Lagrangian panel we have marked three caustic progenitors, 
two $\alpha$ caustics (red contours) and one $\beta$ caustic (blue contours). In the 
corresponding Eulerian panels, we have indicated them at the times they become caustics. In 
panel(1), we see the formation of two pancakes, each with a characteristic $A_3$ cusp at 
their tip. In panel~(2), we see the merging of the two pancakes into one structure and the 
annihilation of the two cusps. The merger location is the Eulerian location of the 
$A_3^-$ saddle, whose Lagrangian position in the $\alpha$ landscape between the 
two hills around the $A_3^+$ points is clearly visible in the topleft panel. 

The structure evolving in figure~\ref{fig:alphaA3m} also reveals the formation of 
a $\beta$-caustic. Its progenitor in the Lagrangian map is the hill around the 
$A_3^+$ maximum of the $\beta$ eigenvalue landscape, indicated by means of blue 
solid contours. The related $A_3^{\beta}$ line is also visible in the same panel, 
as the blue solid line running approximately orthogonal to the (red) $A_3^{\alpha}$ 
line. In the final (bottom righthand) panel, the corresponding $\beta$ pancake 
is seen to emerge in Eulerian space. It is marked by the two $A_3$ cusps at its tip.  

An additional feature visible is the curvilinear triangle visible in the bottom 
lefthand corner of panel~(3). It is the result of a $D_4$ metamorphosis. It will 
be discussed later, in section~\ref{subs:d4}. 

\subsection{$A_4$ points \& swallow-tail singularities}     
\label{sec:A4}
The $A_4$ singularity completes the list of generic singularities of the A-series 
occurring in two-dimensional potential type maps. The $A_4$ singularities are also 
known as swallow tail singularities (for the corresponding normal form illustration, 
see fig.~\ref{fig:a3}). 

$A_4$ singularities are defined as the points where the eigen vector $\vec{n}^{\lambda}$ is 
tangent to the $A_3$ lines. A specific example of an $A_4$ point in its Lagrangian 
environment can be seen in the top left panel of figure~\ref{fig:A4}. $A_4$ singularities 
are special in that they are the locations in the eigenvalue field $\lambda$ at which 
three particular related vectors are colinear. These are the (i) the local eigenvector 
$\vec{n}^{\lambda}$, (ii) the local tangent to the $A_2$ contour and (iii) the 
local tangent to the $A_3$ line. From the definition in equation~\ref{eq:a3}, regarding $A-3$ lines 
we know that by definition the eigenvector is always tangent to the $A_2^{\lambda}$ 
contour. 

The fact that at $A_4$ the eigenvector $\vec{n}^{\lambda}$ is tangential to both the 
$A_2$ and $A_3$ line implies that the magnitude of the eigenvalue - along the $A_3$ 
line on which the $A_4$ point is located - reaches a local maximum. This may be directly 
appreciated from the example in figure~\ref{fig:A4}. The Lagrangian map shows that the 
$A_4$ singularity marks the location where the lower righthand $A_3$ line reaches its 
highest point on the slope of the hill surrounding maximum $A_3$. The dashed 
isocontour marks the points where the eigenvalue $\lambda$ has the same value 
as at $A_4$. Along the lower righthand $A_3$ line, in both directions from $A_4$, 
the value of $\lambda$ is lower than at $A_4$. Physically, at the points along the 
tangent $A_3$ line we see a confluence of mass elements that almost simultaneously 
pass the $A_3$ {\it cusp} singularity. This catastrophe leads to the emergence 
of a special kind of singularity, more complex than that of the $A_3$ cusp, the 
$A_4$ swallow tail. 

An important observation is that the $A_4$ singularity is a metamorphosis. The singularity 
exists only at one single moment of time, the moment $t$ at which the value of $\lambda$ at 
$A_4$ is equal to $\lambda=1/D(t)$. At this instant the $A_2^{\lambda}$ line, whose 
identity is continuously evolving, touches the $A_3$ line at $A_4$. In the situation of 
fig.~\ref{fig:A4} (top lefthand panel), at the moment of the $A_4$ metamorphosis is 
the dashed isocontour the $A_2$ line.  
 
The significance of the $A_4$ metamorphosis within the context of structure 
formation can be best understood by following the development of a representative 
example in Eulerian space. The three Eulerian panels (1) to (3) in figure~\ref{fig:A4} show 
three stages around the instant of $A_4$ metamorphosis. In panel~(1) we see the pancake 
that has formed around the maximum $A_3^+$. At this instant, isocontour~(1) in 
Lagrangian space is the $A_2$ line, which in Eulerian space maps into the {\it fold} 
caustic of the pancake. At the next timestep, we observe the sudden appearance of a 
{\it swallow tail} $A_4$ singularity at the tip of the pancake (panel~2). This instant 
of $A_4$ metamorphosis is the moment at which the Lagrangian dashed eigenvalue 
isocontour~(2) has become the $A_2$ line and touches singularity point $A_4$. 

Following this instant of metamorphosis, the swallow tail disappears again and morphs   
into the complex four-edged configuration visible in panel~(3). At the tip of the pancake, we 
find a two-wing protrusion defined by three $A_3$ singularities. To understand this configuration, 
we should observe that there are three intersections of $A_3$ lines with the $A_2$ line, i.e. 
isocontour~(3). There is one such intersection with the upper $A_3$ line and two with the 
lower righthand one. These three are the progenitors of the three $A_3$ {\it cusp} singularities 
that we see in Eulerian panel~(3). The other points on the $A_2$ isocontour~(3) map into the 
characteristic connecting caustic outline of panel~(3). The first part of the Lagrangian 
isocontour~(3), running from the leftside box boundary to the $A_3$ point near the lower 
box boundary, maps into the Eulerian caustic's top edge and ends at the lower $A_3$ {\it cusp}. 
The caustic's lower edge is the result of the mapping of the top section of the Lagrangian 
$A_2$ line. The $A_3$ point at the intersection of isocontour~(3) with the upper $A_3$ line is 
the progenitor of the $A_3$ cusp at the tip of this caustic's branch. The two winglike caustic edges 
connecting to the third $A_3$ {\it cusp} are stemming from the two righthand sections of Lagrangian 
contour~(3), interspersed by the $A_3$ point which is the progenitor of the connecting cusp. 

The above example illustrates the instrumental role of $A_4$ swallow-tail singularities in 
establishing the global connections between different sections of the cosmic web. The $A_4$ 
singularities are the junction points at which disjoint pieces of Lagrangian $A_3$ lines get 
connected. Observing this on a somewhat larger scale underlines the significance of 
this role. Figure~\ref{fig:A4connectivity} depicts a Lagrangian region (top panel) and 
its Eulerian equivalent at the instant at which the dashed contours have become $A_2$ lines 
(bottom panel) . At this moment, we may identify three independent sections of the 
cosmic web. The large lefthand structure has already passed the stage of $A_4$ 
metamorphosis. It has resulted in the formation of a typical winglike configuration,  
established by the $A_4$ singularity which in the Lagrangian map is located near the inner 
$A_2^{\beta}$ line. The second, central, structure is a typical pancake. It is about to merge 
with the neighbouring complex once the $A_2$ lines reach the $A_3^-$ saddle point positioned 
along the connecting $A_3$ line. The second $A_4$ singularity, close to the $A_2$ line at the 
edge of the lower righthand island, is on the verge of its metamorphosis. Once it materializes, 
it will produce two cusps that merge with the cusp that is already inside the lower 
righthand pancake visible in the Eulerian map.  

\subsection{$D_4$ points: purses \& pyramids} 
\label{subs:d4}

The final class of singularities to investigate in their field setting are the $D_4$ singularities. These 
belong to a different family than the $A$-type singularities that we have discussed in the previous 
subsections. 

The principal difference between the two families of singularities, the $A$- and $D$-type singularities, 
is that of the number of eigenvalues on which they depend. Singularities in the $A$-family depend on 
only one eigenvalue, $\alpha$ or $\beta$. $D$-type singularaties, on the other hand, involve both 
eigenvalues $\alpha$ and $\beta$. 

$D_4$ singularities are the points at which both eigenvalues have the same value, $\alpha=\beta$. In 
Lagrangian space, the singularities occur at isolated points. At first, this might occur as 
counterintuitive. In 2-dimensional space, one would expect the solution of an equation
\begin{equation}
\alpha(\vec{q}) = \beta(\vec{q})
\label{eq:d4}
\end{equation}
to define a continuous line. However, upon closer inspection one may realize that - at any particular 
instant $t$ - equation~\ref{eq:d4} is equivalent to two independent equations, yielding a single point 
as solution. Within this context, it is good to realize that we are dealing with a highly 
constrained situation, in which the two eigenvalues $\alpha$ and $\beta$ are strongly 
correlated and are conditioned by the requirement that $\alpha \geq \beta$. 


\begin{figure}      
\centering
\centerline{\includegraphics[width=0.48\textwidth]{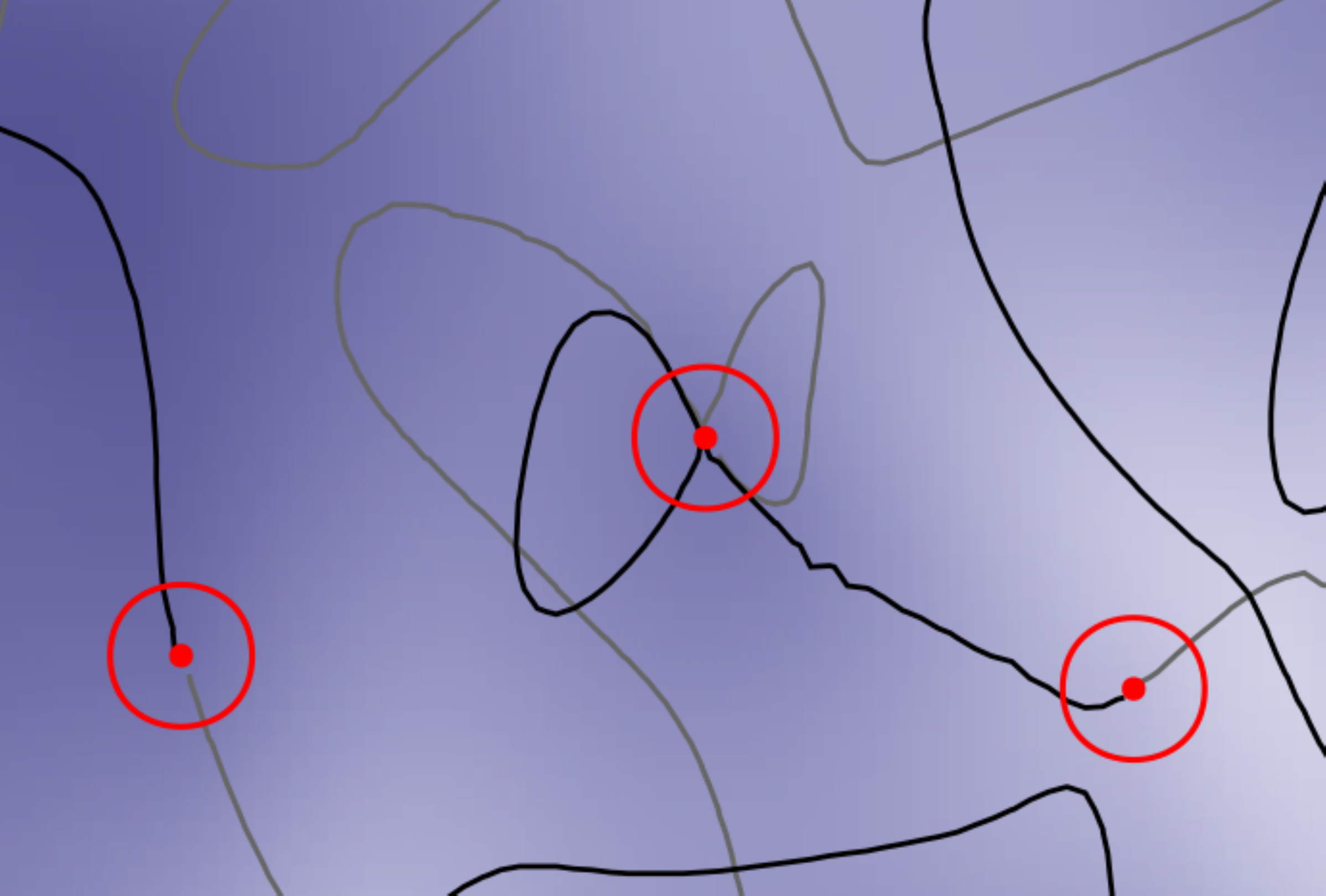}}
\caption{$D_4$ singularities and $A_3$ line connections. The figure shows the location of 
three $D_4$ singularities in Lagrangian space. The $D_4$ locations 
are indicated by a red dot inside a red circle. The black lines 
are $A_3^{\alpha}$ lines, while the grey lines are $A_3^{\beta}$ 
lines. The panel contains two {\it purse} singularities ($D_4^+$)
and one {\it pyramid} singularity ($D_4^-$).}
\label{fig:D4}
\end{figure}
\subsubsection{The $D_4$ equation}
Given the deformation tensor $d_{ik}=-\partial s_i/\partial q_k$, its eigenvalues $\lambda$ are 
the solution to the secular equation,
\begin{equation}
\begin{pmatrix}
d_{11}-\lambda&d_{12}\\
d_{12}&d_{22}-\lambda
\end{pmatrix}
\,=\,0\,.
\end{equation}
This translates into the corresponding secular quadratic equation,
\begin{equation}
\lambda^2-(d_{11} + d_{22}) \lambda + d_{11}d_{22}-d_{12}^2\,=\,0 \,,
\label{eq:sec-eq}
\end{equation}
whose two solutions are given by 
\begin{equation}
\lambda_{1,2} = {1 \over 2} \left( d_{11}+d_{22} \pm \sqrt{(d_{11}-d_{22})^2 + 4 d_{12}^2}\right)\,.
\label{eq:eig-val}
\end{equation}
In order to have $\alpha = \beta$, two independent conditions need to be satisfied simultaneously, 
\begin{eqnarray}
d_{11}(\vec{q})&\,=\,&d_{22}(\vec{q})\,,\nonumber\\
d_{12}(\vec{q})&\,=\,&0\,. 
\end{eqnarray}
Each of these two equations describes a line. Their combined solution consists of the 
discrete points ${\vec q}$ at the intersection of these lines. The $D_4$ singularities 
are therefore located at a set of isolated points (see appendix~\ref{appendix:d2}). 
Notice that the identity of these isolated points is changing as a function of time. Throughout 
this study, we have used reverse triangles to indicate the location of the $D_4$ singularities in the 
various figures (e.g. fig.~\ref{fig:dfield}).

Because of the equality of the two eigenvalues, the eigenvectors at $D_4$ are degenerate: 
any arbitrary 2-D vector may function as eigenvector. It is yet another manifestation of the 
non-Gaussian nature of the eigenvalue $\lambda$ fields. 

\subsubsection{the $D_4$ location:  $A_3$ lines.}
The role of $D_4$ singularities in outlining the cosmic web becomes clear when 
evaluating their lo cation in the Lagrangian eigenvalue field. They have a special 
position on the $A_3^{\alpha}$ and $A_3^{\beta}$ lines. While located on both lines, 
they are not found at the crossing of two such lines. Instead, they are connection 
points between two such lines, located at the tip of an $A_3^{\alpha}$ line and 
a connecting $A_3^{\beta}$ line. 

The first aspect of importance is the fact that also $D_4$ points are located on 
$A_3$ lines. This may be directly understood from the degeneracy of the corresponding eigenvectors. 
Without loss of generality, the eigenvector ${\vec n}^{\lambda}$ at $D_4$ points can always 
be oriented such that it is orthogonal to the gradient of the field. It therefore fulfils the 
$A_3$ condition, $\vec{n}^{\lambda} \cdot \nabla_q{\lambda} = 0$ (eqn.~\ref{eq:a3}). Because 
this is true for both eigenvalues $\alpha$ and $\beta$, any $D_4$ point is located 
on $A_3^{\alpha}$ and $A_3^{\beta}$ lines. 

It may be instructive to look at the formal derivation of the above. To this end, 
we evaluate the inner product $I_{\lambda}({\vec q})$ of the 
normal vector ${\vec n}({\vec q})$ and the gradient of the eigenvalue field, 

\begin{eqnarray}
I_{\lambda}({\vec q})\,=\,\vec{n}^{\lambda} \cdot \nabla_q{\lambda}&\,=\,& 
\begin{pmatrix}
n_1\\n_2
\end{pmatrix}
\cdot
\begin{pmatrix}
\partial \alpha/\partial q_1\\
\partial \alpha/\partial q_2
\end{pmatrix} \nonumber \\
&\,=\,& n_1 \frac{\partial \alpha}{\partial q_1}+n_2 \frac{\partial \alpha}{\partial q_2}
\end{eqnarray}

\noindent To calculate $I_{\lambda}({\vec q})$, we recast its expression in terms of the deformation 
tensor $d_{ij}$ and its derivatives 
\begin{equation}
t_{ijk} \equiv \frac{\partial d_{ij}}{\partial q_k}\,,
\end{equation}
and the eigenvector $\vec{n}^{\alpha}$, whose components are given by

\begin{equation}
\vec{n}^{\alpha}\,=\,
\begin{pmatrix}
d_{12}\\ \alpha-d_{11}
\end{pmatrix}\,.
\label{eq:eigenvd}
\end{equation}

\noindent The tensor $t_{ijk}$ has only four independent components: $t_{111}, t_{112}, t_{122}$ and 
$t_{222}$. The other components $t_{ijk}$ can be inferred from these on the basis of symmetry 
considerations. 

\begin{figure*} 
\centering
\centerline{
	\includegraphics[width=0.95\textwidth]{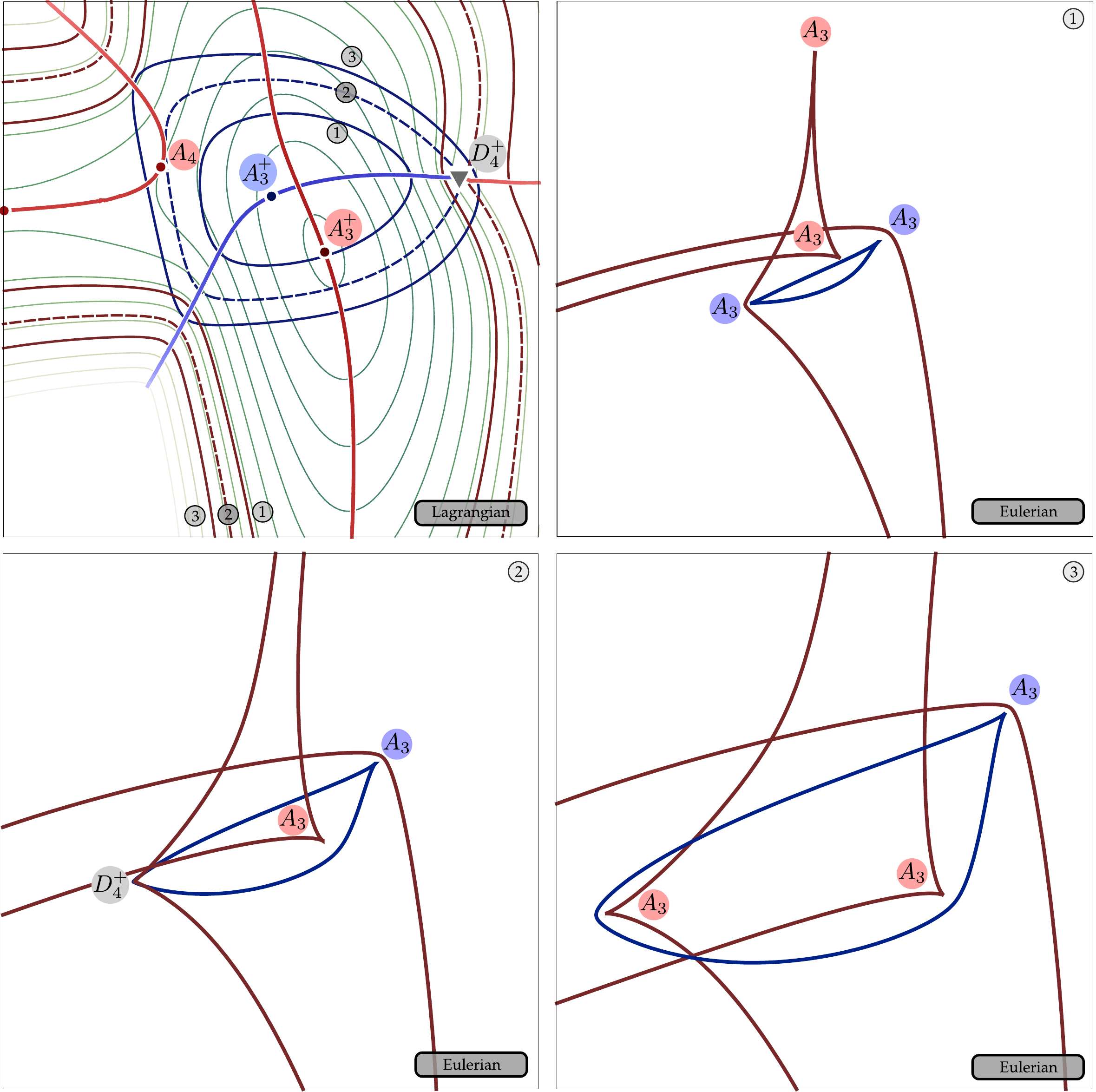}}
\caption{The purse. The setup of this figure is the same as in 
figure\ref{fig:A4}. The three sets of contours in the top left image are shown in Eulerian
view in the other three subfigures. We start at an evolved stage, the
red contour has already passed the $A_3^+$ and $A_4$ points, so the
separate $A_3$ lines are now a single structure. The second eigenvalue (blue)
has formed a pancake inside this structure. This type of conjunction
is often encountered. As the contour of the second eigenvalue is moving out
it passes the $D_4^+$ point. Note that in the Eulerian view this event is mirrored,
it happens on the left-most cusp, due to the change of parity inside the
three-stream region. The cusp changes its colour to red. The second blue
cusp is also nearing the red caustic, but as the corresponding $D_4^+$ point
lies in a region where both eigenvalues are negative, they will never join.}
\label{fig:D4purse}
\end{figure*}
\begin{figure*} 
\centering
\centerline{
	\includegraphics[width=0.95\textwidth]{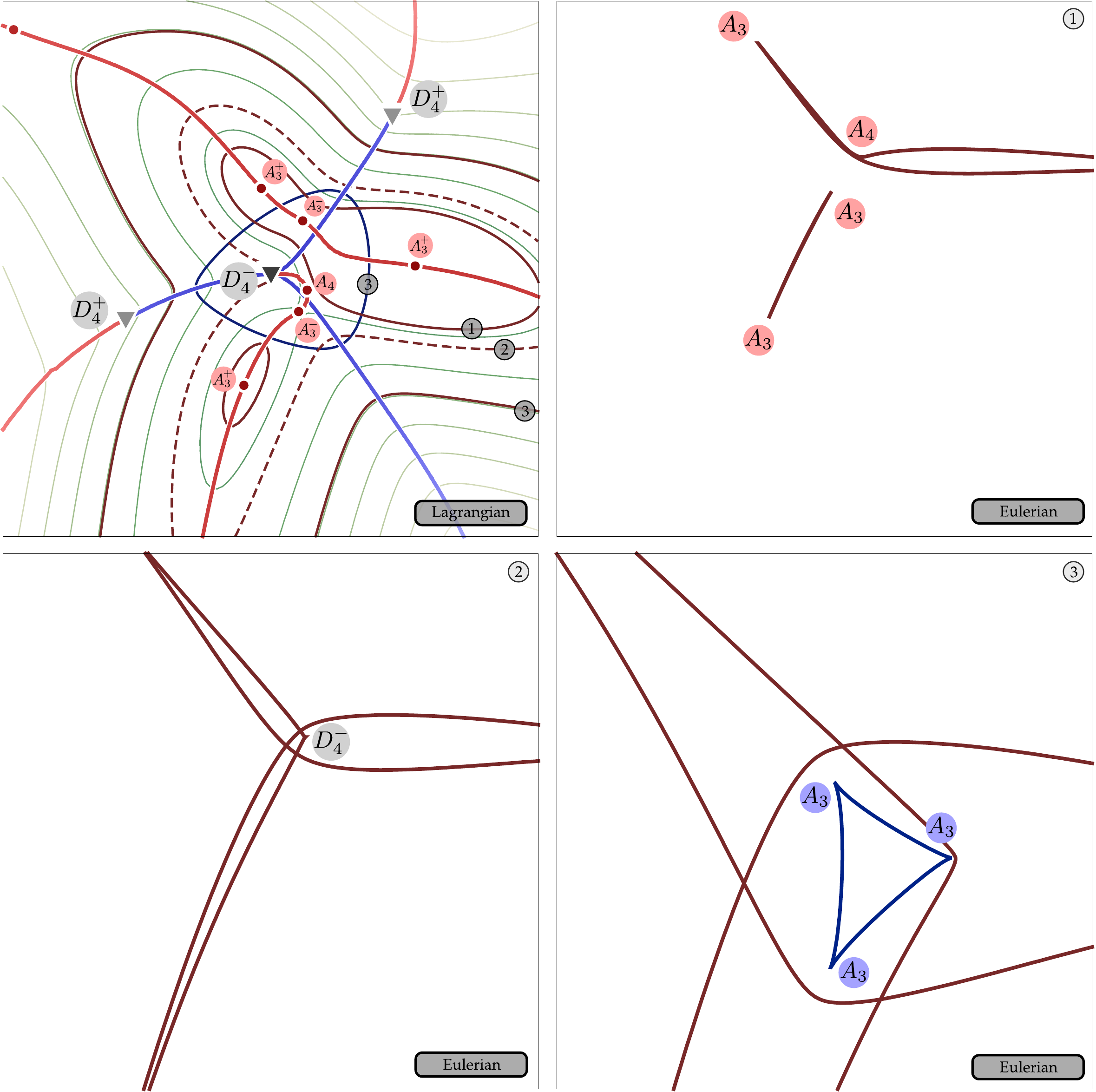}}
\caption{The pyramid. We return to the setting illustrated in figure \ref{fig:dfield}.
Almost all singularities that we know of in 2D are represented in this figure.
In the first image we see two pancakes, one of which is close to getting
a swallow-tail. The resulting cusp merges soon at the bottom most $A_3^-$
point, while the other end heads for the pyramid $D_4^-$. The $D_4^-$ is approached
from three sides, but those are in a very tiny loop that we don't resolve
in this image. The resulting three $\beta$-cusps expand outwards. Two of them
will turn to $\alpha$-cusps at the purses $D_4^+$.}
\label{fig:D4pyramid}
\end{figure*}

\medskip
Differentiation of the secular equation~\ref{eq:sec-eq}, for eigenvalue $\lambda=\alpha$, yields 
an expression for the components of the gradient vector $\nabla_q {\alpha}$,
\begin{eqnarray}
(2 \alpha - d_{11} - d_{22}){\partial \alpha \over \partial q_i} -\alpha(t_{11i} + t_{22i})&\,& \nonumber\\
 + (d_{22}t_{11i}+d_{11}t_{22i}) - 2d_{12}t_{12i} &\,=\,& 0\,.
\label{eq:A3a}
\end{eqnarray}
\noindent Multiplication of these expressions for the gradient components by the corresponding components of the eigenvector 
$\vec{n}^{\alpha}$ (see eqn.~\ref{eq:eigenvd}) leads to the following expression for the inproduct, 
\begin{eqnarray}
I_{\alpha}({\vec q})&\,=\,&d_{12}[(\alpha -d_{22})t_{111} 
+ (\alpha - d_{11})t_{122} + 2d_{12}t_{112}]\nonumber\\
&\,+\,&(\alpha-d_{11})[(\alpha -d_{22})t_{112}\nonumber\\
&\,+\,&(\alpha - d_{11})t_{222} + 2d_{12}t_{122}]\,.
\end{eqnarray}
\noindent To obtain this expression, we have used the fact that the trace of the deformation tensor is 
equal to the sum of the eigenvalues, so that 
\begin{eqnarray}
2 \alpha - d_{11} - d_{22}&\,=\,&\alpha-\beta\,,\nonumber\\
2 \beta - d_{11} - d_{22}&\,=\,&\beta-\alpha\,.
\end{eqnarray}
By means of some algebraic manipulation, we may rewrite the expression for $I_{\alpha}({\vec q})$ into a
more convenient form, 
\begin{eqnarray}
I_{\alpha}({\vec q})&\,=\,&(\alpha-d_{22})\left[d_{12}t_{111}-(\alpha-d_{11})t_{112}\right]\nonumber\\
&\,+\,&(\alpha-d_{11})[d_{12}t_{122}-(\alpha-d_{11})t_{222}]\,\nonumber\\
&\,+\,&2d_{12}[d_{12}t_{112}-(\alpha-d_{11})t_{122}]\,.                      
\label{eq:a3cond2}
\end{eqnarray}
Given that $I_{\alpha}({\vec q})$ is a scalar quantity, we may evaluate it in the eigenvector coordinate 
system, for which $d_{11}=\alpha$, $d_{22}=\beta$ and $d_{12}=0$. The situation in which both eigenvalues 
are equal, $\alpha=\beta$, then leads to the conclusion that 
\begin{equation}
I_{\alpha}({\vec q})\,=\,0\,.
\end{equation}

\medskip
\noindent In other words, given that the condition that $\alpha=\beta$ implies that $I_{\alpha}({\vec q})=0$ and 
thus fulfils the $A_3$ line condition stated in equation~\ref{eq:a3}, we are lead to the conclusion that 
the $D_4$ points are located on $A_3$ lines. 

\subsubsection{The $D_4$ location: $A_3$ termination points}
Given that $D_4$ points belong to both $A_3^{\alpha}$ and $A_3^{\beta}$ lines, 
one might at first expect them to mark the crossing point between these lines. 
However, this turns out not to be the case. Instead, they mark the connection  
points of $A_3^{\alpha}$ and $A_3^{\beta}$ lines, both terminating at that 
specific $D_4$ point. Thus, $D_4$ are points where $A_3^{\alpha}$  and $A_3^{\beta}$ 
lines meet. 

We may understand this by considering the three-dimensional space $(q_1,q_2,\lambda)$ 
defined by the Lagrangian plane and the eigenvalue $\lambda$. This would be the altitude 
map shown in maps like that of figure~\ref{fig:dfield}. When looking at the 2-D map in the 
Lagrangian plane $(q_1,q_2)$, the $A_3^{\alpha}$ and $A_3^{\beta}$ lines may appear to 
intersect. However, this is merely an artefact of the projection of both lines 
on to the $(q_1,q_2)$ space. In reality, they cannot intersect because 
in general $\alpha \geq \beta$: when a given Lagrangian position $(q_1,q_2)$ is 
located on both an $A_3^{alpha}$ line and a $A_3^{\beta}$ line, at that location the 
$A_3^{\alpha}$ line will lie at a higher altitude in the 3D $(q_1,q_2,\lambda)$ 
space. The only exception is when $\alpha=\beta$, i.e. at the $D_4$ points.

The image is one where the $A_3^{\alpha}$ line and $A_3^{\beta}$ line meet smoothly 
at $D_4$ points. One one side of a $D_4$ point we find the $A_3^{\alpha}$ line, 
on the other the $A_3^{\beta}$ line. 

\subsubsection{Purses and Pyramids}
There are two classes of $D_4$ points. One class is the {\it purse} singularity $D_4^+$. 
These points connect one $A_3^{\alpha}$ with one $A_3^{\beta}$ line. The structure of 
caustics and their progenitors in the vicinity of $D_4^+$ singularity is shown in 
figure~\ref{fig:D4purse}.

The second class is the {\it pyramid} singularity $D_4^-$. The $D_4^-$ points form the 
connections of six lines $A_3$, three $A_3^{\alpha}$ and three $A_3^{\beta}$ lines. We may appreciate this in 
figure~\ref{fig:dfield} and ~\ref{fig:D4pyramid}, which shows the caustics and their 
progenitors around a $D_4^-$ singularity. Typically, in the $D_4^-$ situation, 
two of three connecting $A_3^{\lambda}$ lines are small loops. In figure~\ref{fig:D4}, 
we find a small loop of $A_3^{\alpha}$ lines (black) and an even smaller loop 
of $A_3^{\beta}$ lines (grey). Quite often, these loops are so small that an image 
hardly manages to resolve them (see the $D_4$ point in e.g. fig.~\ref{fig:dfield} and 
fig~\ref{fig:D4pyramid}, see appendix~\ref{appendix:d2}). 

The type of the $D_4$ type is determined by the sign of the quantity $s_D$, 
\begin{eqnarray}
s_D &\,\equiv\,& t_{111}^2 t_{222}^2 -3 t_{112}^2 t_{122}^2 - 6t_{111}t_{112} t_{122} t_{222} \nonumber \\
&\,+\,& 4t_{111} t_{122}^3 +4t_{222}t_{112}^3\,.
\end{eqnarray}
If $s_D$ is positive, the singularity is a $D_4^+$ purse singularity. Otherwise, 
it is a $D_4^-$ pyramid singularity \citep{rs-84}.

\subsection{Metamorphoses \& transient singularities}
\label{sec:metamorphosis}
The inventory of singularities in section~\ref{sec:theoryofsing} and the inventory of their 
position within the cosmic deformation field in the previous subsections forms the basis 
for our understanding of the evolution of of structure in terms of that of the network of 
caustic structures. 

As we may already have appreciated, the evolution proceeds via a series of {\it metamorphoses}. A metamorphosis 
-- perestroika in Russian -- is a transient singularity that may still be considered stable in 
the sense that a perturbation of the manifold will only change the time and location of the event: 
a metamorphosis cannot be removed by perturbation. 

Each metamorphosis involving singularities from the $A$-series results in a change in the global topology of the 
network. The systematic inventory of the fundamental singularities in terms of their normal forms that we present 
below allows us to appreciate how a system of folds and cusps changes topology as it passes through the transient 
singularities of metamorphoses. 

A nice illustration of the latter is that in terms of the only two stable types of singularities, stable 
in the sense that they persist in time. These are the \emph{fold} ($A_2$) and the \emph{cusp} ($A_3$) 
singularity (see fig.~\ref{fig:alphaA3p}). Their time evolution is  straightforward, in that it only 
involves their movement. In section~\ref{sec:A3} we have seen that the combination of fold and cusp singularities 
defines the formation of a pancake. Starting from its tenuous embryonic state, a pancake grows in length as the 
bounding cusps at its tip move apart and the pancake region, in between the two cusps and enclosed between two 
fold caustics, gradually thickens. This growth gradually slows down. 

In the physical reality of a gravitational system, these transitions work to increase the complexity of the 
system. A direct manifestation of the process in which matter passes through the shocks of singularities 
is that matter passes into regions with a larger number of streams than before. 

\begin{figure*}      
\centering
\centerline{
  \includegraphics[width=2.1in]{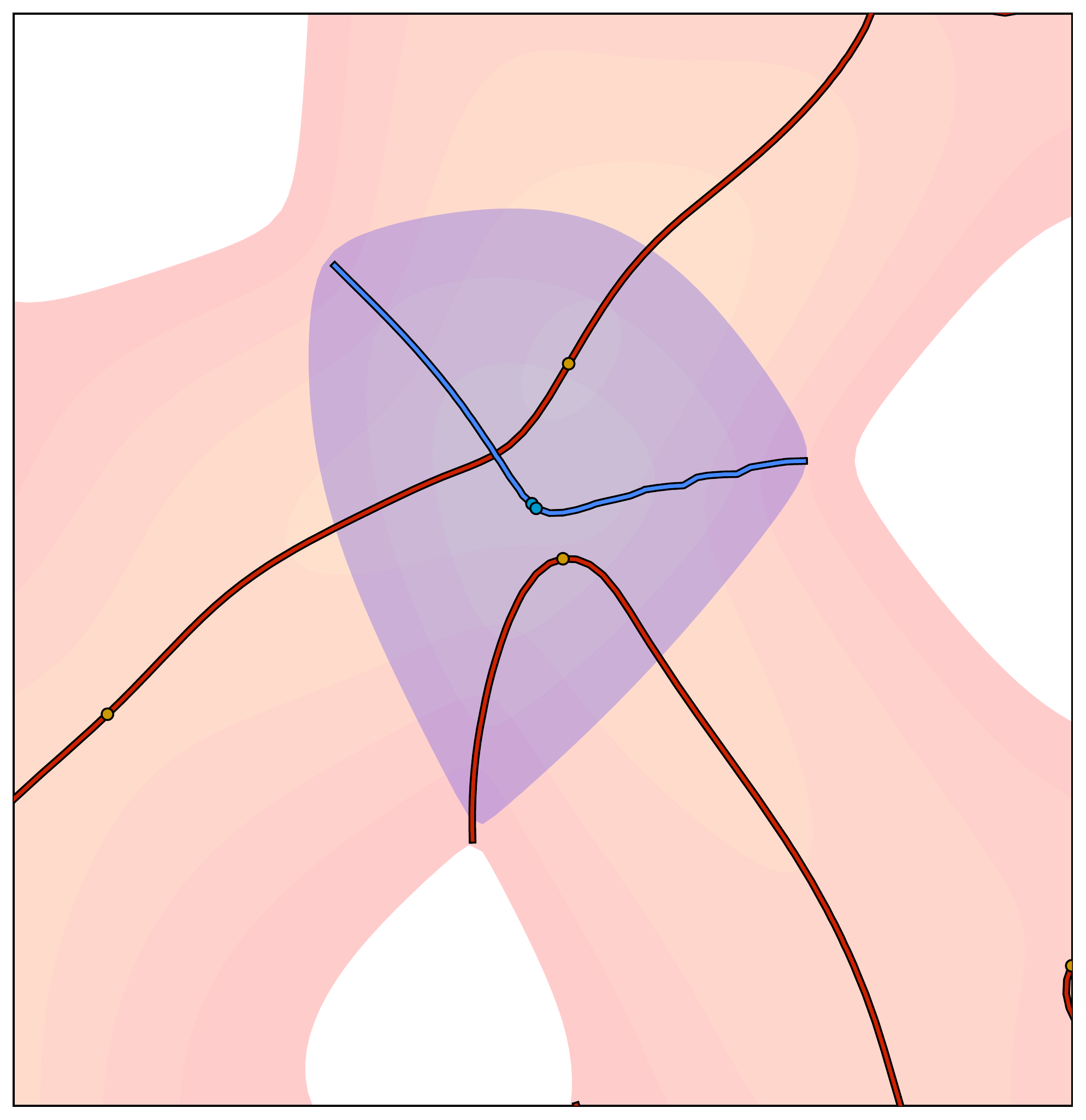}
  \includegraphics[width=2.1in]{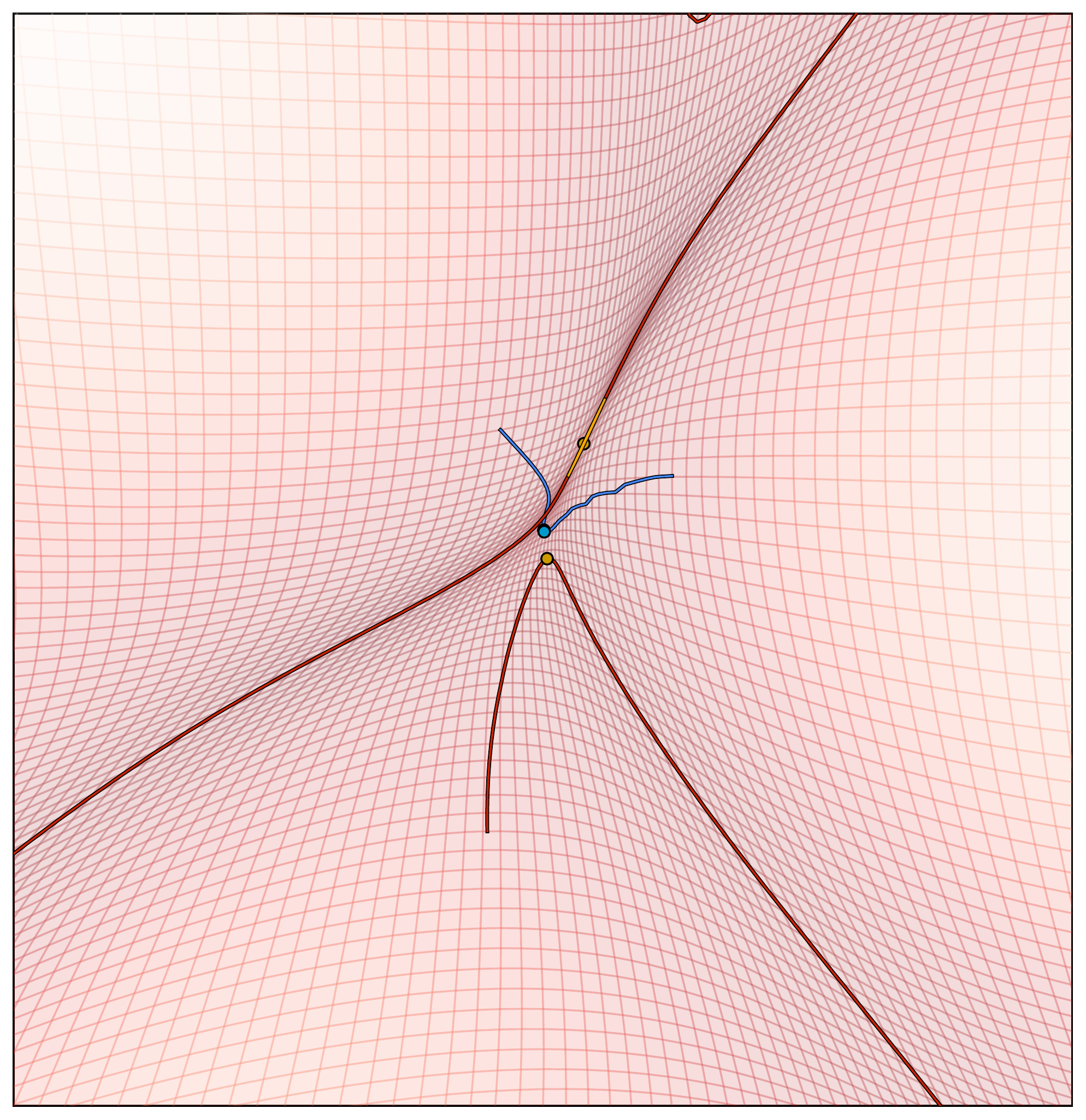}
  	\includegraphics[width=2.1in]{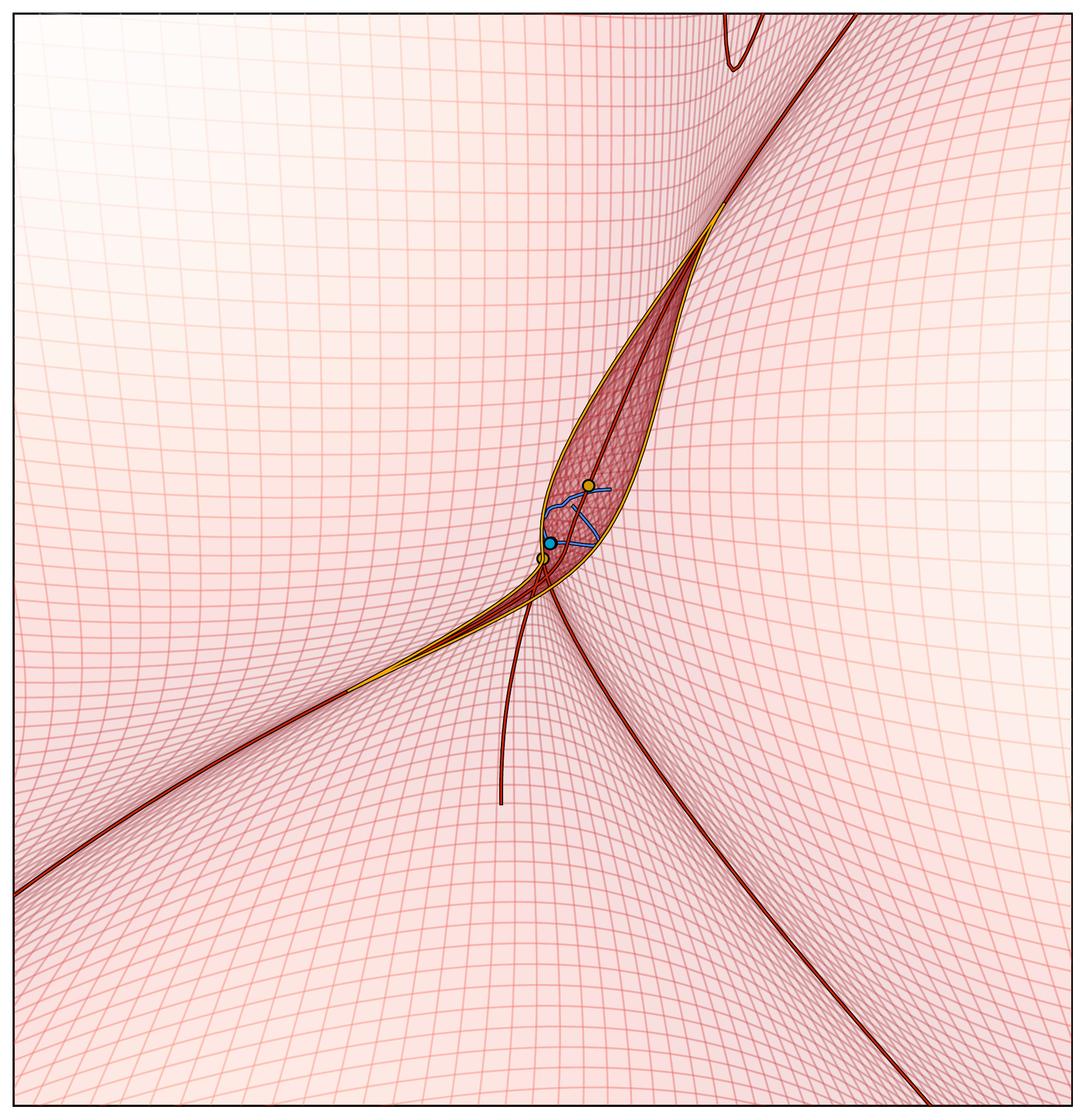}}
\centerline{ 	
  \includegraphics[width=2.1in]{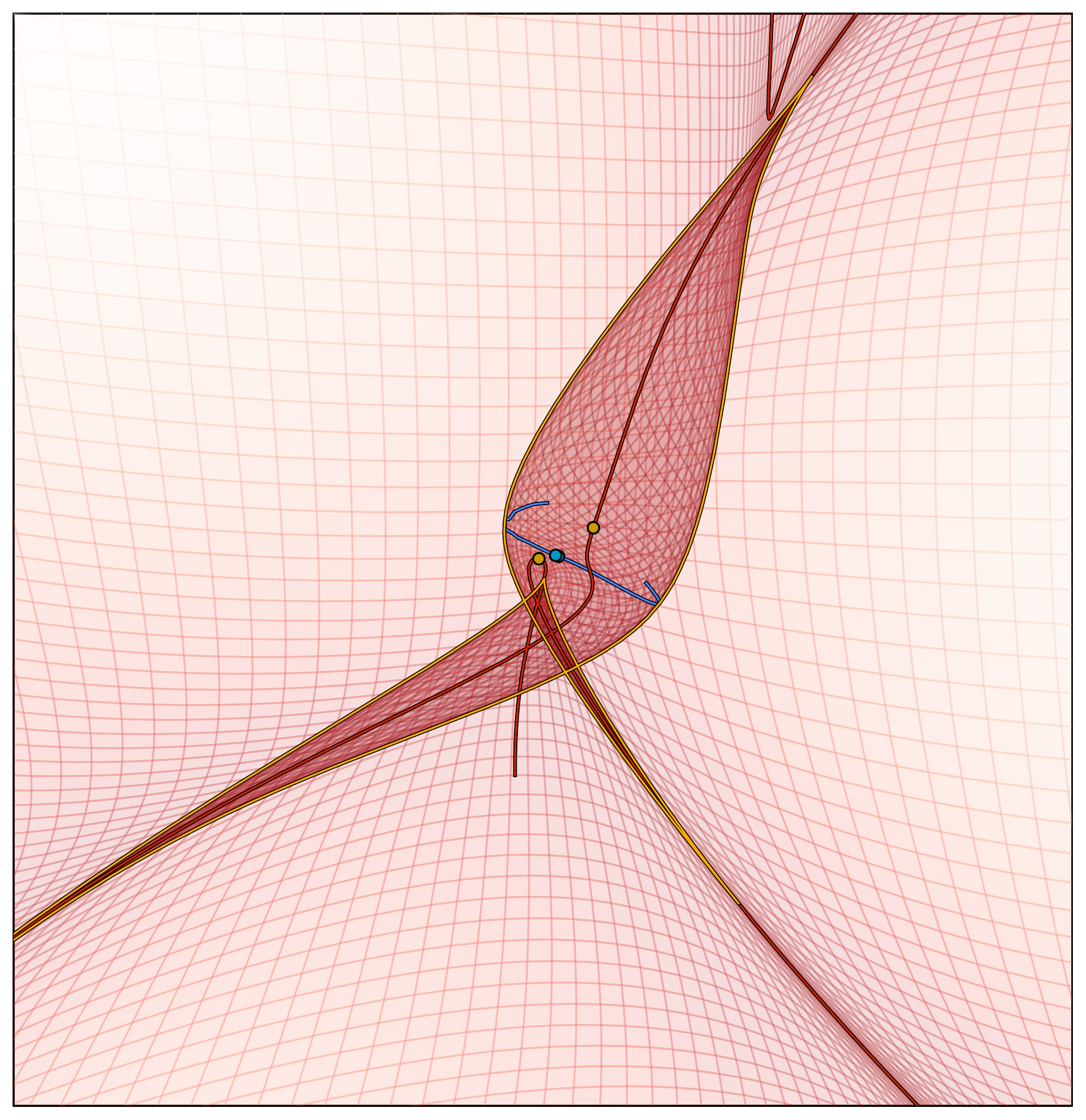}
	\includegraphics[width=2.1in]{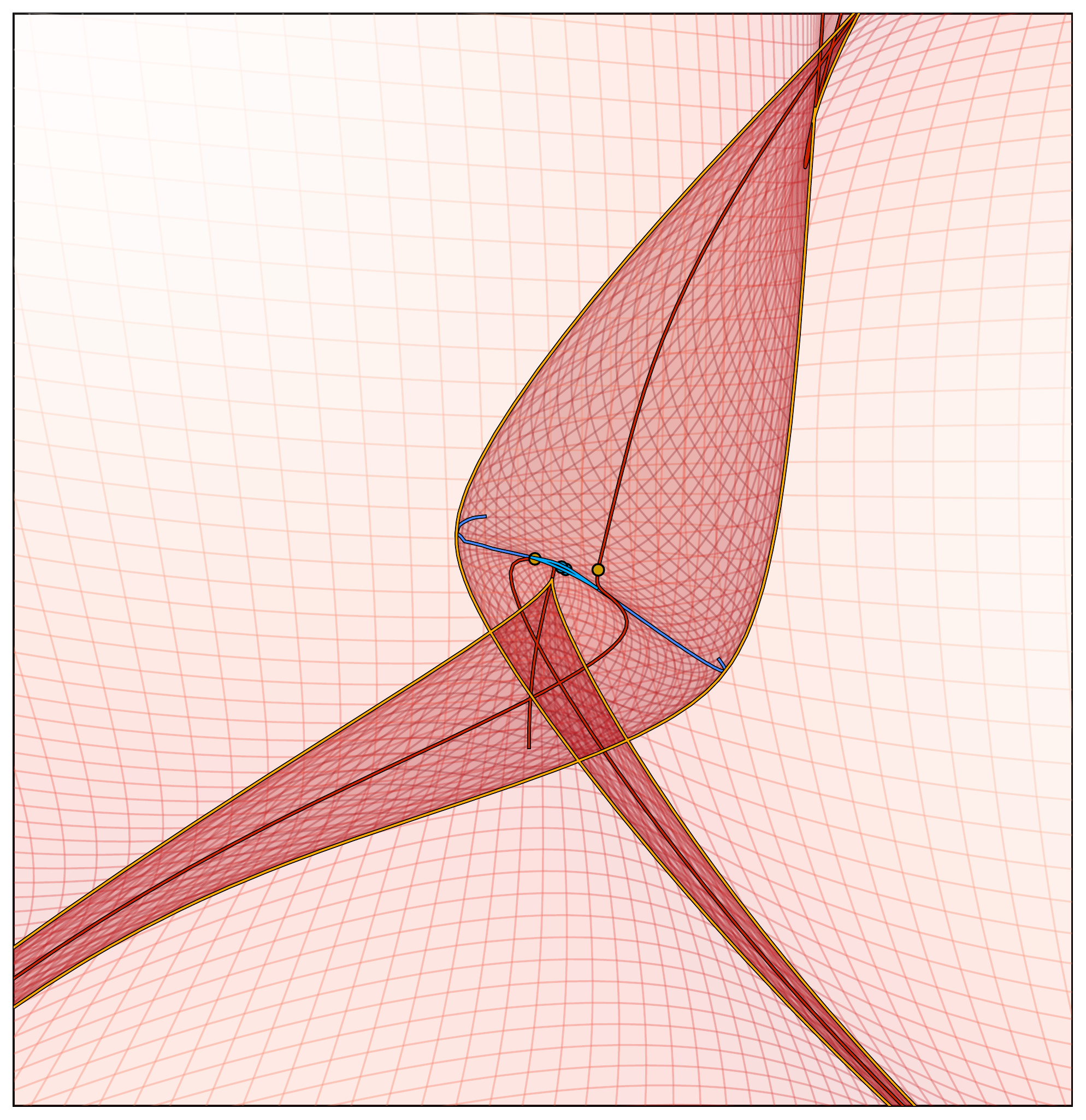}
  \includegraphics[width=2.1in]{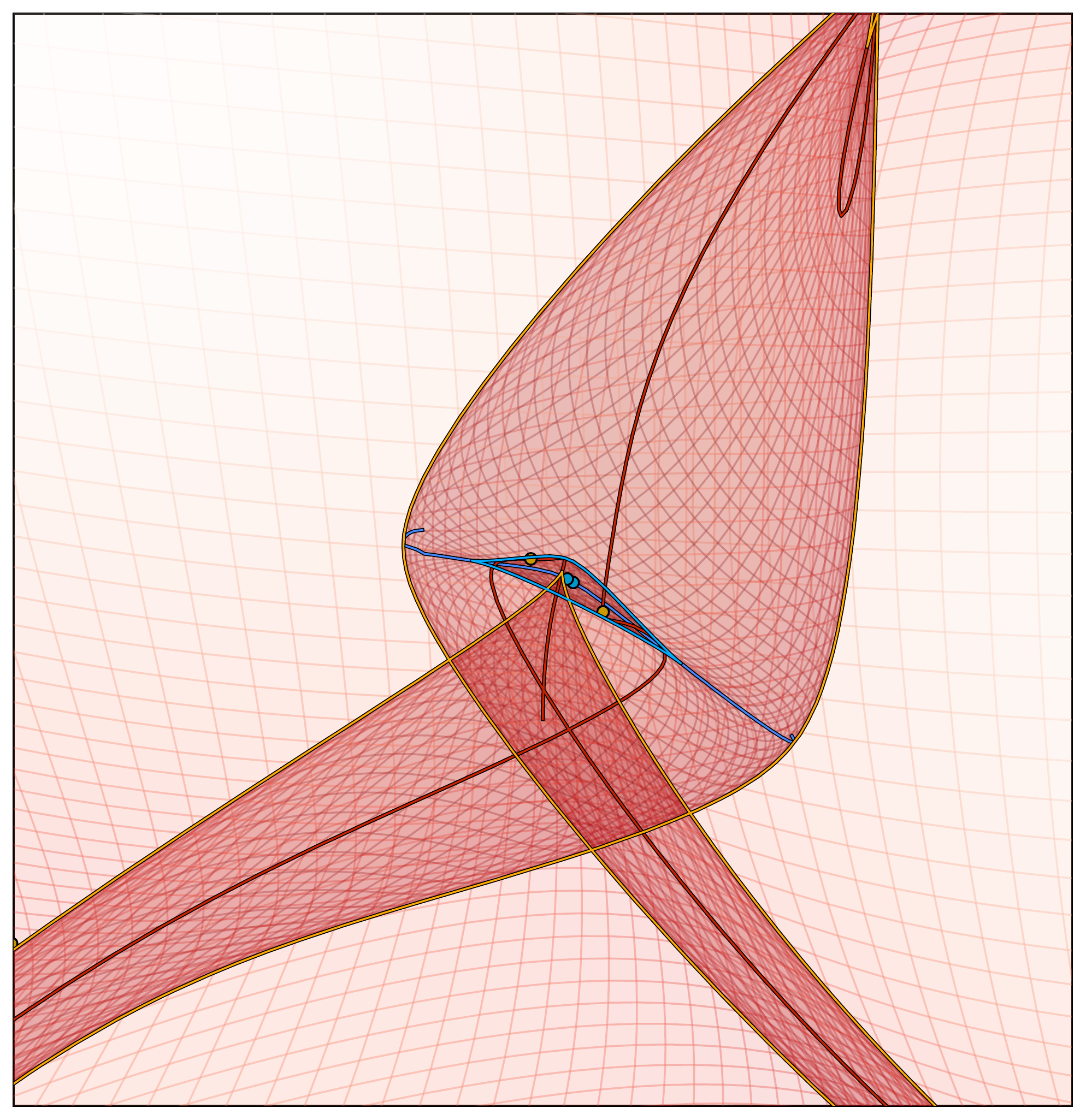}}
\caption{Evolving caustic structure around a cosmic web node. 
Time evolution of the Lagrangian region shown in the top left panel. 
Left top panel: Lagrangian space. $A_3^{\alpha}$ lines -- red, 
$A_3^{\beta}$ line -- blue.  Light brown region shows $\alpha > 0$, blue 
is the region where $\beta > 0$. 
The top yellow dot shows a maximum of $\alpha$, the bottom yellow dot shows $A_4$
and a dot on the left marks a saddle point. 
A blue dot shows a maximum of $\beta$.
Middle top panel: Eulerian space at the time when $\alpha$-pancake just have emerged as 
 a yellow thick line on the top yellow dot above the blue line. The remaining
four panels show the region at different stages of caustic evolution, representing
a highly common sequence of events.}
\label{fig:evolution}
\end{figure*}

\section{the Cosmic Web: \\ 
\ \ \ \ Assembly \& Lagrangian blueprint}
\label{sec:blueprint} 

In the previous section~\ref{sec:singinza}, we have noticed the central role of 
$A_3$ lines in outlining and defining the structure that emerges as a result of 
gravitational evolution. $A_3$ lines form the Lagrangian blueprint of structures 
found in the Cosmic Web, and may be regarded as its spine. 

In this section we investigate the connection between the $A_3$ spine and the 
emerging Cosmic Web. As a step-up towards the systematic discussion, we first 
look at a case study of the structure evolving in a typical specimen of a
key region in the cosmic web. Having established in the previous section the 
identity and role of the various $A$- and $D$-class singularities, it is 
instructive and helpful for our intuition to observe the time evolution of the 
caustic structure in a small and characteristic Lagrangian region. 

\subsection{Caustic Synergy: \\ \ \ \ \ \ \ \ assembly of structure around a node} 
\label{sec:assemblingweb}
Figure~\ref{fig:evolution} shows five stages during the evolution of the 
caustic structure around a small Lagrangian region, shown in the top lefthand 
panel. Following the Lagrangian panel is a sequence of five panels showing 
the evolving structure in the corresponding region in Eulerian space. The 
sequence runs from a very early epoch, in the top central panel, to a 
highly advanced one in the bottom righthand panel. 

The four-dimensional Lagrangian submanifolds that define the phase space 
density field get deformed in time as matter elements flow pursue their 
path $({\vec q},{\vec x})$. The Eulerian position $\vec{x}$ of each mass element is 
dependent on time, while the projection of their position in Lagrangian space, 
${\vec q}$, remains fixed. The projection of the progenitors of caustics 
and $A_3$ lines on to the Lagrangian plane is shown in the top lefthand 
panel of figure~\ref{fig:evolution}, while the changing projection 
to the Eulerian plane is the subject of the additional five panels. 

The projection to Lagrangian space is nonsingular and geometrically 
simpler and more transparent. Moreover, knowledge of the Lagrangian 
projection helps substantially in understanding and making sense of 
the Eulerian projection. 

\subsubsection{The Lagrangian Spine}
In the Lagrangian panel, the light brown shaded region demarcates the regions of 
positive deformation tensor eigenvalue $\alpha$. The light blue shaded 
region has a positive eigenvalue $\beta$. 

Three $A_3$ lines form the spine of the emerging web in this region. The two red 
lines are $A_3^{\alpha}$ type lines, the blue line a $A_3^{\beta}$ line. On the upper 
$A_3^{\alpha}$ line, the yellow dot indicates the position of an $A_3^+$ maximum. Note 
that it is also located within the light blue region. Another dot is a $A_3^-$ minimum 
along the $A_3$ line, a saddle point in the eigenvalue $\alpha$ field. A blue dot 
on the $A_3^{\beta}$ line marks the location of the $A_3^{\beta}$ maximum. 

The lower $A_{\alpha}$ line does not run throughout the entire region: it stops at the 
contour $\alpha=0$. As has been discussed in section~\ref{sec:A3line}, structure does 
not develop in regions with negative $\alpha$. Of high relevance for the emerging 
pattern in Eulerian space is that the dot at the tip of the lower $A_3^{\alpha}$ line 
is an $A_4^{\alpha}$ point.  

The identity of the various lines and points in the Lagrangian region determine 
the nature of the emerging structure in Eulerian space. The five Eulerian panels 
show the evolving density field. In the maps, the density is represented by the 
colour intensity, running from white in the underdense regions to dark red in 
high density collapsed regions. 

The black solid lines are the mapping of the Lagrangian $A_3^{\alpha}$ lines to Eulerian space, 
while the blue solid line represents the corresponding mapping of the $A_3^{\beta}$ line. It is 
very instructive to follow the displacement and gradual merging, morphing, stretching and 
twisting of the $A_3$ lines as we see the development of the structure in the density field. It 
immediately reveals the essential nature of these lines in the formation of the 
geometric pattern: they clearly function as the - flexible - spine of the web. 

\subsubsection{The Eulerian Assembly}
The earliest depicted timestep is that shown in the top central panel. At this stage, 
we observe the emergence of the first $\alpha$ pancake. It forms at and immediately around the 
$A_3+$ point on the upper $A_3^{\alpha}$ line in the Lagrangian panel. At this early stage, the 
pancake is so thin that its two caustic edges lie indistinguishably on top of each other. The 
next stage is shown in the top righthand panel. The central pancake has grown in extent, and 
has attained its classical morphology in which its inner region is enclosed by two 
{\it fold} caustics, which are connected into a single closed contour by the two {\it cusps} at 
its tip. The interior region is a three-stream region. The panel shows the structure 
just before before the emergence of the $A_4^{\alpha}$ {\it swallow-tail} singularity. 

Already just before the $A_4^{\alpha}$ singularity, we can see that the two $A_3^{\alpha}$ lines 
are on the verge of connecting into one structure around a characteristic three-edged backbone. 
At the next stage, shown in the bottom lefthand panel, we find the complex caustic structure 
that formed in the aftermath of the $A_4^{\alpha}$ singularity. The characteristic swallow-tail 
pattern consisting of three protrusions is defined by four caustic edges connected by 
four cusps. Three cusps connect the four caustics at the lower end, while the fourth one 
closes the caustic outline at the top of the configuration (also compare the structure in 
the bottom righthand panel of figure~\ref{fig:A4}). While most of the interior of 
the caustic configuration involves a three-stream region, the small internal zone 
enclosed by four caustics is a region with a five-stream flow field. 

During the most advanced stages of development, shown in the bottom centre and 
righthand panel, we see the growth and expansion of the $\alpha$-caustic 
structure that we found in the third timestep. In addition, we find the 
emergence and formation of $\beta$-pancake, within the confines of the large 
$\alpha$ pancake and around the $A_3^{\beta}$ maximum (blue dot, Lagrangian map). 

The emergence of the $\beta$ pancake goes along with the formation of another 
five-stream flow in its interior. At the last timestep, we find that even a seven-stream 
region has formed in a small part of the $\beta$ pancake. It is the result of the 
intrusion of the inner $A_3$ cusp into the interior of the $\beta$ pancake. This 
leads to an intersection of the two five-stream regions, which produces a 
seven-stream region. 

\subsection{The role and significance of $A_3$ lines}
All point-like singularities are located on $A_3$ lines. The direct implication of this 
is that all dynamically significant processes and events take place in the vicinity of 
the $A_3$ lines. Singularities in the evolving density field are associated with the 
maxima, saddle points and minima of the deformation tensor eigenvalues $\lambda$. In 
addition, also the special $A_4$ and $D_4$ points are located on $A_3$ lines. 

Each $\alpha$--pancake originates at a maximum of $\alpha$. The pancakes is enclosed 
by two {\it fold} caustics, whose progenitors are $A_2^{\alpha}$--lines. Two cusps 
at the tips of the pancake connect the {\it fold} caustics into a closed pancake 
boundary. The progenitors of the cusps are $A_3$ points which traverse 
the $A_3$ line in opposite direction as the pancake grows in time.  In Eulerian space, the 
pancake enclosed by two caustics is a three-stream region, where mass elements 
with from a different Lagrangian origin are crossing each other. 

The saddle points corresponding $A_3^-$ minima along an $A_3$ line are the progenitors 
of the points at which the cusps of two pancakes meet and annihilate, leading to the 
merging of the two pancakes into one larger one.
 

\subsection{Global geometry and topology of $A_3$ lines}
\label{sec:global}
The global geometry and topology of the caustic structure emerging as the density field evolves is to a 
large extent determined by the characteristics of the $\alpha$-field, ie. of the spatial distribution of 
the first eigenvalue $\alpha$ of the deformation field. 

The previous section has taught as that in  particular the $A_3^{\alpha}$ lines are instrumental in setting 
the spatial outline of the cosmic web. 
The evolution of the cosmic web, and in particular that of the connectivity between its 
main elements, can be largely deduced from the run of the $A_3$ lines and the distribution of 
singularity points along their edge. Mainly because collapse along the $A_3^{\beta}$ lines is always 
happening - usually substantially - after that along the $A_3^{\alpha}$ lines, their effect on the 
percolation and topology of the network is at most marginal. 

In principle, an $A_3$-line runs from one $D_4$ point to another. However, the corresponding 
eigenvalue may drop below zero. In reality, structure evolution will not proceed to this 
point as it would never reach it within a finite time. 
One might therefore limit the dynamical evolution to a maximum cosmic time. The limiting time would 
limit the range of values for the 
eigenvalues $\alpha$ and $\beta$. For example, one may limit the time over which the density field 
has evolved to a level at which the density perturbations have a $\sigma_{\delta}(t^{\star}) \sim 1$. 
At this stage the ZA would still give a crude but qualitatively reasonable approximation to the
true evolution of a gravitating medium \citep{cms-93,mps-94}. If at this point of time the 
linear growth factor $D(t^{\star})=D^{\star}$, the minimum eigenvalue up to which we can still 
follow the evolution would be $\lambda^{\star}=1/D^{\star}$. For lower values of $\lambda$ we 
would no longer trust the approximation. 

The segments of $A_3$--lines that satisfy the condition that $\alpha \ge \lambda^{\star}$ 
form a set of disconnected lines in L. These segments correspond to values of 
the eigenvalue $\alpha$ that will be reached within the time in which the 
Zel'dovich approximation remains a good description of the full gravitational 
evolution. The isolated $A_3$ line segments often contain small loops and 
bifurcation pointss. However, in general they are not connected to other 
$A_3$ line segments. By implication, the set of $A_3$ segments will not be 
able to form a percolating network in Lagrangian space.

The situation is somewhat different when we turn towards the emerging 
structure in Eulerian space. When mapped from Lagrangian to Eulerian 
space, different structures that are disjunct in the Lagrangian 
context get connected. The $A_4$ singularities are instrumental 
in settling these connections. We saw a nice illustration of this 
in figure~\ref{fig:A4connectivity}. It shows a pancake which has 
been mapped into the interior of another pancake. The latter 
is on the verge of passing through an $A_4$ singularity. This will 
establish a connection between the corresponding pancake and the 
small central one. In turn, the latter will merge by means 
of an $A_3^-$ saddle point with the nearby swallow tail configuration, 
thereby establishing a pervasive connection between all 
structureal features in the region.

\begin{figure*}    
\centering
\centerline{
  \includegraphics[width=0.95\textwidth]{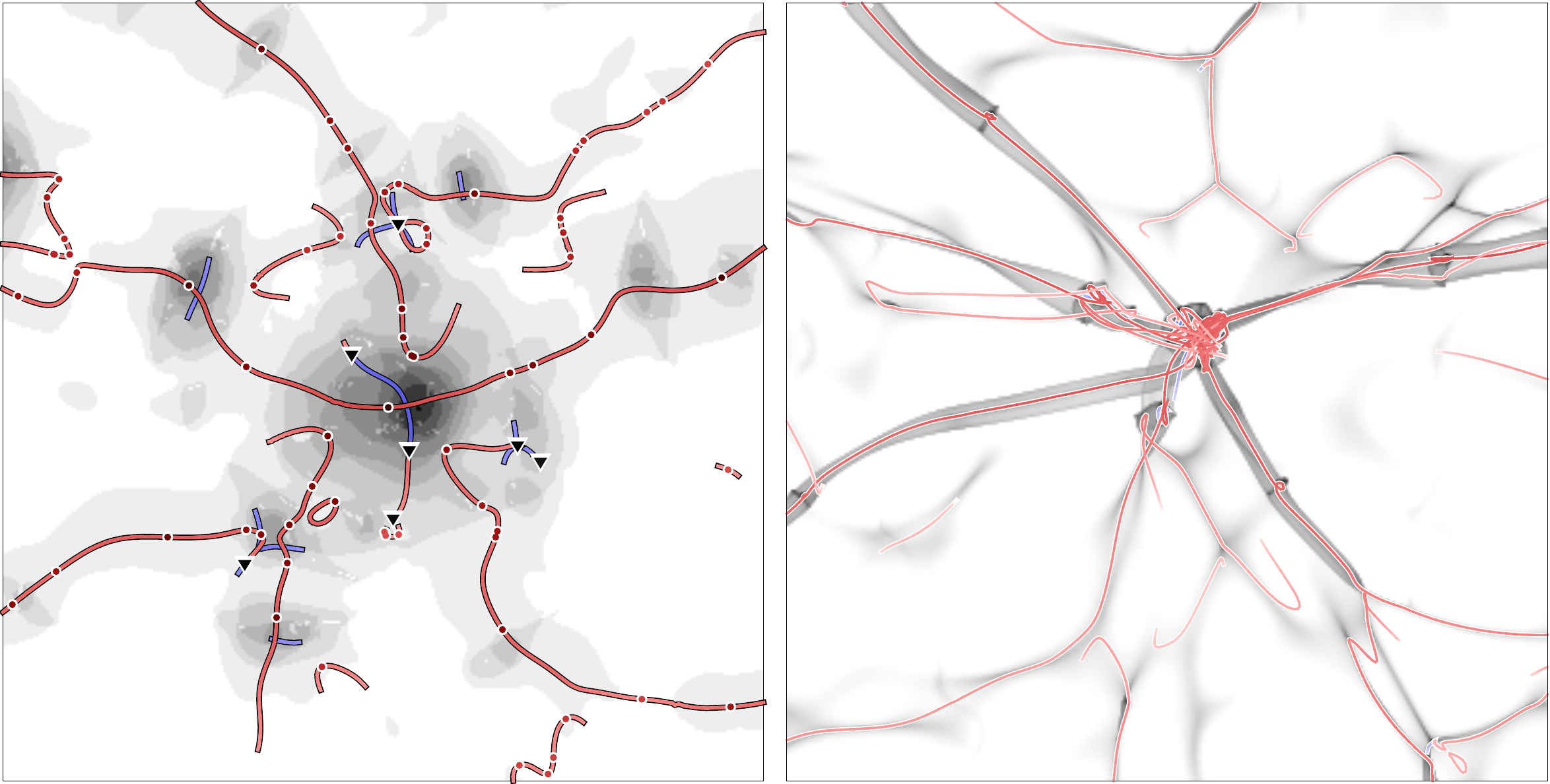}}
\caption{Relating singularities and singularity lines to N-body mass distribution. 
Left panel: the flip-flop field in Lagrangian space is shown
by gray filled contours. $A_3$ lines are shown by red lines. Points on $A_3$ lines
mark the extrema of the $\lambda_1$ field. Triangles mark $D_4$ points. Right panel:
the structure formed at the nonlinear stage in 2D N-body simulation is shown in gray
in Eulerian space. The $A_3$ shown in the left panel are mapped to Eulerian space
using the displacement field obtained in the N-body simulation.}
\label{fig:compare-nonlin}
\end{figure*}
\subsection{Density profiles}
For a complete inventory of the cosmic mass distribution in and around 
the cosmic web, we not only need to understand the connectivity of the 
various components but also the density distribution. The density 
distribution in the vicinity of all singularities is always highly anisotropic.  
Nonetheless, we obtain an adequate first order description of the mass 
distribution around a singularity by evaluating its radially averaged 
density profile.

Analytical expressions for the radial density profile around 
singularities are known. In particular, this concerns the averaged 
profile ${\bar \rho}(r)$ as a function or radius $r$ (in Eulerian space),  
\begin{equation}
\bar{\rho}(r)\,=\,\frac{3 M(r)}{4 \pi r^3}\,,
\label{eq:denprofiledef}
\end{equation}
where $M(r)$ is the mass contained within a distance $r$ from the 
centre of the singularity. 

All singularities have a power-law average density profile,
\begin{equation}
\bar{\rho}(r)\,\propto r^{-n}\,,
\label{eq:denprofile}
\end{equation}
in which the power law index $n$ is dependent on the identity of the 
singularity. Singularities $A_2$,  $A_3$,  $A_4$, and $D_4$ have power law 
slopes $n$ = 1/2, 2/3, 3/4, and 1 respectively \citep{vas-77}. 

The $D_4$ class singularities, at which the two eigenvalues $\alpha$ and 
$\beta$ have the same value, have the steepest density profiles. Notwithstanding 
this observation, it is good to emphasize that the collapse at these points 
is highly anisotropic. This may be immediately inferred from figures~\ref{fig:D4purse} 
and \ref{fig:D4pyramid}.

\subsection{Relation to structure in N-body simulations}
\label{sec:n-body}
Having developed the formalism for the geometrically outlining the backbone 
of the cosmic web, and the identification of the defining singularities and 
their connections, for assessing its relevance with respect to the fully developed 
cosmis mass distribution we resort to a comparison with the outcome of 
N-body simulations. 

The Zeldovich approximation succesfully describes the first stages of the 
development of the nonlinear mass distribution. These are the stages at which 
we see the transition of the cosmic density field from its primordial Gaussian 
origins to the far more complex and rich structure that we observe in the 
present-day Universe. The cosmic web is the manifestation of structure evolution 
at that quasi-linear stage. The subsequent gravitational evolution of the 
cosmic mass distribution is far more complex, and involves the hierarchical 
buildup of structure resulting in a rich multiscale spatial pattern of 
voids, walls, filaments and clumps characterized by a large range of scales 
and densities. 

Here, we show how $A_3$ lines relate to the structure that developed in the  
equivalent two-dimensional N-body simulation. The initial conditions of the 
N-body simulation consist of the same Lagrangian density field used for 
the singularity analysis. The simulation was carried out with a standard 
PM N-body code. The mass resolution and force resolution are equal. The 
simulation consists of 1024$^2$ particles, and forces are computed 
on a uniform 1024$^3$ grid. The initial power spectrum is a power law 
spectrum, 
\begin{equation}
P(k)\,\propto\,k^{-1}\,,
\end{equation}
which is statistically equivalent to the three dimensional power law 
power spectrum $P(k) \propto k^{-2}$. The fluctuation field realization 
has been Gaussian smoothed on a scale of "ngrid" gridcells. 

\subsubsection{Phase Space Sheet \& Mass Cell Parity Flips}
The identification of the progenitors of structures in Lagrangian space is 
based on a novel technique that has been developed within the context of the 
phase space sheet description of cosmic structure \citep{tremaine-99,s-11,shh-12,
abel2012,fns-2012,ney-12,sh-hid-13}. The phase space sheet description looks at the full 
6-dimensional phase space structure of the cosmic mass distribution. Instead 
of following the mass distribution by means of a discrete number of particles, 
it follows the development of spatially extended mass elements. 

Initially, the mass cells define a space-filling spatial tessellation, which is the 
tiling of Lagrangian space. The subsequent gravitational evolution is not 
only followed on the basis of the displacement of the cells but also on the 
basis of their deformation. Deformed cells of the evolving tessellation start 
to overlap as matter streams in the evolving mass distribution start to cross. 
At any one location in Eulerian space, one may determine at each epoch 
the number of overlapping cells as a direct reflection of the number of 
streams. 

Alternatively, one may assume the Lagrangian view and determine the 
number of flips of parity experienced by mass cells. As they get deformed 
to the stage at which the cells pass through a singularity, they undergo 
a flip. The front- and backside of the cells get interchanged. The number of 
flips that the cells undergo corresponds directly to the number of times the 
corresponding mass element has been passing through a singularity. In turn, this  
directly reflects the number of resulting streams \citep{sh-hid-13}. 
For example, mass elements that have experienced singular collapse only once 
form the inner stream  of pancakes. 

As a means of connecting the dynamical state of evolution to the primordial 
mass distribution, our analysis follows the number of flips of mass cells and relates 
this to the number of times the mass element has been passing through 
a singularity. By following the number of parity flips over time, we count 
the number of times a mass element passes through a fold. This {\it fold} field 
is discrete, given the number of singular 
collapses can only be discrete. The morphology of the fold field is similar 
to the ORIGAMI formalism and the concept of parity fields introduced by \cite{ney-12}. The major 
differences with ORIGAMI and parity fields is that the flip-flop field values are not limited 
to two - as in the case of parity fields - or to three - as in the ORIGAMI formalism. Also, 
we should point out that ORIGAMI is based on a different computational technique. 

\subsubsection{Case Study}
Figure~\ref{fig:compare-nonlin} makes a comparison between the Lagrangian 
constellation around a primordial density peak and the resulting nonlinear 
mass distribution in Eulerian space. 

The central peak in the Lagrangian image (lefthand panel) depicts the major 
progenitor of the halo. The greyscale field represents the flip-flop field, with 
darker shades corresponding to a higher number of flip-flops and white 
representing the areas that still did not experience a state of 
singularity. The lightest shade of grey, to be found in the outer 
parts of the central halo and its nearby neighbouring regions, 
indicates the mass elements that have passed through a singurlarity 
only once. 

\begin{figure*}
	\centering
	\centerline{\includegraphics[width=\textwidth]{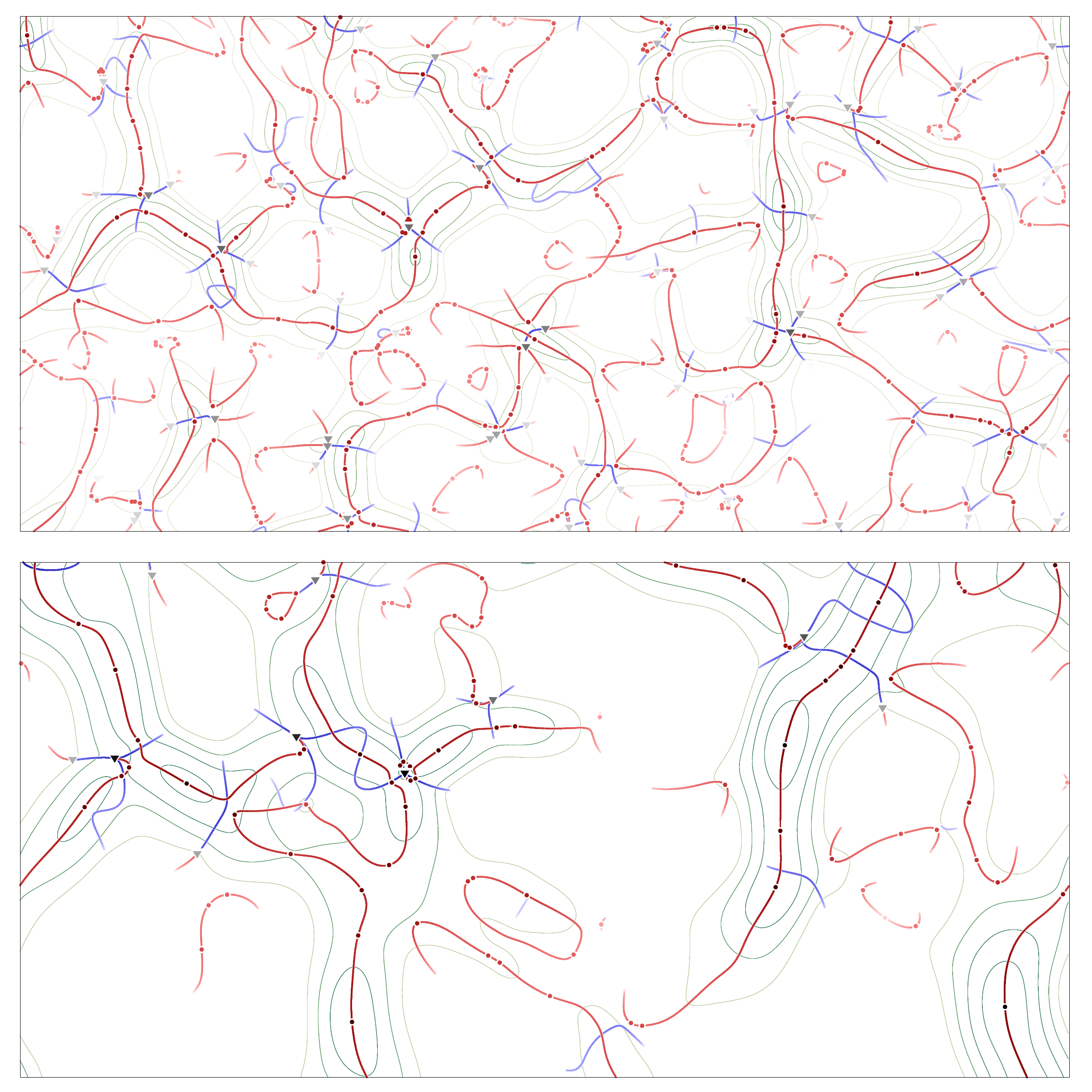}}
	\caption{Singularity structure as function of power spectrum. $A_3$--lines for two different 
        powerspectra, $P(k) = k^n$, $k = 4$ (top), $k=-1$ (bottom) both smoothed at the same scale. 
        As expected for power spectra with a positive index, the pattern
	is much more regular. What is even more intersting is that the clustering statistics of the singularities
	seem to change in a non-trivial way. This will be the topic of a follow-up paper. }
	\label{fig:power}
\end{figure*}
Superimposed on the greyscale image are the thick solid red lines that 
indicate the $A_3$ lines connecting to the peak and the ones that are to be 
found in its surroundings. Seven major $A_3$ lines are seen to exit from the 
central peak. In the Lagrangian plane, the $A_3$ lines pass through progenitors 
of other halos that merged wiith the central halo. In Lagrangian space, two of 
the $A_3$ lines are connected at the center, while the remaining five are disconnected. 
While not connected in Lagrangian space, they will get intertwined through $A_4$ singularities 
as they are mapped to Eulerian space (sect.~\ref{sec:assemblingweb} and 
sect.~\ref{sec:global}). 

In the Eulerian image (righthand panel), the central halo very much resembles 
a "knitting ball". This is an immediate consequence of the multiple collapses 
of the mass elements making up the halo, and is a reflection of this process 
being directed by the $A_3$ lines. 

Surrounding the halo, we find six pancakes - of a variety of lengths - connecting 
to the central halo. Overall, we may recognize and identify the various pancakes 
with the set of $A_3$ lines. These appear to form a reasonably good skeleton for the 
structure that we see emerging in Eulerian space. The pancake in the upper righthand part 
of the panel is the result of a merger of two pancakes, each corresponding to a different $A_3$ 
line in the same region of the Lagrangian plane. The bifurcation of the $A_3$ line is clearly 
standing out, and reflects the fact that the merger of the pancakes has not yet been completed, 

\subsection{Perturbation Power Spectrum and \\ \ \ \ \ \ \ \ Skeleton Singularity Structure}
Having obtained insight into the identity, location, connections and role of 
singularities in the buildup of the cosmic web, we may turn towards the issue 
of the dependence of the spatial configuration of $A_3$ lines and 
the related singularities on the cosmological scenario. 

There have not been many systematic studies that address the dependence 
of the cosmic web characteristics on the underlying cosmology. The one 
exception are several recent studies that have looked at the properties 
of voids as probes of the nature of dark matter \citep{hellwing2010, li2011} and the 
nature of dark energy \citep{parklee2007,lavaux2010,lavaux2012,bos2012}. One of 
the problems for using filaments and walls as probes for the cosmology, is our 
still existing lack of a real quantitative understanding of their structure and 
connectivity. 

In this study we have demonstrated that most of the structure of the cosmic 
web can be condensed in terms of well established and defined 
mathematical concepts, $A_2$ and $A_3$ lines and the classes of 
$A$- and $D$- singularities. In an follow-up publication we will 
present a detailed and systematic study of the spatial distribution 
of singularities and $A_2$ and $A_3$ lines as a function of the 
underlying cosmology. Here we provide a quick impression of what we 
may expect. 

A good impression of the high sensitivity of the cosmic web 
singularity structure in two different cosmologies can be 
obtained from figure~\ref{fig:power}. It shows the location of 
$A_3$ lines and $\alpha$ eigenvalue contours, along with the 
corresponding singularity points, for two realizations 
of the mass distribution in the Lagrangian plane. The top panel 
concerns the pattern for a scenario with a $P(k)\propto k^4$ power law 
power spectrum, dominated by high frequency components, and 
one with a much more large scale character, $P(k)\propto k^{-1}$. 
Both realizations have been smoothed on the same scale. 

The spine in the $P(k)\propto k^{-1}$ cosmology (lower panel) is 
dominated by some very long and stretched $A_3$ lines. The weblike 
pattern in the $P(k)\propto k^{4}$ cosmology seems to have a 
much more  unordered complexion. The cosmic web in that scenario is 
assembled out of many short $A_3$ lines. Still, it has a 
clearly outlined cellular structure, in which the $A_3$ 
lines appear to loop around areas of low $\alpha$ values 
(usually low density areas). 


In summary, we may conclude that the description of the 
Lagrangian field structure in terms of singularity lines and points 
is a very promising and potentially powerful way of exploiting 
the cosmological information content of the cosmic web.


\begin{table*}
	\centering
	\begin{tabular}{| l | l | l | l |}
		\hline
		Family		& type  & name & Eulerian dimension\\
		\hline
		$A$--series	& $A_2$ & fold & surface \\ 
				& $A_3$ & cusp & curve \\ 
				& $A_4$ & swallow-tail & point \\ 
				& $A_5$ & butterfly & transient \\
		\hline
		$D$--series	& $D_4^-$ & elliptic umbilic (purse) & point \\ 
				& $D_4^+$ & hyperbolic umbilic (pyramid) & point \\ 
				& $D_5$ & parabolic umbilic & transient \\
		\hline
	\end{tabular}
	\caption{Inventory of 3D singularity classes.}
	\label{table:catastrophes}
\end{table*}
\section{The third dimension}
Following the extensive two-dimensional analysis described in the previous sections, 
we need to discuss how it relates to the structure we find in the fully three-dimensional 
world. While a comparable fully detailed analysis of the three-dimensional singularity 
structure in the Zel'dovich approximation will be the subject of a separate upcoming 
publication, we here discuss some relevant observations. These pertain mostly to the 
relevance of the 2D discussion to the three-dimensional situation.
 
It may not come as a surprise that the three-dimensional singularity structure of the 
patterns generated via the Zeldovich approximation is considerably more complex than in 
the two-dimensional situation. Nonetheless, the detailed assessment of the two-dimensional 
singularity structures presented in this study includes several important and relevant 
aspects that can be directly related to the equivalent three-dimensional situation. 

A range of key observations on singularities, caustics and catastrophes in 3D may be 
made. In the following, we shortly discuss the most important ones.

\subsection{Caustic Families}
In the two-dimensional situation, the 2 eigenvalues of the deformation tensor 
field express themselves in the existence of two families of caustics, the 
$\alpha$-- and $\beta$--caustics. Given the three eigenvalues of the corresponding 
deformation tensor in 3D, there are three families of caustics, the 
$\alpha$--, $\beta$-- and $\gamma$--caustics.

On the basis of the corresponding transformation, the global geometry and topology of 
the cosmic web in three dimensions is determined by the set of $A_3^{\alpha}$--surfaces. 
Other singularities live on the set of $A_3^{\alpha}$--surfaces. They are the equivalent of the 
set of $A_3^{\alpha}$--lines, and $A_3^{\beta}$--lines, that we found in this study as the features 
that define the spine of the cosmic web. 

\subsection{2D singularities and their 3D equivalents}
For those 3D singularities that have an equivalent singularity in 2D, 
we find - taking into account trivial generalizations - that the structure 
of the caustics remain largely equivalent. One particular aspect for which 
this is always true is the density profile near singularities. These turn 
out to remain the same for equivalent singularities in different dimensions. 

A particular example is that of the simplest caustic, the $A_2$. In 1D, the $A_2$ caustic 
is a point and in 2D it is a line. In the equivalent three-dimensional situation, it 
is a surface. As was pointed out above, the $A_2$ caustic retains the same 
density profile near the singularity as in two dimensions, $\rho(r) \propto r^{-1/2}$. 

The other type of singularity which also exists in 1D is the $A_3$ singularity. 
In one-dimensional space $A_3$ singularities are located at a specific set of singular 
points. At these points, they will only exist as singularity at one particular 
instant of time. In two dimensions, they are located on $A_3$ lines and at any 
one instant materialize as {\it cusps} at isolated points along those lines. In 
three-dimensional space, they are located on surfaces and at any specific instant 
they materialize on a set of lines. In both 1D, 2D and 3D, the mean density profile 
of these singularities is the same, $\rho(r) \propto r^{-2/3}$. 

\subsection{New singularities}
In addition to the classes of singularities that we could also identify in the 
two-dimensional situation, there are new types of singularities in the 3D 
situation. The emergence of new types of singularities in higher dimensional 
situations is not unexpected. For instance, the $A_4$ and $D_4$ singularities that 
we found in this 2D study do not exist in 1D. 

In three dimensions, the list of singularity classes is extended by that of 
$A_5$ and $D_5$ singularities. They are the only new singularities emerging 
in the three-dimensional situation. The $A_5$ and $D_5$ singularities 
have a considerably more complex structure than the singularities which 
have two-dimensional equivalents. This is not unlike the substantially more complex 
structure of the $A_4$ and $D_4$ singularities with respect to the structure 
of the $A_2$ and $A_3$ singularities. The latter exist also in 1D and have a more 
basic geometry. 

\subsection{A and D singularities: 2D versus 3D}
In the previous sections we had observed that there is a principal difference between 
$A$- and $D$ singularities. While the $A$-singularities are determined by only one 
deformation tensor eigenvalue, $D$ singularities are determined by the values of two eigenvalues. 
Note that this also implies that $D$-singularities cannot exist in 
one-dimensional space. 

In three-dimensional space, up to three eigenvalues may be involved in 
determining a singularity. In principle one might therefore expect the 
existence of an additional new class of singularities. This turns out not 
to be the case. There is no additional new singularity class, with the 
exception of the trivial case of $D_{\beta = \gamma}$--singularities. These 
have exactly the same topology as $D_{\alpha = \beta}$--singularities. 

The reason that there are no additional singularities of a new type is 
quite straightforward. In a generic deformation field in 3D there are no 
points where $\alpha = \beta = \gamma$.

\subsection{The three-dimensional singularity inventory} 
To summarize, in table~\ref{table:catastrophes} we list the full inventory of catastrophes 
that exist in 3D space. 

Also in three-dimensional space it is possible to extract a skeleton 
of surfaces, lines and nodes on the basis of a singularity analysis equivalent 
to the one discussed for the two-dimensional situation. 

Filamentary structures are traced by swallow-tail caustics, as well as by 
$D_4$ umbilics. From the experience in the two-dimensional situation, we know that 
umbilics are never far from a swallow-tail. However, the inverse need not be 
true: there are swallow-tails that emerge without any umbilic in their 
vicinity. Given the fact that the 2D case is to some extent a section of the 
3D situation, the same is still true in three dimensions. 

From this, we may conclude that in three dimensions filaments will have 
a diversity of morphologies, each marked by a highly complex structure.

\begin{figure*}    
\centering
\centerline{
  \includegraphics[width=0.95\textwidth]{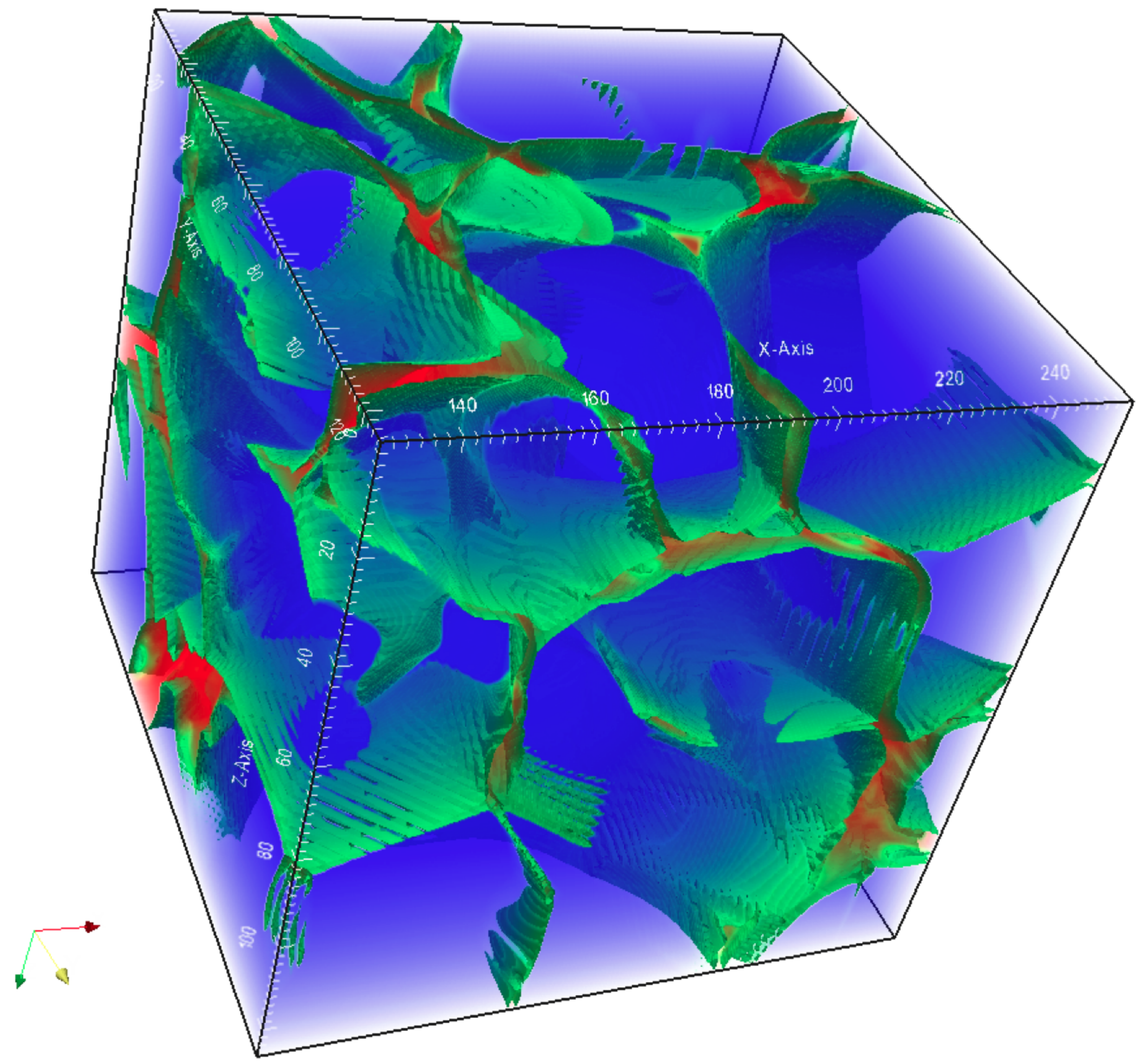}}
\caption{Caustic structure of ZA evolved mass distribution in 3D. This is a three--dimensional analogy of 
the top right panel in figure \ref{fig:overview}. The $\alpha$--caustics are shown in green and 
$\beta$-- caustics in red. The $\beta$--caustics are seen only through
 the openings in the surface of $\alpha$--caustics by the box faces.
This is the reason why one cannot see $\gamma$--caustics.}
\label{fig:pancakes-voids3d}
\end{figure*}

\subsection{Singularities in 2D and 3D: differences}
An illustration of three dimensional structure predicted by the the ZA in 3D is
shown in figure~\ref{fig:pancakes-voids3d}. The caustic surfaces are computed 
using the tessellation of the Lagrangian submanifold and subsequently 
projected to Eulerian space, following the description by \citet{shh-12}. 
The resulting figure is a three--dimensional analogy of the top right panel of 
figure~\ref{fig:overview}, including the $\beta$--caustics shown in red. 

Comparing the characteristic weblike structure in figure~\ref{fig:pancakes-voids3d} with that 
obtained by topological techniques for the dissection of the cosmic mass distribution on the 
basis of the watershed transform \citep{platen2007}. Telling examples for related results are 
those that are obtained by the spineweb technique of \cite{aragon2010} and the skeleton 
technique descibed in \cite{scp-09}. Also other methods for the dissection of the cosmic 
spatial mass distribution into its individual structure components, such as the 
multiscale morphology MMF/Nexus formalism \citep{aragon2007,cautun2013} and the 
tidal and shear tensor methods, advocated by \cite{hahn2007,forero2008} and \cite{libeskind2012}, 
produce spatial patterns that resemble the one seen in figure~\ref{fig:pancakes-voids3d}. 
For example, figure~8 in \cite{aragon2010} (also see fig.~15 in \cite{weygaert2009}) and 
figure 4(a) in \citet{scp-09} reveal a striking qualitative similarity of the weblike structure 
obtained by the topological spineweb and skeleton techniques and the geometric singularity 
structure in the Zel'dovich approximation shown in figure~\ref{fig:pancakes-voids3d}. 

In one sense, there is a profound link between topology-based pattern 
identification and the description of the dynamically evolving mass 
distribution in terms of its singularity structure. Both are intimately 
linked to Morse theory. In terms of the structural analysis of the cosmic 
mass distribution, Morse theory was first invoked by \cite{colombi2000}. 
They used it to characterize the structure of the primordial field of 
Gaussian density fluctuations. However, a complete characterization in  
terms of the full array of complex structural components seen in the 
observed pattern of the cosmic mass distribution became only possible 
after \cite{platen2007} had introduced the watershed transform. 

The similarity between the topology-based identification of the components of 
the cosmic web and the structure encountered in the Zeldovich approximation 
demonstrates that the latter represent a viable framework for the analysis
of N-body simulations. In particular, this concerns the relation between the 
initial density and potential field in Lagrangian space and the resulting 
web in Eulerian spacee. 

We should emphasize that the topological structure identification advocated by 
\cite{aragon2010} and \cite{scp-09} is applied to the evolved matter distribution 
at a particular cosmic epoch. They are not dynamical descriptions that invoke the 
initial field of density fluctuations. Instead, they are more intent on connecting 
the outcome of N-body simulations to the structure seen in the observational 
reality. 

Instead, when the analysis described in this paper is extended to 
its three-dimensional setting, it will provide a dynamical formalism for 
connecting the primordial Lagrangian density field to the weblike structure 
obtained in simulations. We expect that the combination of the topological 
structure identification methods and our dynamical formalisms will lead to 
a set-up in which observational structures can be more directly and 
significantly related to the initial cosmological conditions than 
hitherto possible with existing techniques. 

\section{Summary \& Outlook}
\label{sec:summ}
In this study, we have presented and extensively discussed a comprehensive geometric model 
for the formation of the caustic structure in the frame of Zeldovich approximation \cite{z-70,sz-89}. 
Via the identification of singularities appearing in the evolving mass distribution, and 
by establishing the geometric nature of their connections, our model represents 
a direct analytical description of the dynamical origin of the spine of the cosmic 
web \cite{bkp-96,wb-08}. 

In this study we have restricted ourselves to the two-dimensional equivalent of the 
full three-dimensional system. Because of the complexity of the mathematics involved, 
catastrophe theory of Lagrangian singularities, it is necessary to build an intuition 
for the relevant mathematical and physical context before one can embark on the 
considerably more complex three-dimensional situation. Moreover, although the 
two-dimensional case can not really be directly compared with the full three-dimensional 
structure of the cosmic web, it turns out to provide us with most of the concepts, 
principles and language that pertain in 3D space. 

The Zeldovich approximation \cite{z-70,sz-89} has been demonstrated to be a good theoretical 
model for the evolution of cosmic structure into the mildly nonlinear regime of structure 
formation. It crudely, yet conceptually correctly, describes the onset of the non-perturbative 
quasi-linear regime.  This includes the emergence of multistream flow regions and 
caustics \citep{drs-73,sz-89,buchert89,b-92,cms-93,mps-94,msw-94,cs-95,mbw-95,ymm-98,ymgm-06}. 
Its main virtue is that of a good approximate description of structures on scales 
larger than the scale of nonlinearity, while erasing structures on smaller scales. 
In the currently popular hierarchical scenarios of structure formation, such as the 
$\Lambda$CDM standard cosmology, small scale structure does not heavily influence the 
structure that emerges on larger scales \citep[see e.g.][]{little-91}. This implies  
the Zel'dovich approximation to remain a reasonably good description for the 
large scale Universe. 

We have put the Zeldovich approximation into the geometric framework used in the 
mathematical theory of singularities and catastrophes \citep{t-69,zeeman77,a-82,a-83}. 
Our analysis is directed towards understanding the structure and projections of the 
Lagrangian submanifold, the four-dimensional phase-space ($\vec{q},\vec{x}$). From 
catastrophe theory, it is known that there are only a limited number of different 
singularity classes. We commence our study with a systematic identification of 
these singularities. We follow the well-known ADE classification of catastrophes 
introduced by Arnold, yielding the cryptic notation of $A_2$ singularities for 
\emph{fold} singularities, the $A_3$ \emph{cusp} singularities, the $A_4$ 
\emph{swallow-tail} singularity and the $D_4$ umbilic singularities. Central in 
this characterization is the analysis of the Lagrangian spatial structure of the 
two fields of eigenvalues $\alpha$ and $\beta$ of the deformation tensor that 
characterizes the displacement field in the Zeldovich approximation. 
We study the singularities in Lagrangian space, in their identity as the 
progenitors of caustics in Eulerian space. We combine this with an assessmen 
of the corresponding evolution of the Eulerian environment of the emerging caustics. 

A crucial and fundamental singularity is the $A_2$ \emph{fold} singularity, which 
is the progenitor of the well-known \emph{pancake} that has become one of the most 
familiar features in the structure that we see emerge in Zeldovich descriptions 
of the evolving cosmic mass distribution. In addition, we provide an impression 
of the additional point--like singularities - $A_3^+, A_3^-, A_4, D_4^+$ and $D_4^-$ - 
on the basis of illustrations of three-dimensional projections of the corresponding 
phase-space manifolds. Crucial in the dynamical context that we are considering is that 
they are stable under perturbation. This implies that they remain in existence, in 
one form or the other, into the nonlinear regime of gravitational instability. 

The key finding of our study is that in Lagrangian space all point--like 
singularities - $A_3^+, A_3^-, A_4, D_4^+$ and $D_4^-$ - are located on characteristic 
one-dimensional lines, the $A_3$-lines. In Lagrangian space, the outline of these 
lines is defined by the fundamental structure equation~\ref{eq:a3}. This definition is 
entirely motivated by the dynamics of the Zeldovich Approximation. 

The $A_3$ lines delineate the sites where all dynamical activities involved in the buildup of the cosmic 
web are taking place. They form the site around which we find the emergence and merger of pancakes, 
the bifurcation of cusps in Eulerian space, and the merger of caustics. The most complex dynamical 
processes take place at the crossings of pancakes. These usually involve singularities of different 
types and usually are the places where most halos are formed. It is worth stressing that the collapse 
is highly non-spherical. This is true for intersections of pancakes and for regions around $D_4$ 
singularities, where $\alpha=\beta$.

When mapped to Eulerian space the $A_3$ lines outline the spine of the cosmic web. In essence,  
as the structure equation~\ref{eq:a3} only involves quantities at one point it 
represents a local definition of the skeleton of the cosmic web. Geometrically it
can be defined either as a set of points where the gradient of one of the eigenvalues is
orthogonal to the corresponding eigen vector or, equivalently, as a set of points where the
eigen vector is tangential to the level contour of the corresponding eigen value. 

In summary, we find that the complex geometry of cosmic structure can be understood to a large extent by the 
$\alpha$-fields of the principal eigenvalue $\alpha$ of the deformation tensor, and through these 
specifically through the outline of the $A_3^{\alpha}$-lines. The role of $\beta$ field is considerably more 
modest. This is largely due to the fact that the first collapse of mass elements is 
directed towards $A_3^{\alpha}$--lines. The $A_3^{\beta}$--lines tend to bend and 
wrap around $A_3^{\alpha}$--lines. Nonetheless, locally the $A_3^{\beta}$--lines
play a significant role in the highest concentrations of mass, ie. the locations 
where halos form.

In this study, we have also made a beginning with the comparison of the spinal 
structure outlined by our geometric description and that found in fully nonlinear 
two-dimensional N-body simulations. By identifying the $A_3$ lines in the (Lagrangian) 
displacement field of the 3D N-body simulations, and mapping them to Eulerian 
space we can compare the resulting network with the cosmic web skeleton that 
emerged in the simulation. We find that it indeed encapsulates quite an accurate 
skeleton of the cosmic web. 

As yet, we do not present a statistical study of the two-dimensional 
case. One of the reasons is that the structure in three dimensions will differ 
significantly from that in two dimensions. We therefore refer such a study to the 
upcoming extension of our analysis to three dimensions. A more extended comparison 
with cosmological N-body simulations is also referred to such a 3D study. 

This study has paved the road towards a profound dynamical analysis of the 
detailed and intricate structural aspects of the cosmic web. We expect that 
its combination with the outcome of large N-body simulations will help us 
towards a considerable improvement of our understanding of the processes 
involved in the buildup of structure, not only limited to the highly nonlinear 
formation and virialization of fully collapsed dark halos but also that of the 
prominent filamentary features that appear to dominate the maps produced 
by large galaxy surveys. 

\section*{Acknowledgments}
We gratefully acknowledge numerous helpful and inspiring discussions with Bernard Jones and 
Gert Vegter. S. Shandarin and R. van de Weygaert acknowledge the support from the 
New Frontiers of Astronomy and Cosmology program at the Sir John Templeton 
Foundation. S. Shandarin and R. van de Weygaert also acknowledge the support by the 
NWO visitor program (grant $\#040.11.344$). In addition, S. Shandarin acknowledges 
the scholar visitor program at ANL where the text of the paper was finalized.


\appendix

\section{Lagrangian submanifold in one-dimensional case}
\label{appendix:1d}
\begin{figure*}
	\label{fig:lagrmanifold}
	\centering
	\centerline{\includegraphics[width=\textwidth]{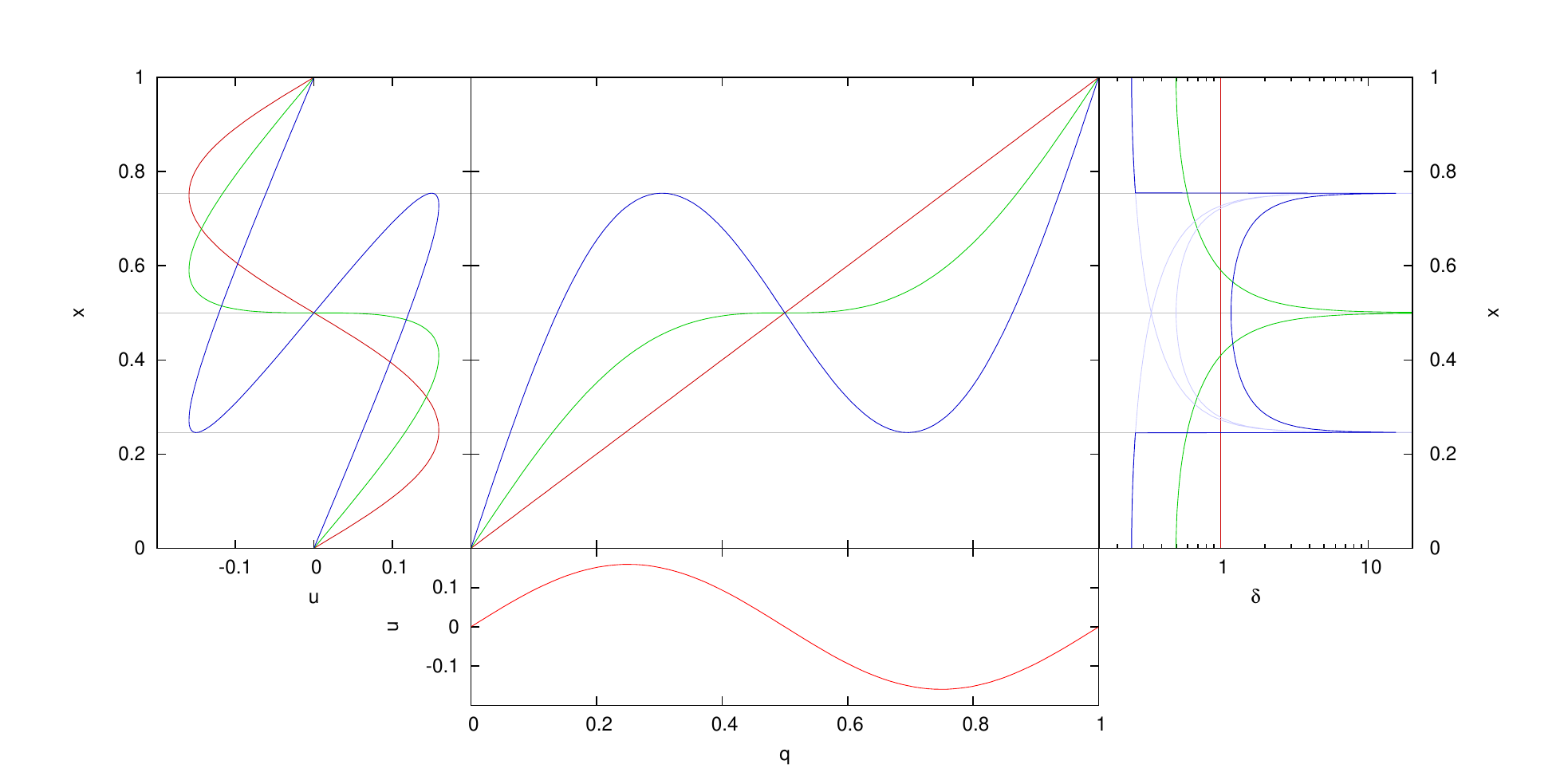}}
	\caption{One-dimensional phase-space manifold evolution. The evolution of 
        singularities in the most simple one-dimensional example of the ZA, a sinusoidal 
        perturbation. In this case we find a single transient $A_3$ cusp singularity (black line), 
        immediately followed by two $A_2$ folds (blue lines). Centre panel: Lagrangian $(q,x)$ manifold 
        at three subsequent timesteps. Left panel: velocity streams, i.e. velocity as function of 
        Eulerian position $x$, around Eulerian catastrophe. Right panel: density structure $delta(x)$ around 
        Eulerian catastrophe. Bottom panel: velocity $u$ of Lagrangian mass elements $q$.}
	\label{fig:zeld1d}
\end{figure*}
The middle panel in the top row of figure \ref{fig:lagrmanifold} shows 
the Lagrangian submanifold embedded in a 
two-dimensional space where the horizontal axis is Lagrangian coordinate $q$ and 
the vertical axis is Eulerian coordiante $x$. Three curves show three qualitatively
different states of the evolution of a sinusoidal perturbation. 
The red diagonal line sows the 
initial state when $x = q$, the green line shows the instant of the origin of
 $A_3$ singularity at the origin. Finally, the blue line shows the state with 
 three streams between two singularities of $A_2$ type. Three panels surrounding
 the central panel shows the initial velocity $ \propto s(q)$ (the bottom panel),
 the evolution of the phase space (the left panel) and the density field 
 in Eulerian space (the right panel). Three density fields are the projections 
 of either the phase-space lines from the left panel or the Lagrangian submanifolds 
 from the central panel on Eulerian coordinate $x$ at the corresponding times. 
 The information
 content of the phase-space and the Lagrangian submanifold is similar but it is 
 greater than that of the density field. Singularities do exist neither in the
 phase space nor in the Lagrangian submanifold, they emerge as a result of the 
 projection to Eulerian space only. The caustics are located at the points where
 the projecting direction is tangent to the phase-space curve or the Lagrangian
 submanifold. All geometrical features just described are similar in two and
 three dimensions. The major difficulty is due to complexity of projection the 
 Lagrangian submanifolds from four- or six-dimentional spaces to two- or 
 three-dimentional Eulerian spaces respectively. As a result the number of 
 archetypical singularities increases with the growth of the number of dimensions.

\section{About loops near $D_4^-$ singularities.}
\label{appendix:d2}
\begin{figure}
	\includegraphics[width=0.48\textwidth]{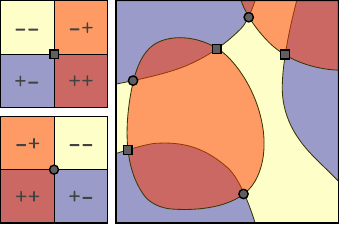}
	\caption{The two different geometries of $D_4$ singularities are
	in the space of $d_{11} - d_{22}$ and $d_{12}$ each others mirrors.
	This illustration shows why topologically there must always be
	as many $D_4^-$ as $D_4^+$ singularities.}
	\label{fig:d4geom}
\end{figure}

\begin{figure*}
	\includegraphics[width=\textwidth]{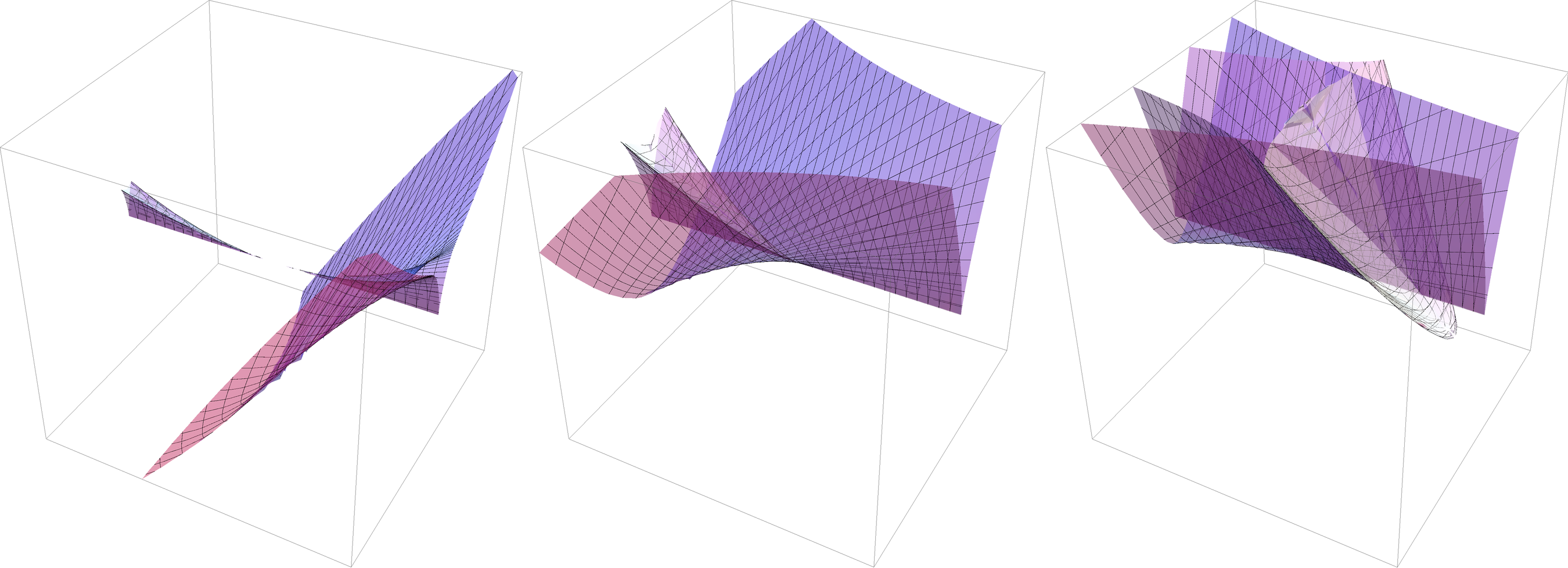}
	\caption{The $D_5$ catastrophe. Here we plotted the caustic surface
	of the catastrophe in 3d. To interpret its significance in 2d, we need
	to take a cross-section of this image, moving in time form the front-right 
	plane to the left-back or vice versa. In this sense each of these three
	images shows an evolution history of caustics in 2d. The 3d evolution shows
	us how this evolution might change if we smoothly change some parameter
	determining the character of the 2d field. For example by slowly changing
	some of the phases in a random realisation of the potential, we can change
	a $D_4^-$ on the left into a $D_4^+$ on the right.}
	\label{fig:d5}
\end{figure*}

The small loops near $D_4^-$ singularities form a salient feature in our results.
In this appendix we show how this result can be understood in terms of the higher
order $D_5$ catastrophe. The $D_5$ catastrophe in 3-D marks the transition from a $D_4^-$ to
$D_4^+$ catastrophe. As we mentioned before the $D_4$ singularity is a location where both
eigenvalues are equal, and we derived the following equation from the deformation tensor
$d_{ij}$, that if $\alpha(\vec{q}) = \beta(\vec{q})$, then
\begin{equation}
	d_{11} = d_{22}\quad {\rm and}\quad d_{12} = 0.
\end{equation}
To determine the type we had to look at the higher derivatives of the potential.

Formally the amount of $D_4^+$ and $D_4^-$ singularities are exactly equal. This can be
seen from the topology of the method by which we find them. The two fields
$U_1 = d_{11} - d_{22}$ and $U_2 = d_{12}$ are Gaussian and uncorrelated. 
The zero-levelset is the boundary between regions that are negative and positive. 
A point where both fields are zero exists then in two kinds, illustrated in figure
\ref{fig:d4geom}. In practice, some of these points have a corresponding eigenvalue
that is negative, so these points are not considered in our figures.

A $D_5$ singularity corresponds to the situation where the zero-sets of 
$U_1$ and $U_2$ touch exactly\footnote{In the language of geometry, 
to have a stable configuration, the sets must cross \emph{transversally}.}
. In general this never happens, however
the probability that a $D_4$ point is \emph{close} to that, is very high.
This means that the global geometry of a $D_4^-$ tends to look a lot like
that of a $D_5$ singularity. 

If we now look at figure \ref{fig:d5}, we can
see what a geometry close to a $D_5$ looks like. In the first of the three images,
if we take an intersection and move it from right to left, we start with a
single cusp. This cusp is then joined by two swallowtails, to form a
triangle of cusps. This triangle contracts to a $D_4^-$, and then expands
again as a triangle of cusps in the second eigenvalue. By the same recipe we can
approach the $D_4^+$ catastrophe, seen in the right-most panel in figure \ref{fig:d5}. 
We start out with an $\alpha$-cusp. This cusp is joined
by a $\beta$-pancake, which is soon thereafter pierced by the $\alpha$-cusp. The
cusp ``changes colour'' and through the $D_4^+$ becomes a $\beta$-cusp.
Both these scenarios are frequently encountered in the ZA, even if they don't
result in an eventual $D_4$ singularity, as is evident from figures \ref{fig:alphaA3m},
\ref{fig:A4}, \ref{fig:A4connectivity}, \ref{fig:D4purse} and \ref{fig:D4pyramid}.

\bsp

\label{lastpage}

\end{document}